\documentclass[fleqn,usenatbib]{mnras}

\usepackage[T1]{fontenc}

\usepackage{xcolor}

\usepackage{graphicx}	
\usepackage{amsmath}	
\usepackage{amssymb}	

\usepackage{newtxtext,newtxmath}

\DeclareSymbolFont{matha}{OML}{txmi}{m}{it}
\DeclareMathSymbol{\varv}{\mathord}{matha}{29}

\def\del#1{{}}

\title[Two striking head-tail galaxies in the galaxy cluster IIZW108]{Two striking head-tail galaxies in the galaxy cluster IIZW108: insights into transition to turbulence, magnetic fields and particle re-acceleration}
\author[M\"uller et al.]{Ancla M\"uller,$^{1}$\thanks{E-mail: amueller@astro.rub.de}
Christoph Pfrommer,$^{2}$
Alessandro Ignesti,$^{3}$
Alessia Moretti,$^{3}$
\newauthor
Ana Louren\c{c}o,$^{4}$
Rosita Paladino,${^5}$
Yara Jaff\'{e},$^{4}$
Myriam Gitti,$^{5,6}$
\newauthor
Tiziana Venturi,$^{5}$ 
Marco Gullieuszik,$^{3}$
Bianca Poggianti,$^{3}$
Benedetta Vulcani,$^{3}$
\newauthor
Andrea Biviano,$^{7,8}$
Björn Adebahr,$^{1}$
and
Ralf-Jürgen Dettmar$^{1}$
\\
$^{1}$Ruhr University Bochum, Faculty of Physics and Astronomy, Astronomical Institute, Universit\"atsstr.150, 44801 Bochum, Germany\\
$^{2}$Leibniz-Institute for Astrophysics Potsdam (AIP), An der Sternwarte 16, 14482 Potsdam, Germany\\
$^{3}$INAF-Osservatorio Astronomico di Padova, Vicolo dell'Osservatorio 5, I35122 Padova, Italy \\
$^4$Instituto de F\'{i}sica y Astronom\'{i}a, Universidad de Valpara\'{i}so, Avda. Gran Breta\~{n}a 1111 de Valpara\'{i}so, Chile\\
$^5$INAF, Istituto di Radioastronomia di Bologna, Via Gobetti 101, 40129 Bologna, Italy\\
$^6$Dipartimento di Fisica e Astronomia, Università di Bologna, Via Gobetti 93/2, 40129 Bologna, Italy\\
$^{7}$INAF-Osservatorio Astronomico di Trieste, via G. B. Tiepolo 11, 34143 Trieste, Italy \\
$^{8}$IFPU-Institute for Fundamental Physics of the Universe, via Beirut 2, 34014 Trieste, Italy\\
}

\date{Accepted 2021 October 6. Received 2021 October 1; in original form 2021 July 28}

\pubyear{2021}

\begin{document}
\label{firstpage}
\pagerange{\pageref{firstpage}--\pageref{lastpage}}
\maketitle

\begin{abstract}
We present deep JVLA observations at 1.4\,GHz and 2.7\,GHz (full polarization), as well as optical OmegaWINGS/WINGS and X-ray observations of two extended radio galaxies in the IIZW108 galaxy cluster at $z=0.04889$. They show a bent tail morphology in agreement with a radio lobed galaxy falling into the cluster potential. Both galaxies are found to possess properties comparable with {narrow-angle} tail galaxies in the literature even though they are part of a low mass cluster. We find a spectral index steepening and an increase in fractional polarization through the galaxy jets and an ordered magnetic field component mostly aligned with the jet direction. This is likely caused by either shear due to the velocity difference of the intracluster medium and the jet fluid and/or magnetic draping of the intracluster medium across the galaxy jets. We find clear evidence that one source is showing two active galactic nuclei (AGN) outbursts from which we expect the AGN has never turned off completely. We show that pure standard electron cooling cannot explain the jet length. We demonstrate therefore that these galaxies can be used as a laboratory to study gentle re-acceleration of relativistic electrons in galaxy jets via transition from laminar to turbulent motion.
\end{abstract}




\section{Introduction}
\label{sec:intro}
Radio galaxies found in cluster environments often show distorted morphologies due to their interaction with the intracluster medium (ICM). In particular, a class of {so-called} head-tail (HT) galaxies {has} been discussed over decades \citep{Odea,Mao,Gregory,Cuciti,Srivastava}. They appear with a very luminous head (the AGN in the {centre} of the hosting galaxy) and extended jets that are bent by the ram pressure experienced by the galaxy due to its motion through the ICM. Typically, the jets are fanning out and expand in the form of lobes in the wake of the galaxy. HT galaxies are generally classified as Fanaroff-Riley I (FRI, see \citealt{Fanaroff}) objects with a radio power $<10^{25}$\,W\,Hz$^{-1}$ at 1.4\,GHz and exhibit a luminous head and fainter jets (for more details, see also \citealt{Saripalli}). These galaxy types can be divided into two subclasses: 1)  {narrow-angle} tail (NAT) jets experiencing high ram {pressure} \citep{Venkatesan} and 2) {wide-angle} tail (WAT) jets that experience a weaker ram pressure, which could be caused by low velocities relative to the cluster {centre} as a result of cluster-cluster mergers \citep{Klamer}. It is also found that the morphology of WATs cannot be explained {only} due to the motion of the host galaxy but by its combination with the bulk motion of the dense ICM in the central cluster regions \citep{Bliton,Giacintucci}. These systems are useful to probe different aspects of the interaction of the radio source with the ICM and to study general properties of the ICM as well as cluster and galaxy evolution in general.

HT galaxies have the same characteristics in rich and poor clusters \citep{Venkatesan}. The velocity ($\sim 600$\,km/s) of them{,} as well as the density of the surrounding medium ($\sim 10^{-4}$\,cm$^{-3}$ in poor clusters){,} is therefore found to be an important factor in shaping such jets. Radio continuum studies have shown a general steepening of the spectral index from the head (showing very flat spectra $j_\nu\propto \nu^\alpha$ with $\alpha = -0.4$) towards the tail (with spectral indices up to $\alpha = -2.5\;\textrm{to}\;-4$, see \citealt{Cuciti,Srivastava}). While the flat spectra at the head signal {can be caused by} electron acceleration at strong shocks, possibly in combination with free-free absorption \citep{McKean2016}, the spectral steepening along the jet is in agreement with radiative {ageing} of the electrons. However, also substructures with high surface brightness within the jet can be found, which may signal shock compression \citep{Cuciti} or re-energization \citep{Gasperin} that influence the electron spectra as well as the magnetic field strength and structure. The NATs exist with a large variety of opening angles between the two jets \citep{Odea} and jet length (30\,kpc up to 900\,kpc, see \citealt{Odea,Srivastava}). \citet{Odea} also found a correlation between the intensity of the head and the total NAT luminosity, which is larger in smaller clusters as inferred from the original richness class of Abell clusters \citep{Abell}.

The radio total power emission traces the total magnetic field. The regular magnetic field component can be investigated using radio polarization measurements. While strength and orientation of the magnetic field component in the plane of the sky {are} directly given by the intensity of the polarised emission and its angle, respectively, the component along the line-of-sight can be estimated from the Faraday rotation measure $RM  = 0.81 \int_0^L n_e(s) \mathbfit{B}(s)\bmath{\cdot} d\mathbfit{s}$ \citep[e.g.,][]{Brentjens}, where $RM$ is the rotation measure in rad m$^{-2}$, $\mathbfit{s}$ is the coordinate along the line of sight, $L$ is the distance from the source to us in parsec, $n_e$ is the electron density along the line-of-sight in particles per cm$^3$ and $\mathbfit{B}$ is the magnetic vector field with the field strength measured in units of $\upmu$G.
Such polarization studies of HT galaxies are still rare but \citet{Miley} already found significant linear polarisation along the jet with a {steady rising in} the degree of polarization. However, no correction for Faraday rotation was applied to the data. More recently, \citet{Klamer} also found the degree of polarization to increase along the jet, reaching values up to 60\,\%, while the core region shows values about 10\,\%. The magnetic field configuration (ordered component) was found to be aligned with the jet flow. Slight disruptions could be measured associated with a bent jet along the line of sight.

On a broader perspective, AGN jet interactions with the ICM have received great interest because the interplay of cooling gas, subsequent star formation, and nuclear activity appears to be tightly coupled to a self-regulated feedback loop \citep[for reviews, see][]{McNamara2007,McNamara2012,2012AdAst2012E...6G,2016NewAR..75....1S}. The physics of this interaction regulates the amount of cooling and star formation via dissipation of mechanical heat by outflows, lobes, or sound waves from the AGN \citep[e.g.,][]{2001ApJ...554..261C, 2002Natur.418..301B,2002ApJ...581..223R,2004ApJ...611..158R,Brueggen2005,2012MNRAS.424..190G} or due to cosmic rays escaping from the jet lobes that resonantly drive Alfv\'en waves, which are damped and thereby heat the ambient cooling ICM \citep{Guo2008,Pfrommer2013,Jacob2017a,Jacob2017b,Ruszkowski2017,Ehlert2018}. Simulations of interactions of AGN jets with a realistic magnetized and/or turbulent ICM \citep{Brueggen2005,Heinz2006,Sijacki2008,O'Neill2010,Mendygral2011M,Weinberger2017,Bourne2017,Bourne2019,Ehlert2021} enable {to probe} the characteristics of ICM turbulence, the filling of the AGN jet lobes through the Sunyaev-Zel'dovich effect \citep{Pfrommer2005,Ehlert2019}, and enable to indirectly infer the dynamical state of the local ICM through observations of the radio morphologies of HT galaxies \citep{PfrommerII,Jones2017}. In particular, simulations of NAT galaxies with passive cosmic ray electrons \citep{O'Neill2019} that encounter the passage of (possibly oblique) shocks \citep{O'Neill2019shocks,Nolting2019a,Nolting2019b} can be used to produce synthetic radio observations and to study turbulence and particle acceleration in the jet lobes.

In this paper{,} we give detailed insights into the radio continuum and polarization properties of two NAT galaxies that are members of the same low-mass cluster IIZW108. We will use this information in addition to archival X-ray data and a detailed optical study to analyse the non-thermal component in these AGN jets experiencing ram-pressure interaction. While a continuous steepening in the spectral index indicates progressive {ageing} of electrons as they propagate along the galaxy jets, a deviation from this {behaviour} indicates a more complex situation. For example, flattening of the spectral index can be caused by internal shocks or turbulent re-acceleration, and a contribution by an external source in projection along the line-of-sight is also a possibility. By carefully analysing the radio morphology and spectral properties of the tails, we study a possible transition from laminar to turbulent motion in AGN jets and assess the consequences for particle (re-)acceleration and comment on the AGN duty cycle.

The paper is structured as follows: Section~\ref{sec:data} gives an overview of the observations and different data products used through this study. In Section~\ref{sec:analysis} we present the continuum and polarization results for the two HT galaxies as well as a detailed study of the spectral index and fractional polarization evolution through the jets. We derive ICM properties via a de-projected X-ray study. In Section~\ref{sec:discussion} we discuss the implications of our results by studying the effect of turbulent motion, the resulting electron cooling length and multiple (two) AGN cycles and conclude in Section~\ref{sec:conclusion}. We assume a standard flat $\Lambda$CDM cosmology with H$_{0}$ = 70, $\Omega_{\Lambda}$ = 0.7 and $\Omega_{M}$ = 0.3. With a cluster central redshift of 0.04889 \citep{Moretti2017} this results in a scale of 0.957\,kpc/$\arcsec$.

\section{Data}
\label{sec:data}

\subsection{Observed targets}
\begin{figure*}
    \begin{minipage}[t]{.49\textwidth}
        \centering
        \includegraphics[width=\textwidth]{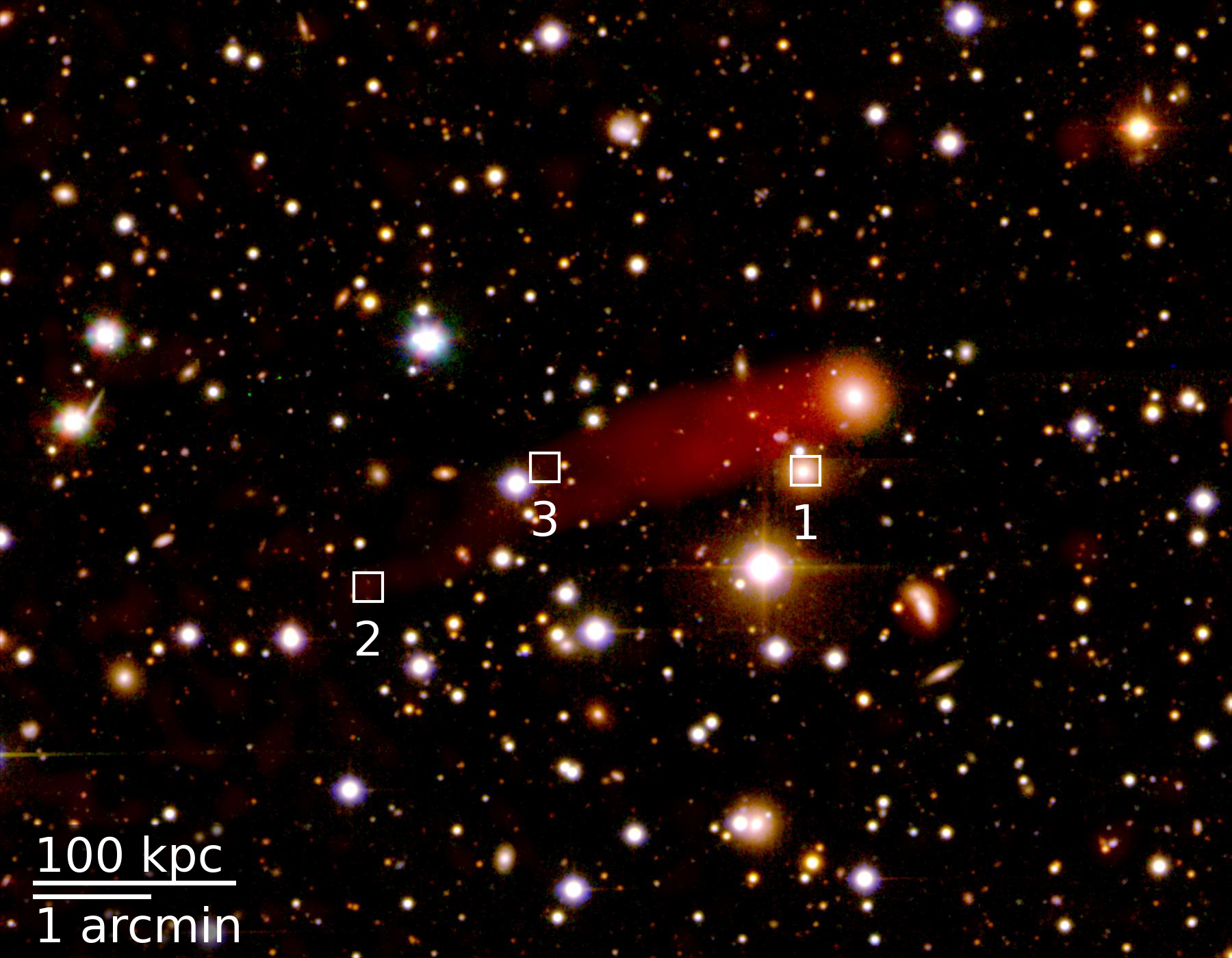}
    \end{minipage}  
    \hfill
    \begin{minipage}[t]{.49\textwidth}
        \centering
        \includegraphics[width=\textwidth]{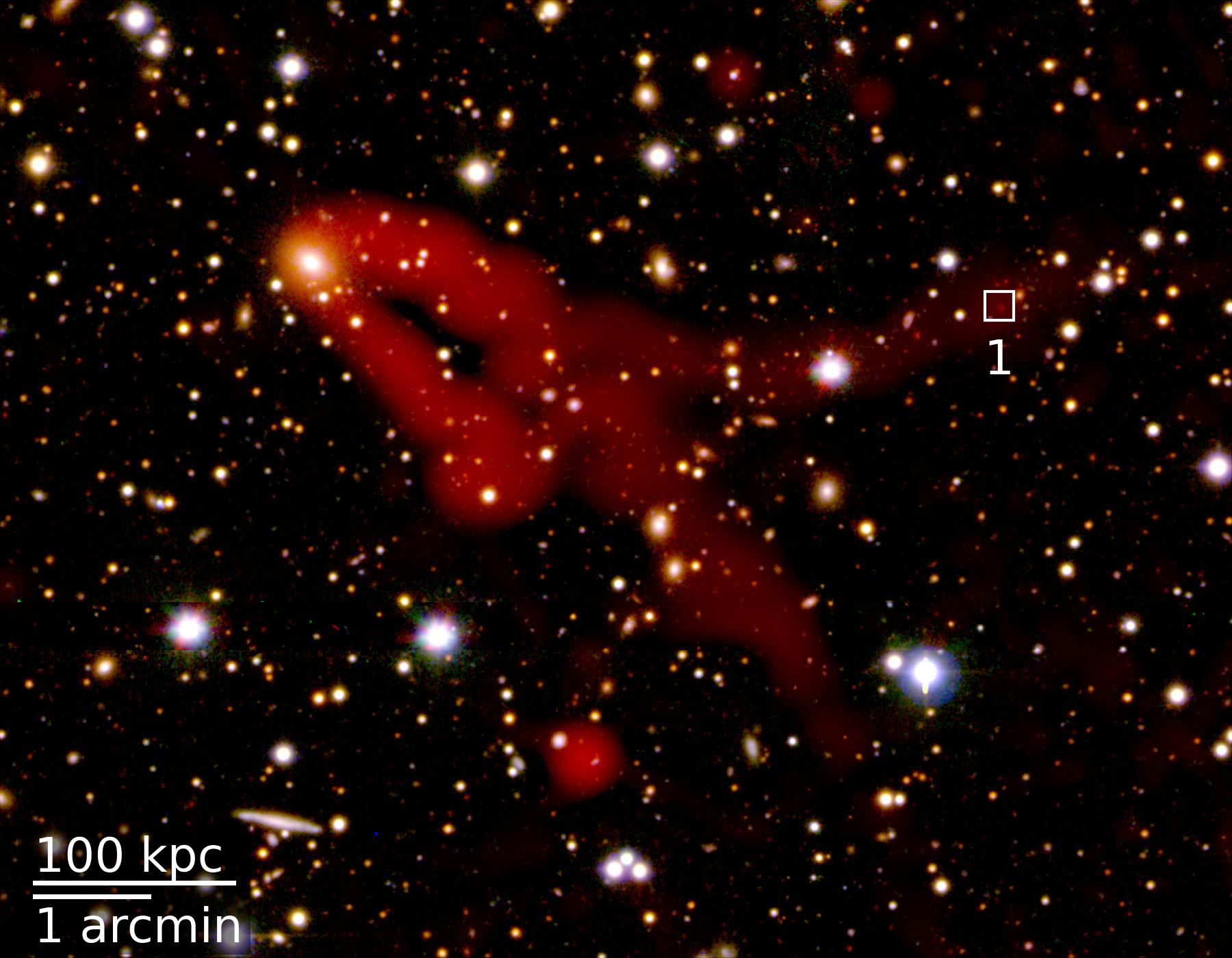}
    \end{minipage} 
    \caption{Radio continuum emission is superimposed in red to RGB images obtained by combining V-, B-, and u-band images from WINGS/OmegaWINGS. HTI and HTII are shown in the left and right panels, respectively. The regions discussed as possible optical sources producing line-of-sight emission are marked with {white} squares.} \label{fig:optical}
\end{figure*}
\begin{table*}
\def\arraystretch{1.2}
    \centering
\begin{tabular}[t]{r|r|r|r|r|r|r|r}
 WINGS ID and name  & RA, Dec & redshift & $\log(M/M{\odot}$)& SFH [M$_\odot$~yr$^{-1}$]& $d_\textrm{jet}$ [kpc] & $d_\textrm{cl}$ [kpc]\\
  \hline
 WINGSJ211406.64+022743 [HTI]  & 21h14m06.629, +2d27m43.48 & 0.04699 & 11.5 & [0; 5.9; 3.9; 47] & 300 & 390 \\
 WINGSJ211326.68+023123 [HTII]  & 21h13m26.669, +2d31m23.36 & 0.04814 & 11.5 & [0; 3.3; 7.8; 43] & 560 & 440 \\
\end{tabular}
    \caption{General properties: WINGS ID + [name used through the text], position of the central AGN ('head'), galaxy redshift, galaxy stellar mass, average star-formation rate in 4 age intervals given from youngest to oldest, projected maximum jet length observed at 1.4\,GHz, and projected cluster-centric distance $d_\textrm{cl}$}
    \label{tab:1}
\end{table*}
We detected two extended HT galaxies in the IIZW108 cluster, which are included in the field of view of two observations targeting a galaxy experiencing extreme ram-pressure stripping, the jellyfish galaxy JO206 \citep{Mpati,Mueller} with the Karl G. Jansky Very Large Array (JVLA). In Table~\ref{tab:1} general information about the two HT galaxies can be found.
IIZW108 is a low mass cluster, characterized by a line-of-sight velocity dispersion of 575\,km~s$^{-1}$ \citep{Biviano2017} that translates into a three-dimensional velocity dispersion of $\sim 1000$\,km~s$^{-1}$. The two HT galaxies have an optical counterpart detected in \citet{Marco2015} from which we take the ID given in Table \ref{tab:1}. For each galaxy we also give the spectroscopic redshift from \citet{Moretti2017}, the aperture corrected, total stellar mass (assuming a Salpeter IMF between 0.15 and 120 solar masses) and the average SFH in 4 age bins (1: 0-20~Myr; 2: 20-600~Myr; 3: 0.6-5.6~Gyr; 4: >5.6~Gyr). The first galaxy presented in Table~\ref{tab:1} (HTI) is moving with a line-of-sight velocity of 480\,km~s$^{-1}$ with respect to the cluster centre at a projected distance of 390\,kpc while the second galaxy shows a line-of-sight velocity component of 140\,km~s$^{-1}$ with respect to the cluster centre, significantly lower than the velocity of HTI, and it is located at a projected distance of 440\,kpc. 
In the following, we name the first galaxy of Table~\ref{tab:1} (in the first row) HTI and the second galaxy HTII.

\subsection{Radio data} 

We obtained HI and L-band (1.4\,GHz) continuum data with a bandwidth of 31.25\,kHz \citep[see][for further details on the L-band data reduction]{Mpati} and S-band (2.7\,GHz) full polarization data with a bandwidth of 2\,GHz both on a single pointing in C configuration. The 2.7\,GHz data reduction was carried out using the Common Astronomy Software Application \citep[CASA,][]{CASA}, flagging was performed to reduce radio frequency interference (RFI) beforehand with AOFlagger \citep{Offringa}. We identified additional RFI during calibration, applied manual flagging techniques and re-calibrated the data. Self calibration phase-only and imaging was performed in Miriad \citep{Miriad} in an iterative manner. We decreased the solution interval and increased the ($u,\varv$)-range (including shorter spacings) and improved the source masks within the self calibration process until a 30\,s interval was reached including all baselines. Self-calibration was carried out using robust weighting -2 to reach the highest resolution possible (see below), however, the lower resolution images ($15\arcsec\times15\arcsec$), shown later on, are finally imaged with a robust weighting 0 (to be comparable to the 1.4\,GHz data images with the same weighting), and convolved to the actual 1.4\,GHz resolution, which is also the best compromise of resolution and sensitivity in polarized intensity. 
For a dedicated polarization study, the 2\,GHz bandwidth of the 2.7\,GHz polarization data were imaged channel-wise in Stokes Q and U. An image cube was created and corrected for Faraday rotation via rotation measure (RM) synthesis \citep{Brentjens}. 
The highest resolution images ($4.1\arcsec\times4.6\arcsec$ at 2.7\,GHz) were obtained using a multi frequency synthesis clean and a robust weighting of -2 including the whole bandwidth at once on the already self-calibrated data. New source masks were created and another self calibration process was performed on the cross-calibrated data. We performed a primary beam correction on all images. 
More details about the 2.7\,GHz observation can be found in \citet{Mueller}. 
High resolution images are only made for the 2.7\,GHz data to study the morphology of the head and tails in more detail later on. Such a high resolution and sensitivity to the jet emission cannot be achieved using the 1.4\,GHz data. 

In the following, we define the central AGN, referred to as 'head', by a beam-sized region centred on the peak flux (the properties are measured therein). The galaxy itself (including the extended jets) is confined by the noise-level (3\,$\sigma$/4.5\,$\sigma$ in total/polarized intensity) given in Table~\ref{tab:3}. Therefore, the jet properties are defined by the corresponding noise-level from which we subtract the results of the head.
\begin{table*}
\def\arraystretch{1.2}
    \centering
\begin{tabular}[t]{r|r|r|r|r}
  name & \multicolumn{1}{c}{$I_{1.4}$} & \multicolumn{2}{c}{$I_{2.7}$} & \multicolumn{1}{c}{$PI_{2.7}$} \\
  \hline
  HTI & $15\arcsec\times15\arcsec$, 130\,mJy & $15\arcsec\times15\arcsec$, 35\,mJy /& $4.1\arcsec\times4.6\arcsec$, 20\,mJy & $15\arcsec\times15\arcsec$, 18\,mJy \\
  HTII & $15\arcsec\times15\arcsec$, 130\,mJy & $15\arcsec\times15\arcsec$, 35\,mJy /& $4.1\arcsec\times4.6\arcsec$, 33\,mJy & $15\arcsec\times15\arcsec$, 45\,mJy \\
\end{tabular}
    \caption{Resolution, noise level (3\,$\sigma$ for total intensity $I$ and 4.5\,$\sigma$ for polarized intensity $PI$) of the final data sets measured in a source-free region close to the most distant jet emission at 1.4\,GHz and 2.7\,GHz.}
    \label{tab:3}
\end{table*}

\subsection{X-ray and optical data}

To study the thermal cluster properties we will use the X-ray observations that we already investigated in \citet{Mueller}. There, we analyzed a 10.1\,ks archival {\it Chandra} observation. In order to infer the local density and temperature of the ICM surrounding JO206, we carried out a de-projected spectral analysis by assuming a spherical symmetry for the system. We refer the reader to the supplementary information of \citet{Mueller} for further details about the data analysis.

IIZW108 has also a WINGS/OmegaWINGS optical coverage (both photometric and spectroscopic) that we used to infer further galaxy properties. 
The WINGS survey is generally described in \citet{WingsI}, and its complete data set is described in \citet{Moretti+14}. It includes optical photometry and spectroscopy, as well as its u-band \citep{Omizzolo+14} and near-infrared \citep{WingsIII} extensions. The OmegaWINGS photometric survey \citep{Marco2015}, covers a wider area around the cluster {centre} (1 deg$^2$), with the optical V and B photometry with a limiting magnitude of V$=$23.1, and has a corresponding u-band coverage \citep{WingsII}.
A combined image of the two HT galaxies in the u, V, and B filter is shown in Fig.~\ref{fig:optical}. Spectroscopic information is derived by means of AAOmega fibre spectroscopy described in \citet{Moretti2017}, which contains redshifts for 597 galaxies in this cluster, out of which 185 are considered cluster members. On the blue part of these spectra, we run the SINOPSIS spectrophotometric code \citep{Fritz2011} deriving the galaxy masses, their spectral type and the Star Formation History (SFH) for 156 galaxies (Moretti et al., in prep.).

\section{Analysis}
\label{sec:analysis}

\subsection{AGN activity and cluster environment in the optical}\label{sec:optical}

\begin{figure}
    \centering
    \includegraphics[width=0.5\textwidth]{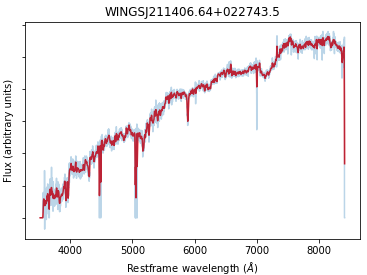}
    \includegraphics[width=0.5\textwidth]{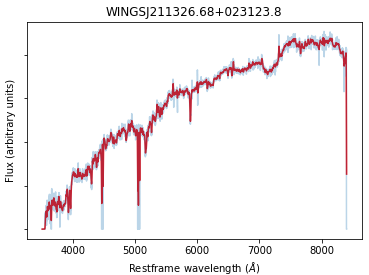}
    \caption{AAOmega optical spectra of the two HT galaxies: WINGSJ211406.64+022743 [HTI] (top panel) and WINGSJ211326.68+023123.8 [HTII] (bottom panel). The red spectra are Gaussian smoothed with a 3-pixel kernel.}
    \label{fig:optspec}
\end{figure}

Based on the optical data, both HT galaxies turned out to be massive (log(M/M$_{\odot})=11.5$), {elliptical galaxies with a k-type spectrum (i.e. } showing no emission lines and weak Balmer lines in absorption), with a mass--weighted age of $\sim 10$\,Gyr.
No sign of ongoing star formation is present, while most of the stellar mass formed in the oldest age bin, i.e., more than 5.6\,Gyr ago.
The two optical spectra can be seen in Fig.~\ref{fig:optspec}.
There is no sign of an ongoing AGN activity (no emission lines), which would leave traces in the optical, though, for a very short time. 
\begin{figure*}
    \centering
    \includegraphics[width=1\textwidth]{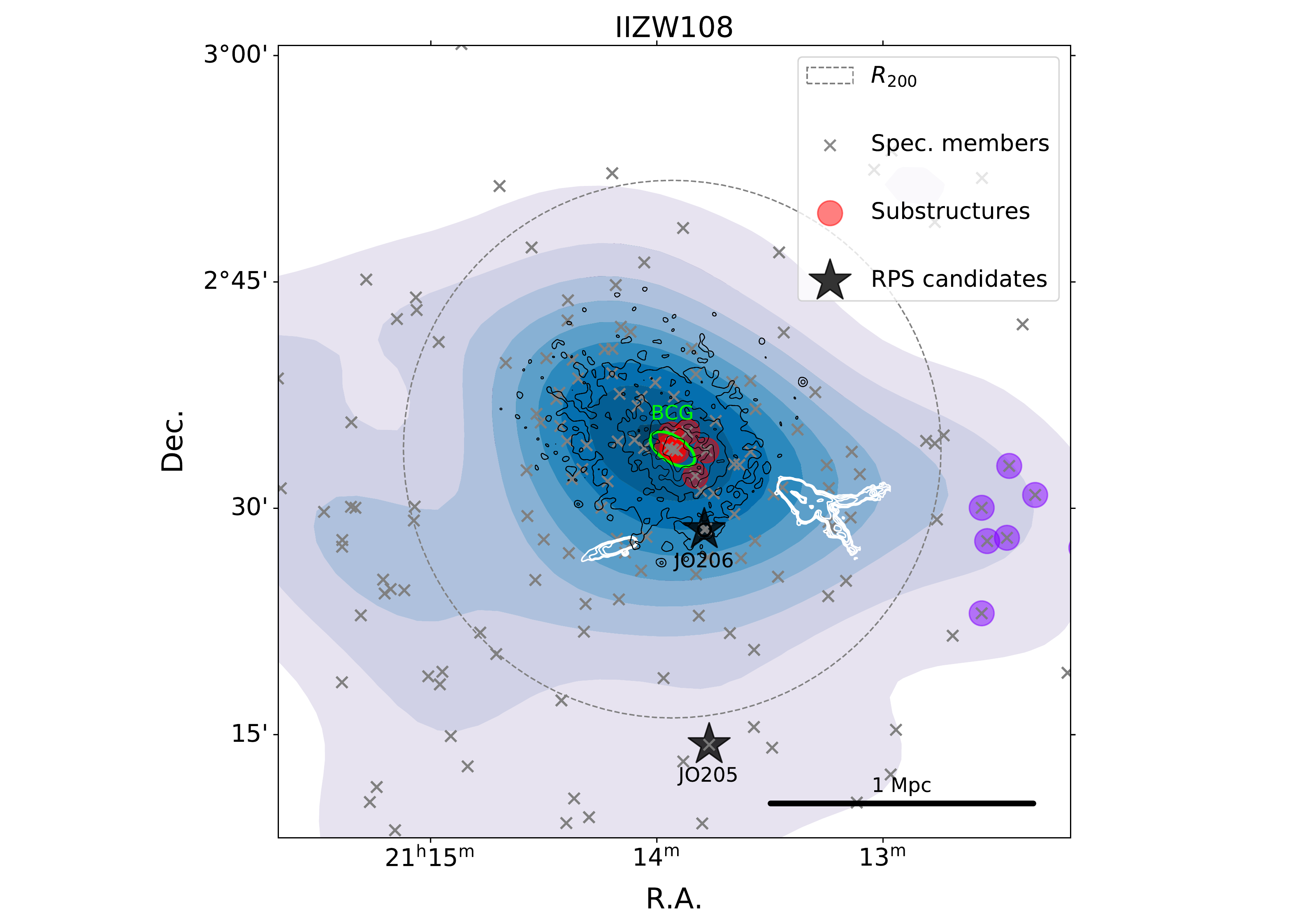}
    \caption{Location of the HT galaxies with respect to the cluster members ({grey} Xs) and optical substructure (coloured dots). The BCG is represented by a green ellipse (with ellipticity and position angle taken from HyperLEDA) and the density of members is shown as blue contour levels. We show the contours of the radio emission of the head-tail galaxies (white), and of the X-ray emission (black). The jellyfish galaxies are indicated by a black star symbol.}
    \label{fig:environment}
\end{figure*}
%

In previous substructure analysis for the WINGS sample, the IIZW108 cluster was found to be fairly relaxed having only one optical substructure identified by DEDICA, an adaptive-kernel technique that searches for substructures in two dimensions \citep{Ramella_2007}. Substructures within clusters are the product of the hierarchical assembly of clusters which can leave imprints on the ICM and affect the evolution of galaxies. When seen in X-rays (see also Sect. \ref{sec:xray}), IIZW108 seems mildly elongated without a particularly strong concentration. Such imprints can also be identified by the analysis of optical substructures.
Figure \ref{fig:environment} shows the position of the HT galaxies in the cluster. The red-filled circles show a small optical substructure near the core of IIZW108 found by a modified \citep{Dressler_Shectman_1988} test detailed in the Appendix A of \citet[][\texttt{DS+} method]{Biviano2017} {and the purple-filled circles correspond to a substructure located more in the filament}. The ICM is represented by the black contours (X-rays) and the galaxy density is seen as blue contours (tracing the spectroscopic members: crosses). The galaxy density shows an elongated profile towards the {southwest} which is also aligned with the slight elongation seen in X-rays and interestingly, also with the BCG major axis. BCGs are believed to grow by accreting galaxies preferentially along the direction of the main cluster filaments as the clusters grow hierarchically, and hence their major axis tends to align with the cluster \citep[e.g.,][]{Binggeli_1982,Durret_1998}.
It is therefore possible that a filament could be feeding the cluster from the south west, in the direction of HTII, which is likely falling into the cluster for the first time. The radio morphology of HTII supports this idea and suggests the galaxy is moving towards the cluster {centre} quite radially with a significant velocity component in the plane of the sky (and so is HTI from the {southeast}). 
%

\subsection{HTI}

\begin{figure*}
    \begin{minipage}[t]{.49\textwidth}
        \centering
        \includegraphics[width=\textwidth]{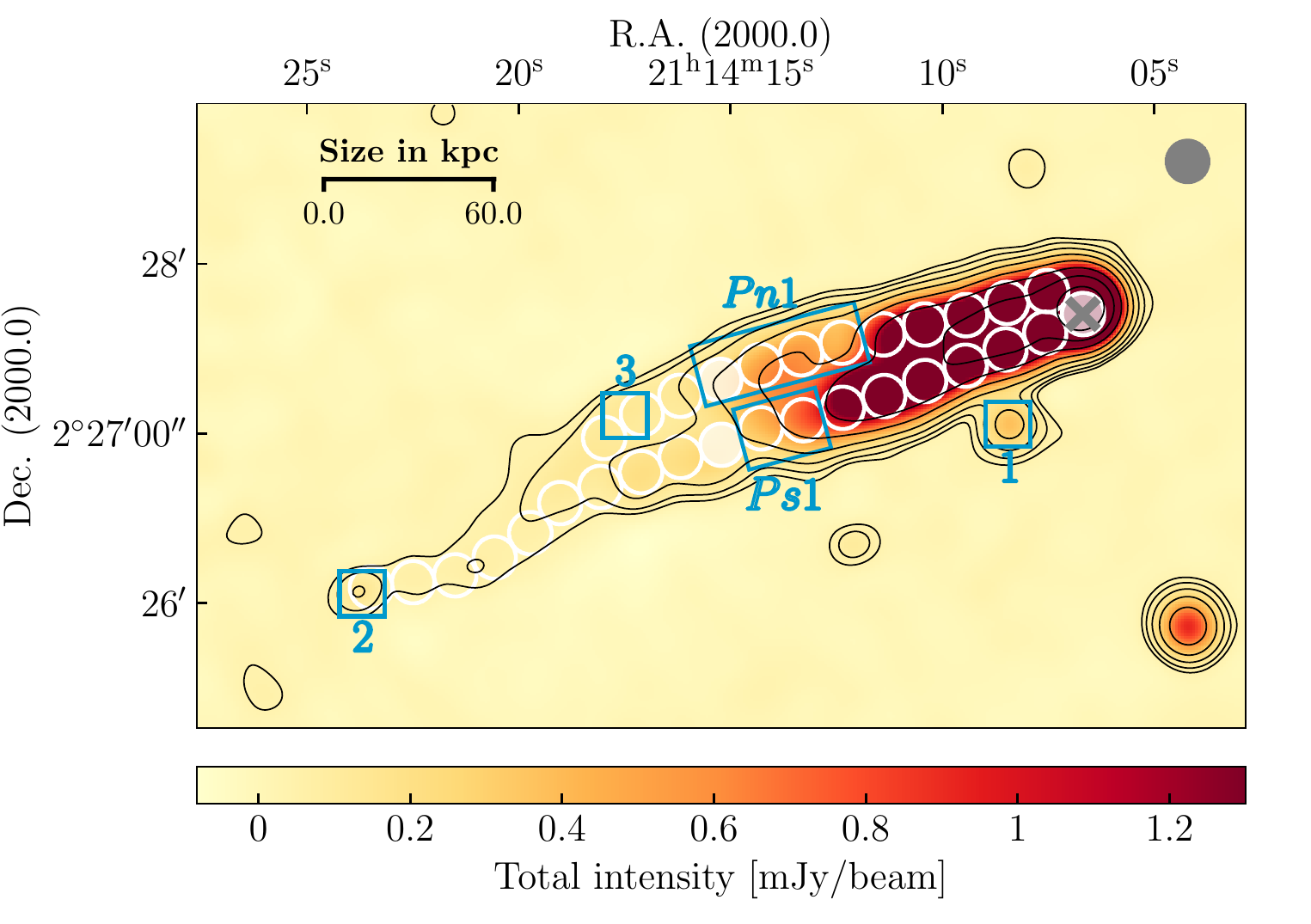}
    \end{minipage}  
    \hfill
    \begin{minipage}[t]{.49\textwidth}
        \centering
        \includegraphics[width=\textwidth]{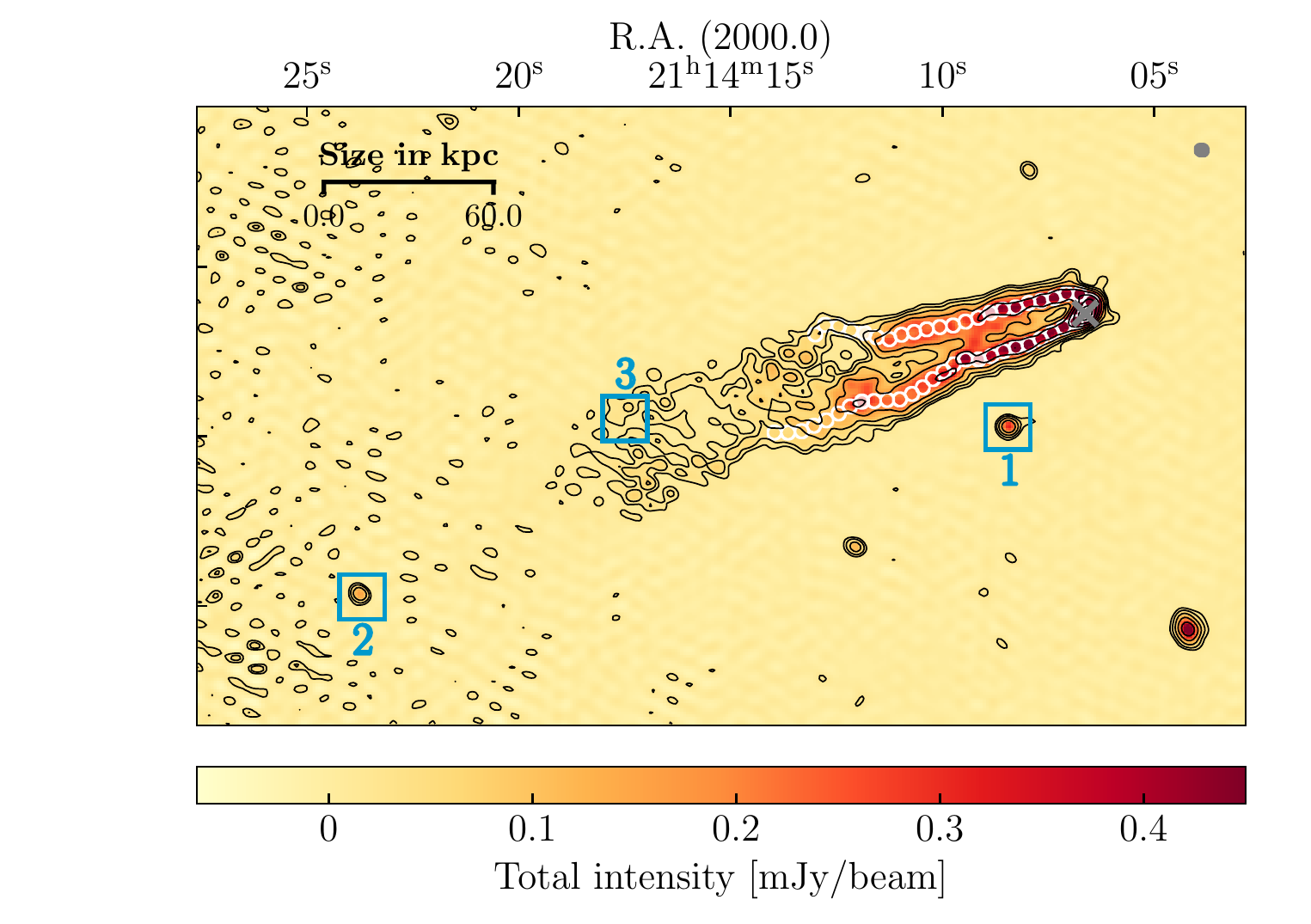}
    \end{minipage} 
     \begin{minipage}[t]{.49\textwidth}
        \centering
        \includegraphics[width=\textwidth]{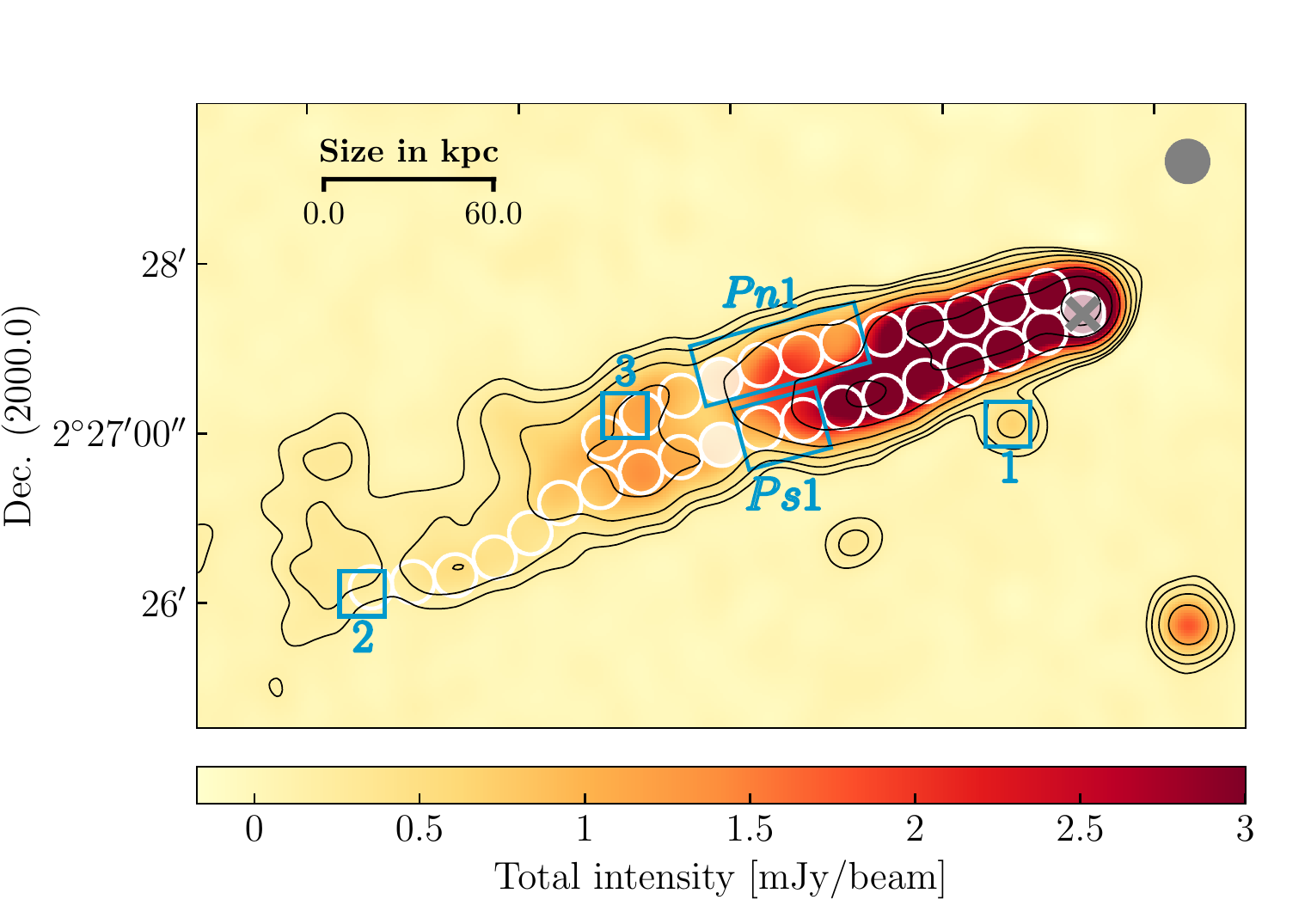}
    \end{minipage}  
    \hfill
    \begin{minipage}[t]{.49\textwidth}
        \centering
        \includegraphics[width=\textwidth]{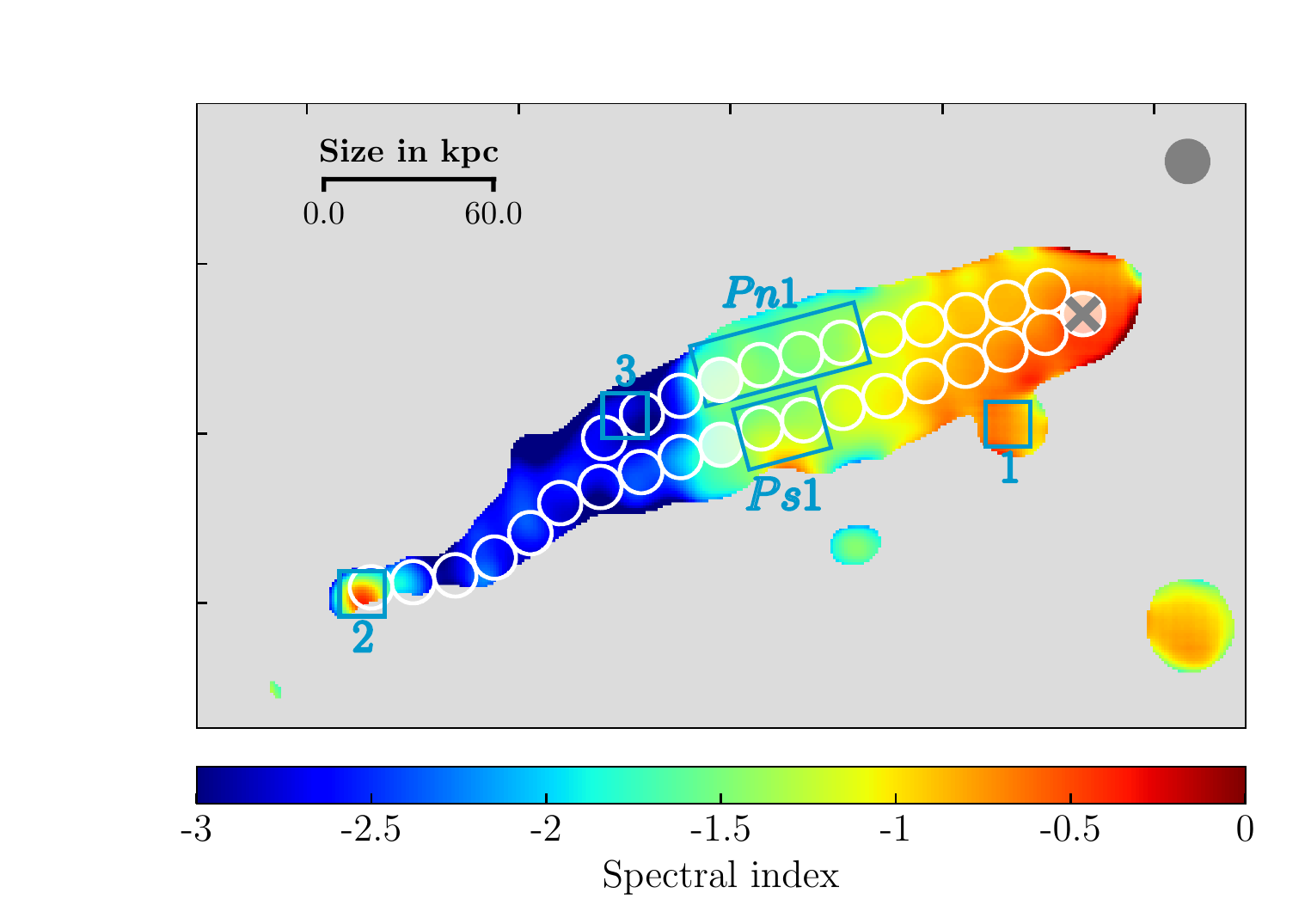}
    \end{minipage}
    \caption{HTI. Continuum results of the 2.7\,GHz (top) and 1.4\,GHz (bottom) data at a resolution of $15\arcsec\times15\arcsec$ if not stated differently. White circular beam-sized regions are superimposed along the jet to track different properties centred on the central AGN marked by a grey cross. Each 10th region is filled in white to ease the comparison with Fig.~\ref{fig:trackHTI}. Regions discussed as possible optical sources producing line-of-sight emission {(`1', `2', `3')} and {plateaus (`Pn1', `Ps1')} are marked with blue {squares and rectangles, respectively}. The beam is shown in grey in the top right corner and {lightgrey} area contains no information. \textbf{Top left}: 2.7\,GHz total intensity map with superimposed contour levels of $\epsilon \times (1,\;2,\;4,\;8,\;16,\;32,\;64,\;128,\;256$) with $\epsilon = 35\,\upmu$Jy/beam. \textbf{Top right}: 2.7\,GHz total intensity map with superimposed contour levels of $\epsilon \times (1,\;2,\;4,\;8,\;16,\;32,\;64,\;128,\;256$) with $\epsilon = 20\,\upmu$Jy/beam at a resolution of $4.1\arcsec\times4.6\arcsec$. \textbf{Bottom left}: 1.4\,GHz total intensity map with superimposed contour levels of $\epsilon \times (1,\;2,\;4,\;8,\;16,\;32,\;64,\;128,\;256$) with $\epsilon = 130\,\mu$Jy/beam. \textbf{Bottom right}: Spectral index map derived from the 1.4\,GHz and 2.7\,GHz data.} \label{fig:TotalIntensityHT1}
\end{figure*}

\begin{figure*}
    \begin{minipage}[t]{.49\textwidth}
        \centering
        \includegraphics[width=\textwidth]{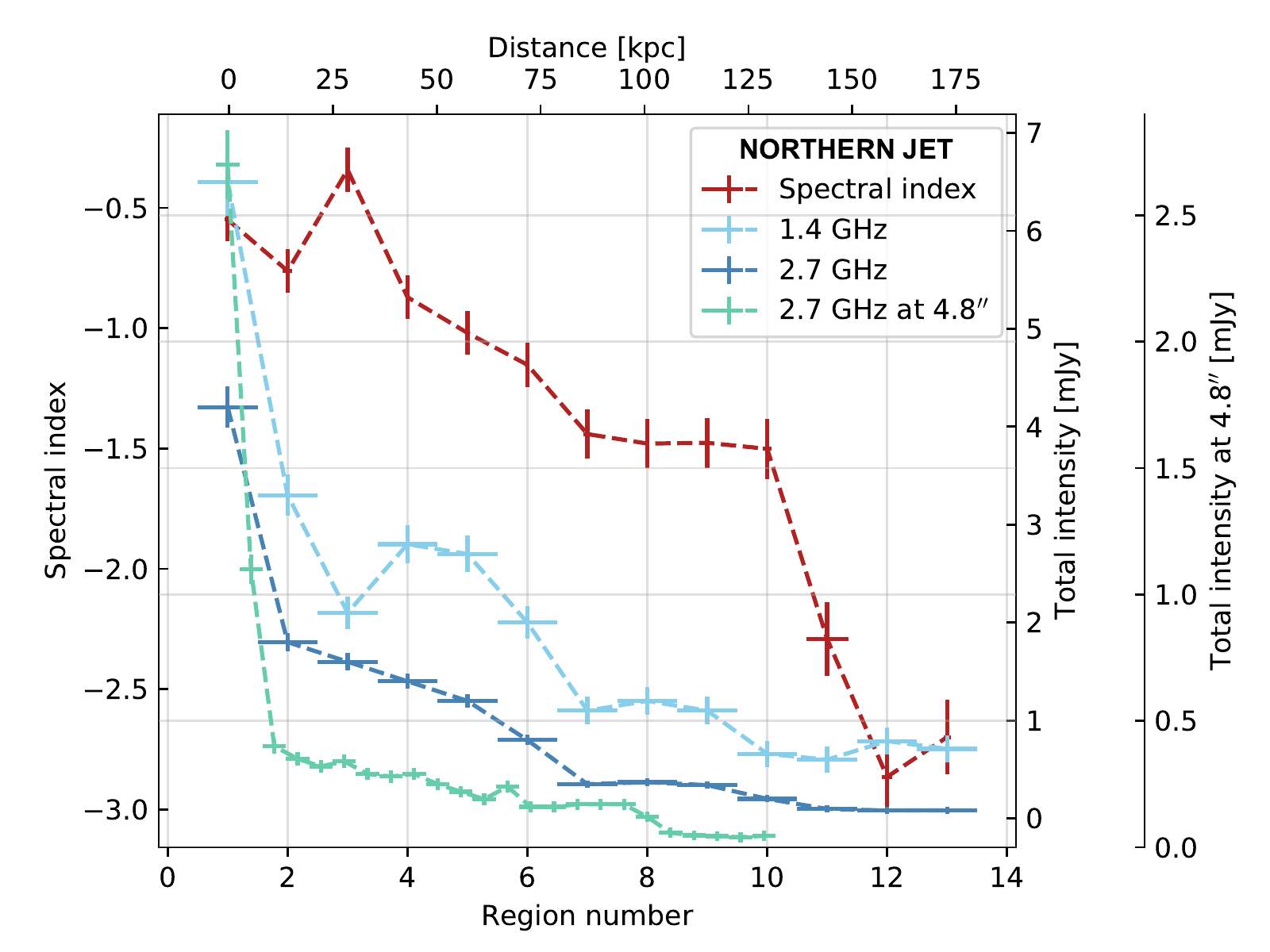}
    \end{minipage}  
    \hfill
    \begin{minipage}[t]{.49\textwidth}
        \centering
        \includegraphics[width=\textwidth]{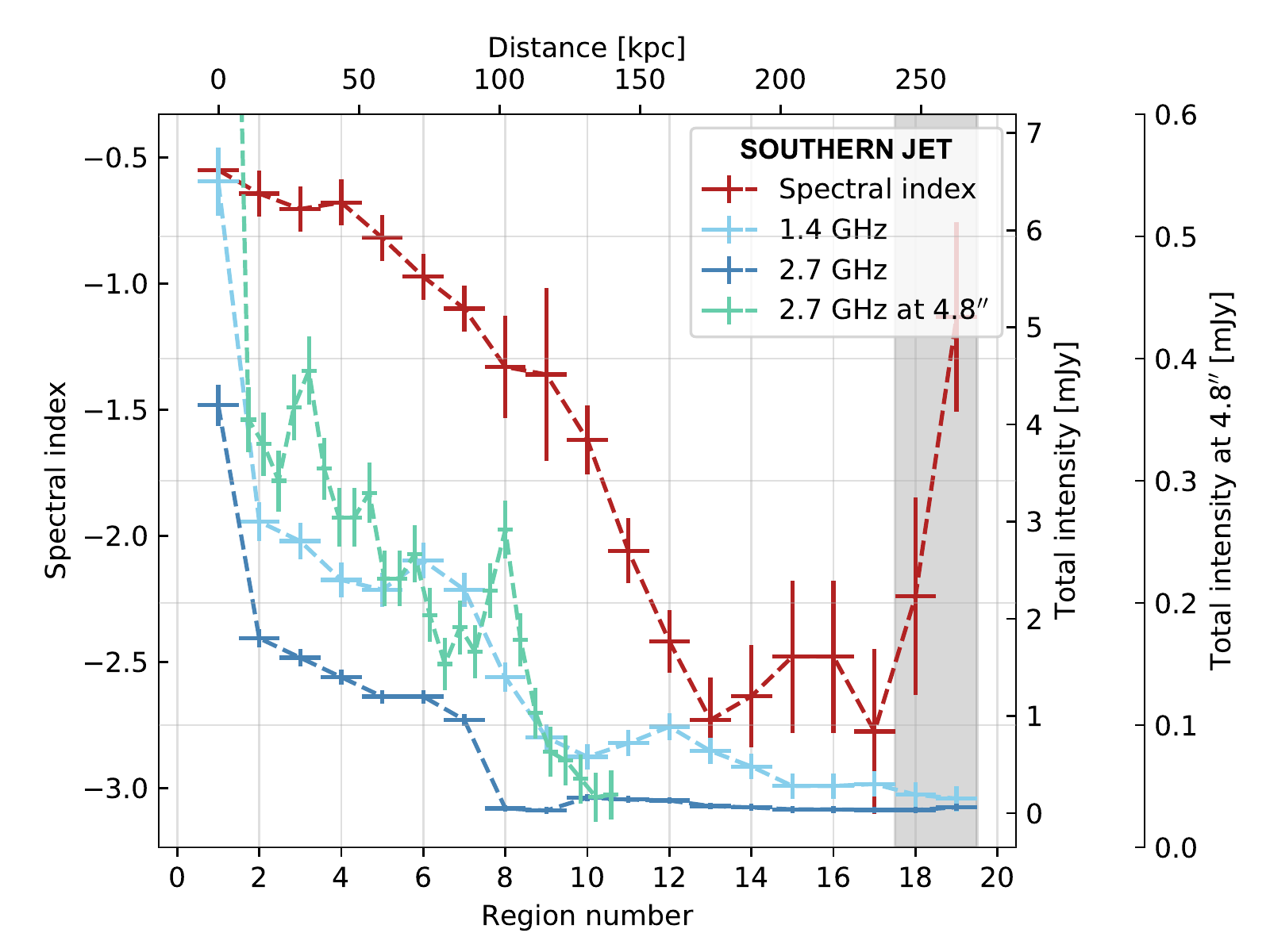}
    \end{minipage} 
     \begin{minipage}[t]{.49\textwidth}
        \centering
        \includegraphics[width=\textwidth]{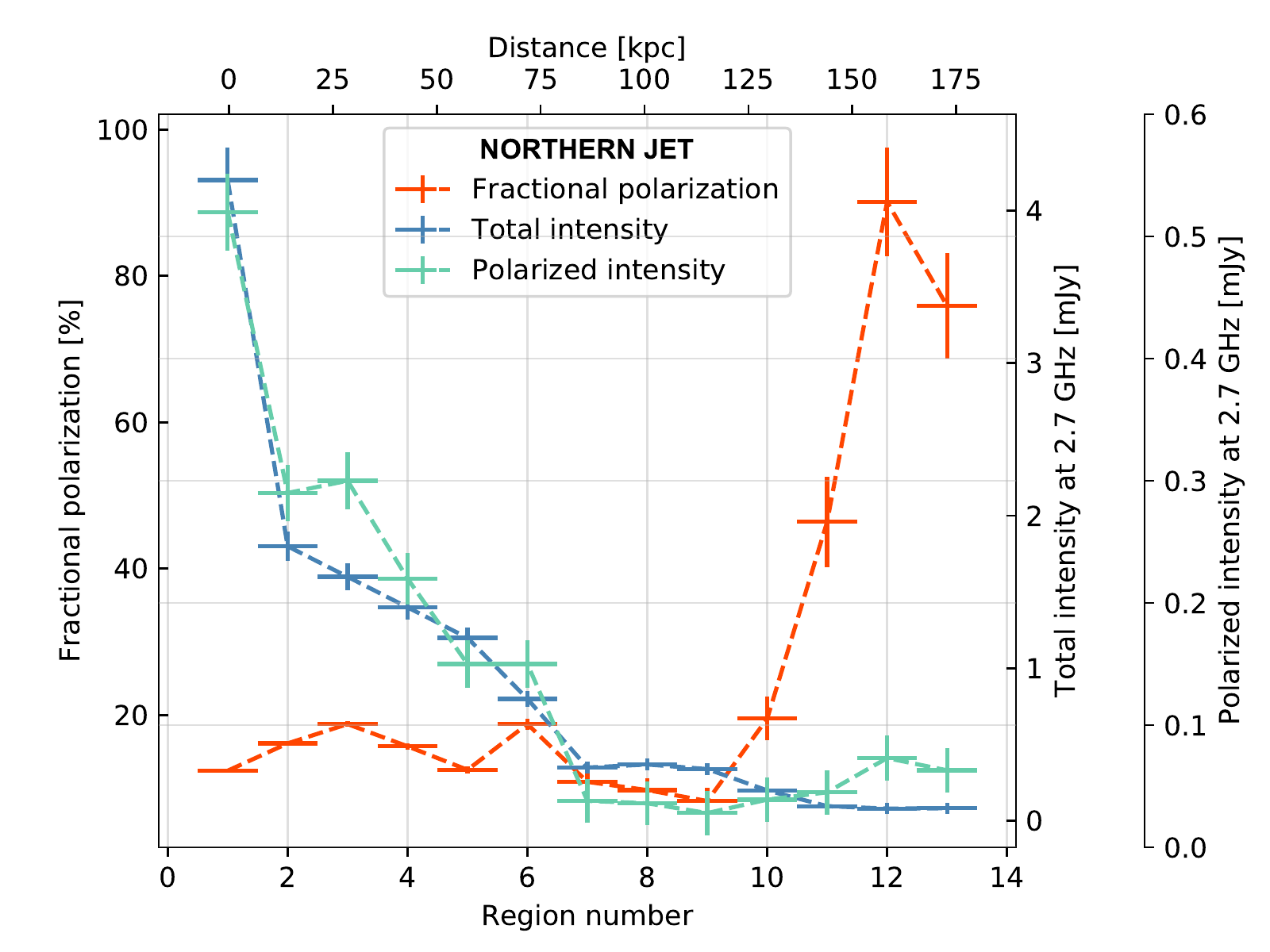}
    \end{minipage}  
    \hfill
    \begin{minipage}[t]{.49\textwidth}
        \centering
        \includegraphics[width=\textwidth]{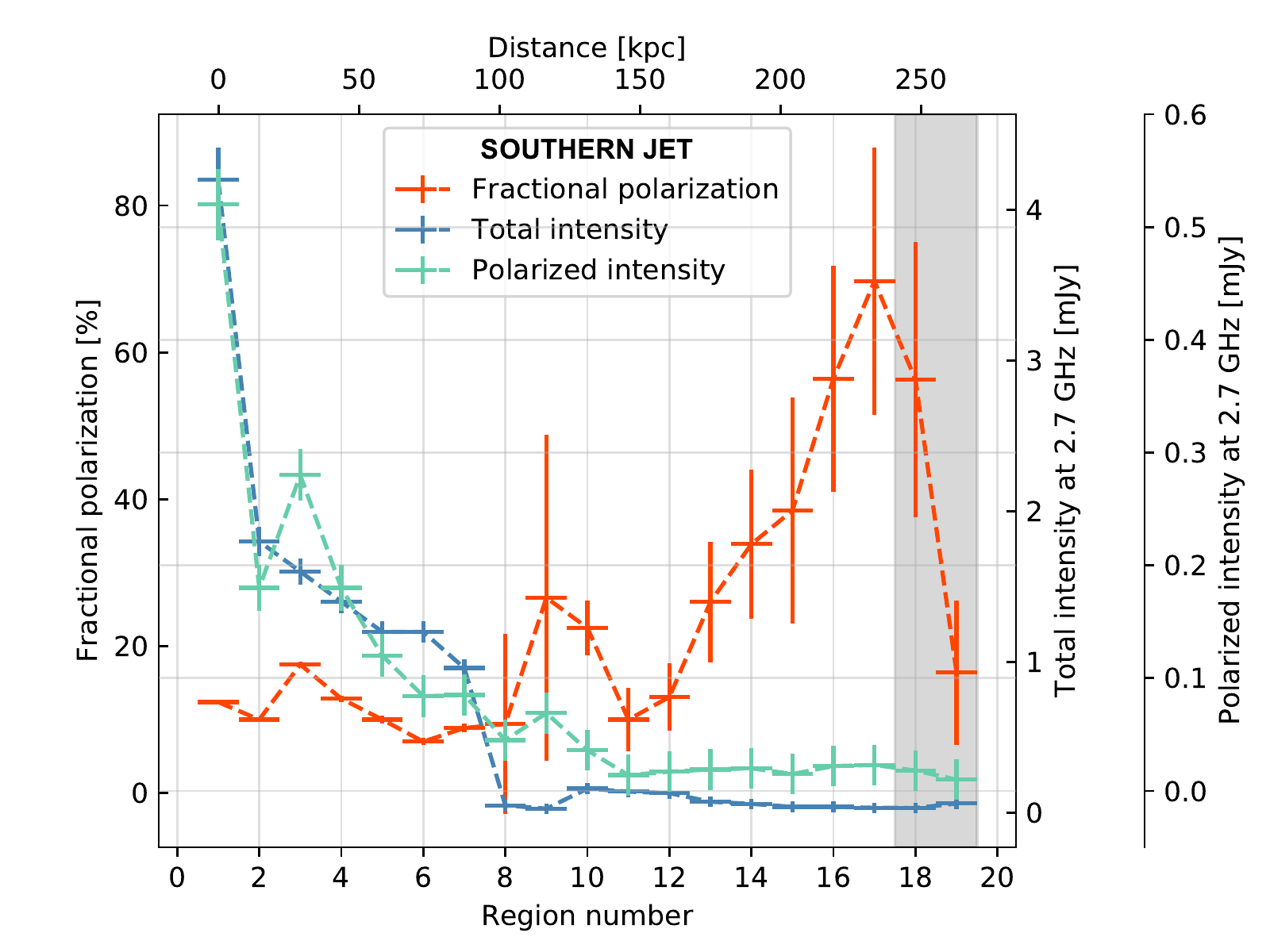}
    \end{minipage}
        \begin{minipage}[t]{.49\textwidth}
        \centering
        \includegraphics[width=\textwidth]{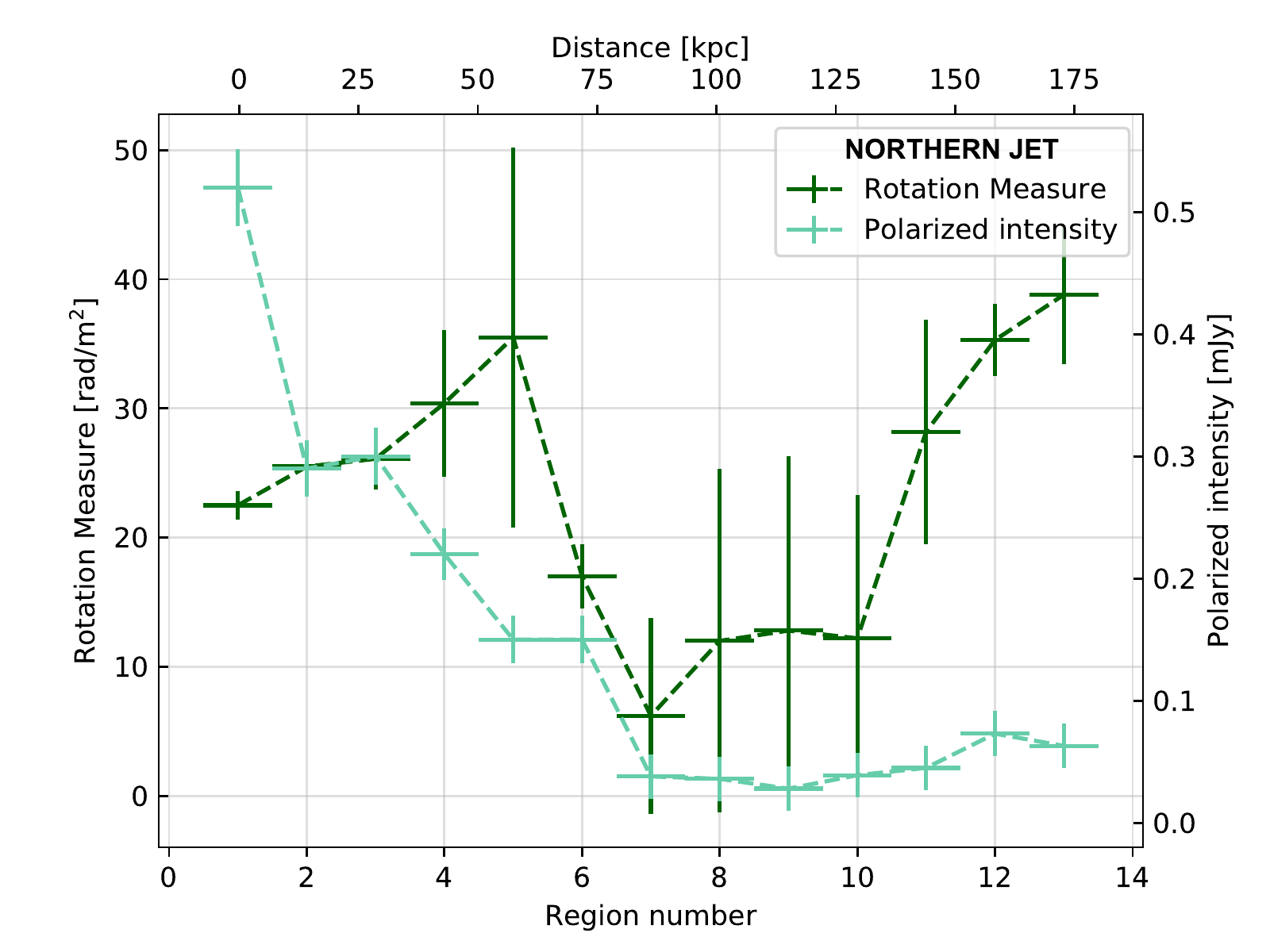}
    \end{minipage}  
    \hfill
    \begin{minipage}[t]{.49\textwidth}
        \centering
        \includegraphics[width=\textwidth]{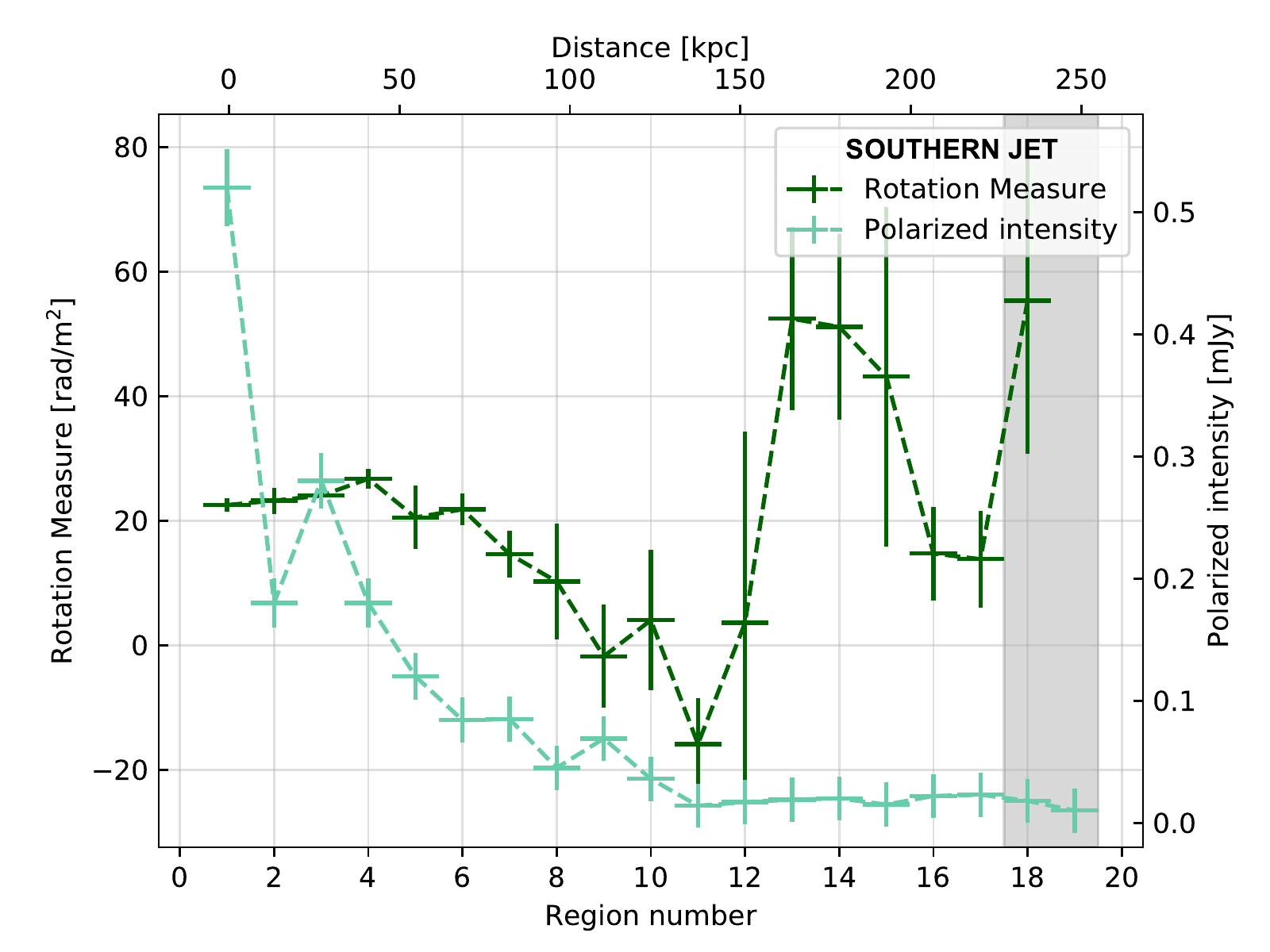}
    \end{minipage}
    \caption{HTI. Tracks of different observables along the galaxy jets with higher region numbers correspond to larger distances to the head. The region number refer to the maps with a resolution of $15\arcsec\times15\arcsec$, the higher resolution regions are converted to that scale by taking the physical distances into account. For comparison, the head property is shown in each graph (1st region). \textbf{Left panels:} Northern tail. \textbf{Right panels:} Southern tail. The values within the grey area are expected to be contaminated by a background source. \label{fig:trackHTI}}
\end{figure*}

\begin{figure*}
    \begin{minipage}[t]{.49\textwidth}
        \centering
        \includegraphics[width=\textwidth]{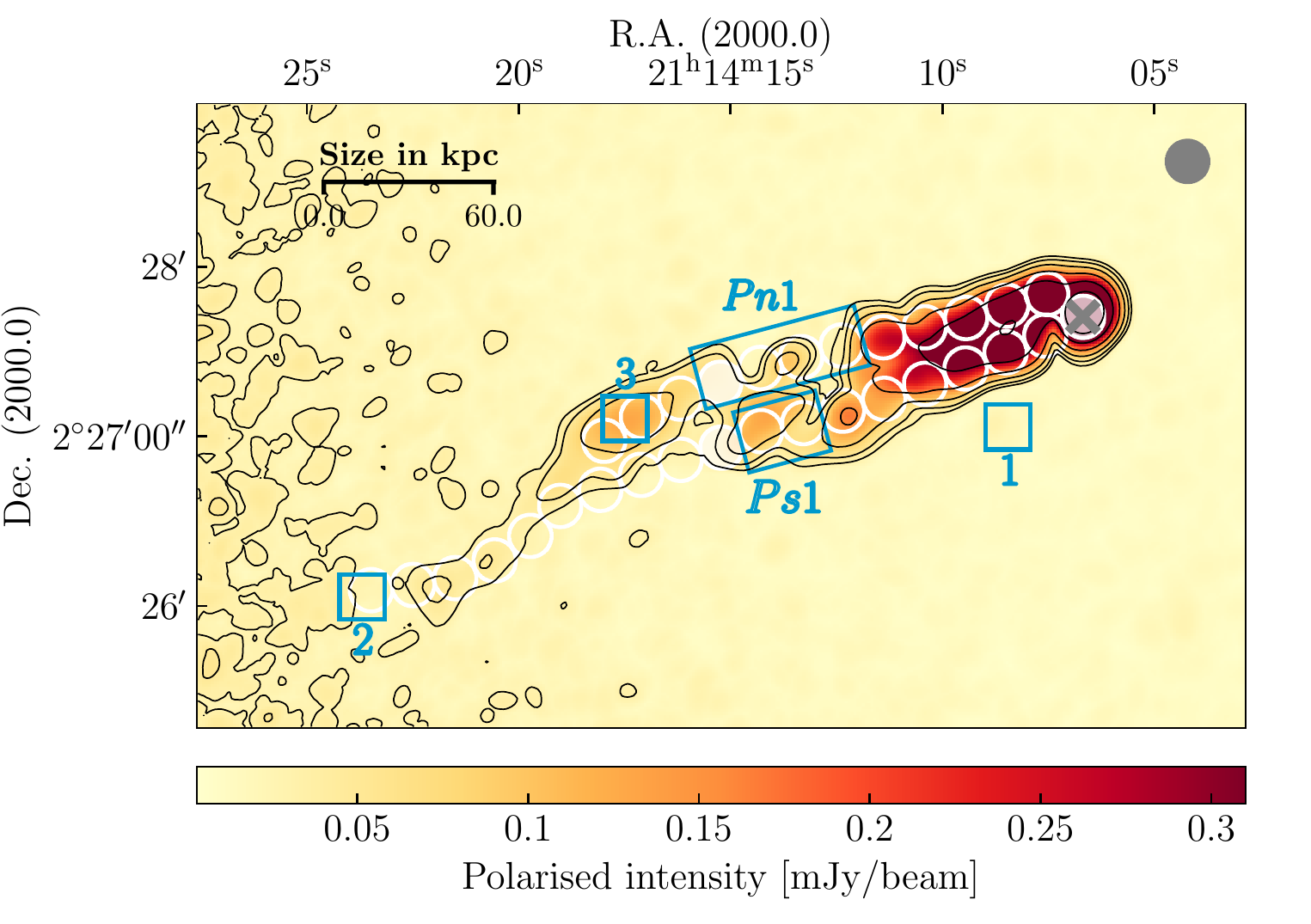}
    \end{minipage}  
    \hfill
    \begin{minipage}[t]{.49\textwidth}
        \centering
        \includegraphics[width=\textwidth]{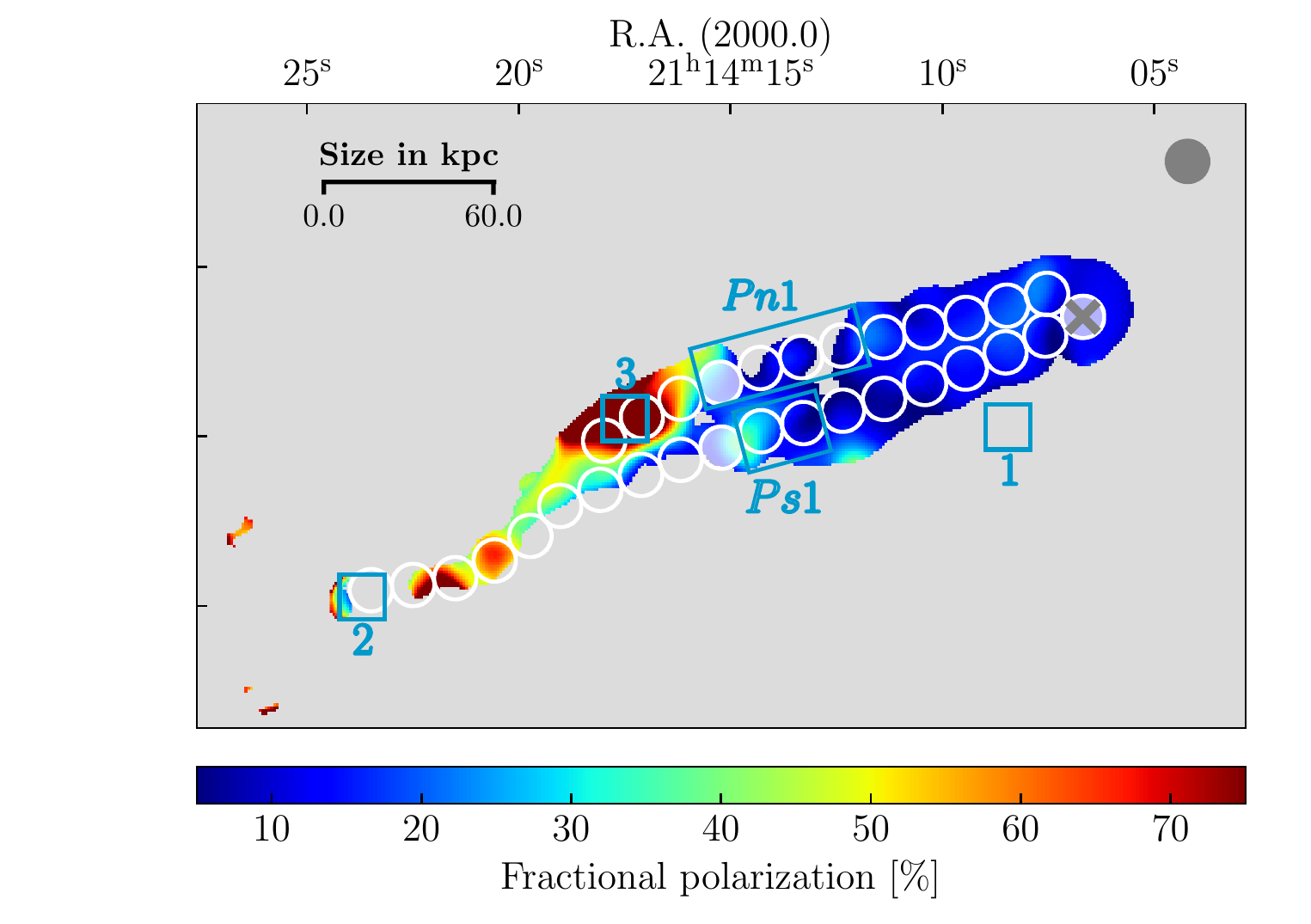}
    \end{minipage} 
     \begin{minipage}[t]{.49\textwidth}
        \centering
        \includegraphics[width=\textwidth]{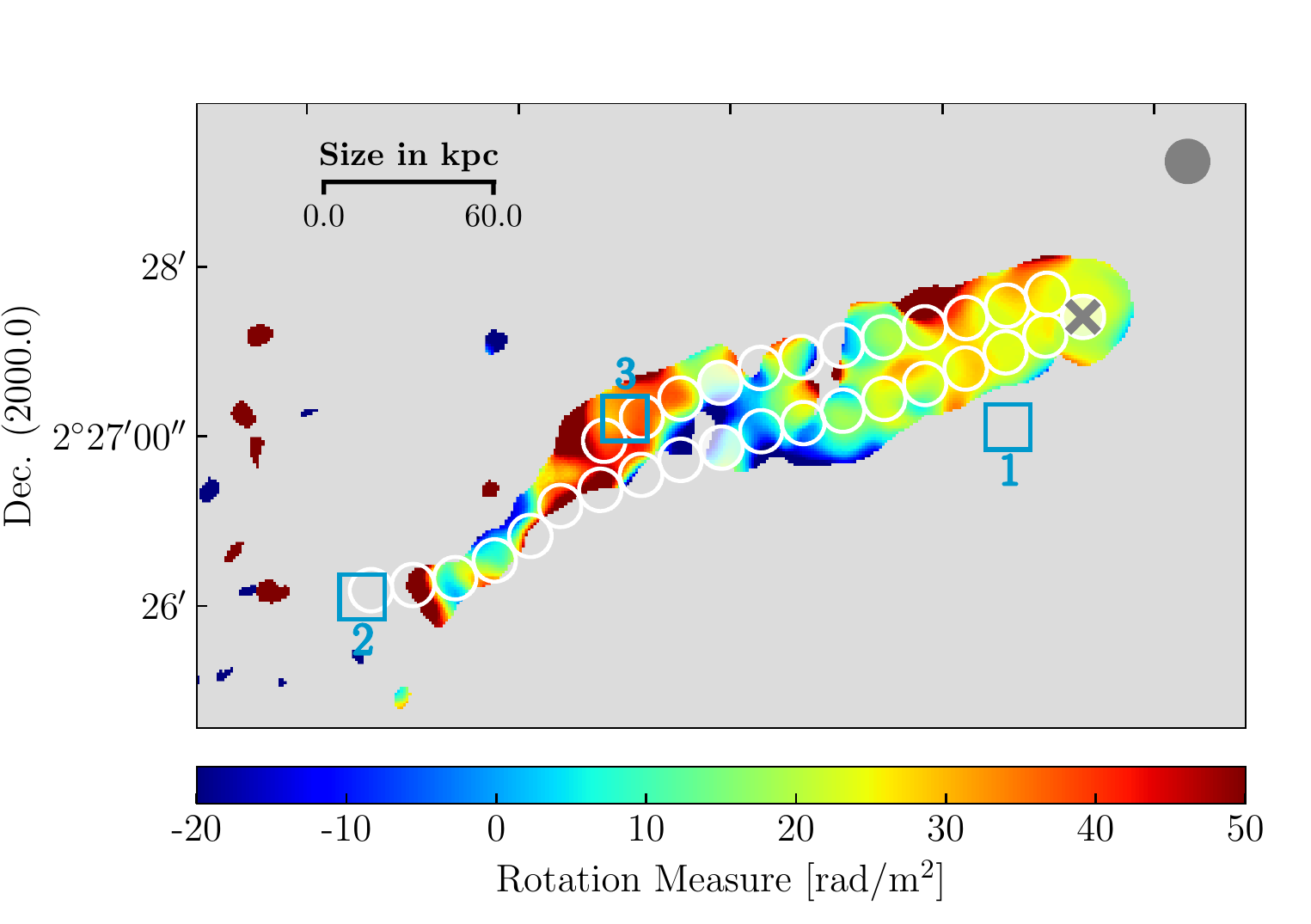}
    \end{minipage}  
    \hfill
    \begin{minipage}[t]{.49\textwidth}
        \centering
        \includegraphics[width=\textwidth]{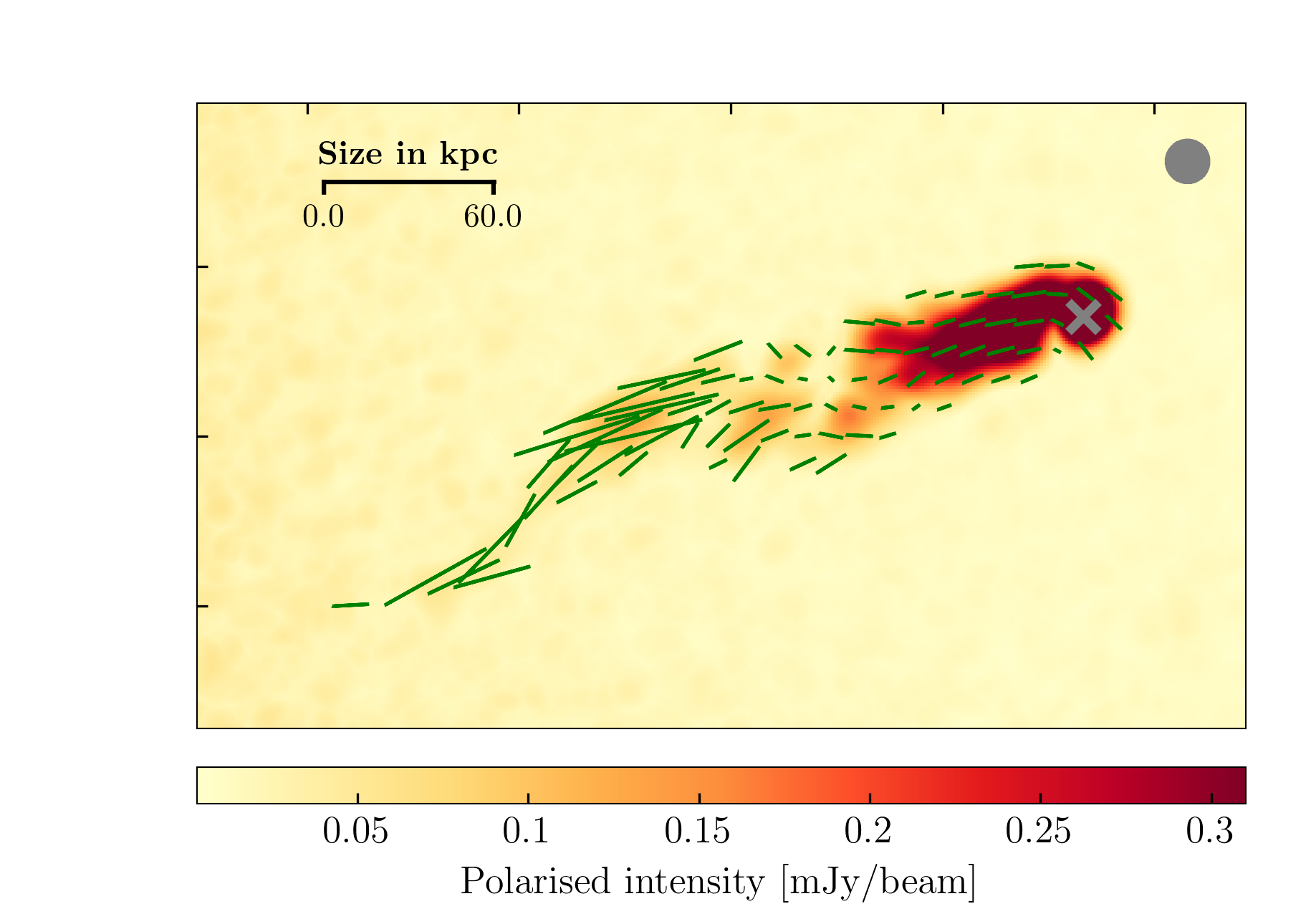}
    \end{minipage}
    \caption{HTI. Polarization results of the 2.7\,GHz data at a resolution of $15\arcsec\times15\arcsec$. White circular beam-sized regions are superimposed along the jet to track different properties centred on the central AGN marked by a grey cross. Each 10th region is filled in white to ease the comparison with Fig.~\ref{fig:trackHTI}. Regions discussed as possible optical sources producing line-of-sight emission {(`1', `2', `3')} and {plateaus (`Pn1', `Ps1')} are marked with blue {squares and rectangles, respectively}.
    The beam is shown in grey in the top right corner and {lightgrey} area contains no information. \textbf{Top left}: Polarized intensity map with superimposed contour levels of bias + $\epsilon \times (1,\;2,\;4,\;8,\;16,\;32,\;64,\;128,\;256$) with $\textrm{bias}=12\,\upmu$Jy/beam and $\epsilon = 18\,\upmu$Jy/beam. \textbf{Top right}: Fractional polarization map. \textbf{Bottom left}: RM map. \textbf{Bottom right}: Magnetic fields vectors weighted by $0.5\times$ (fractional polarization) on top of polarized intensity.} \label{fig:PolarizedIntensityHT1}
\end{figure*}

\begin{table*}
\def\arraystretch{1.2}
    \centering
\begin{tabular}[t]{r|r|r|r|r|r|r|r|r|r|r|r|r}
& & \multicolumn{5}{c}{\textbf{Head}} & & \multicolumn{5}{c}{\textbf{Jet}}\\
 name & \phantom{b} & $I_{\textrm{1.4}}$ & $I_{\textrm{2.7}}$ & $\alpha$ & $\textit{PI}$ & $\textit{FP}$ & \phantom{b} & $I_{\textrm{1.4}}$ & $I_{\textrm{2.7}}$ & $\Delta\alpha/\Delta{d}$ &  $\textit{PI}$ &  $\textit{FP}$ \\
 & & [mJy] & [mJy] & & [mJy] & [\%] & & [mJy] & [mJy] & & [mJy] & [\%]\\
  \hline
 HTI & & {6.5$\pm$0.3} & {4.2$\pm$0.2} & {-0.56$\pm$0.09} & {0.52$\pm$0.03} 
 & {12.4$\pm$4.8} & & {80.8$\pm$4.1} & {31.8$\pm$1.6} & -{0.18$\pm$0.02} & {4.2$\pm$0.2} & {14.0$\pm$5.7} \\
 HTII & & {10.4$\pm$0.5} & {9.1$\pm$0.5} & {-0.17$\pm$0.09} & {0.39$\pm$0.02} & 4{.3$\pm$5.5} & & {599$\pm$30.0} & {294$\pm$15} & {-0.073$\pm$0.006} & {34.0$\pm$1.7} & 1{0.6$\pm$5.2} \\
\end{tabular}
    \caption{Radio properties; name used through the text, total intensity $I$ at 1.4\,GHz and 2.7\,GHz, mean spectral index ($\alpha$) and mean slope along, polarized intensity \textit{PI}, and fractional polarization \textit{FP} measured at 2.7\,GHz {separated for} the head and jet, respectively.}
    \label{tab:2}
\end{table*}

\subsubsection{Radio continuum and spectral index}

HTI is showing two bent jets pointing away from the cluster {centre} (Fig.~\ref{fig:TotalIntensityHT1}).
Due to the narrow appearance and the unknown inclination of the jets{,} it is difficult to distinguish between their individual emissions.
{They appear to have comparable shape and sizes taking the inclination of the systems into account.} 
Both jets appear straight in the plane of the sky and broaden after $\sim 60\,$kpc.
Even at 4.6$\arcsec$ major-axis resolution (Fig.~\ref{fig:TotalIntensityHT1}, top right), the contours along the jets are partially connected. A possible physical origin of this connection can only be studied with higher resolution data, which is therefore postponed to future studies.

At 15$\arcsec$ the two jets are smeared out to form one entity (Fig.~\ref{fig:TotalIntensityHT1}, top and bottom left).
However, the dips in the radio contours along the jet direction still indicate the two different jets highlighted by the white circular regions tracing the jets individually (the track mask is matched to the total intensity map in Fig.~\ref{fig:TotalIntensityHT1}, top left).
At 1.4\,GHz the jets are showing the largest extent up to 300\,kpc (Fig.~\ref{fig:TotalIntensityHT1}, bottom left). The extent is shorter at 2.7\,GHz also affected by a background point source, also visible in the optical, at the very distant part of the jet ({square region `2' of Fig.~\ref{fig:TotalIntensityHT1}, top left and Fig.~\ref{fig:optical}, left; see also Appendix \ref{App:optical} Fig. \ref{fig:app_optical}, top right}) emitting at this frequency which is less bright at 1.4 GHz. 
The extent of the radio contour in both frequencies south of the southern jet (corresponding to the blue square marked with a `1' in Fig.~\ref{fig:TotalIntensityHT1}) can be identified with an optical source and can also be seen in the higher resolution image {(top right)}.
At this position{,} the jet morphology{,} as well as the source contours{,} do not show any indication for interaction.  
The jet emission is broadening with distance {from} the galaxy head showing additional emission in the {southern} jet at 1.4\,GHz that cannot be seen at 2.7\,GHz (compare Fig.~\ref{fig:TotalIntensityHT1}, top and bottom left). Such morphology can be caused by electrons {losing} their energy during propagation and are therefore easier tracked at larger distances by lower frequency data (see also spectral index discussion) and is caused by significant {ageing} of the electrons through the jet.

{In addition to the radio continuum images, in Fig~\ref{fig:TotalIntensityHT1} we show the spectral index image of {HTI}}. We find a total intensity (head + jets) of $(87.3\pm4.4)$\,mJy and $(36.0\pm1.8)$\,mJy at 1.4\,GHz and 2.7\,GHz of which $(6.5\pm0.3)$\,mJy and ($4.2\pm0.2$)\,mJy are generated by the head (corresponding to the white circular region centered on the peak flux indicated by a grey cross, see Fig.~\ref{fig:TotalIntensityHT1}), respectively. The jet properties are defined by the outermost black contour corresponding to the noise level reported in Tab.~\ref{tab:3} from which we subtract the head properties.
These measures are corrected for the contamination of the additional sources along the line of sight (`1' and `2' in Fig.~\ref{fig:TotalIntensityHT1}). Their contribution has been measured in circular beam-sized regions, ($0.70\pm0.04$)\,mJy at 1.4\,GHz and ($0.51\pm0.03$)\,mJy at 2.7\,GHz, and subtracted to the measures.
We find a spectral index between 1.4\,GHz and 2.7\,GHz of $-0.56\pm0.09$ in the galaxy head and a decrease through the jets with a mean slope of $-0.18\pm 0.02$ per circle (with the circles being equidistant), neglecting the 5 most distant regions. An overview of these properties can be found in Table~\ref{tab:2}.

{We measured the observational properties of the jets covering each jet with N circular regions with a radius compared to the major angular resolutions (see Tab. \ref{tab:3}) to ensure that these are independent regions.
The properties of each jet are shown separately in the left and right panels of Fig.~\ref{fig:TotalIntensityHT1}.}
The region numbers correspond to the white circular regions (as defined in Fig.~\ref{fig:TotalIntensityHT1}) starting from the galaxy head. The tracks along the two jets are shown separately.
The total intensity show{s} a {head-dominated} galaxy with a decreasing flux along the galaxy jets thereafter, but with some exceptions at 1.4\,GHz. The first increase in the northern jet can be associated with the broadening of the jets (clearly visible also in Fig.~\ref{fig:TotalIntensityHT1}, top right) and the second increase at the very end of the northern jet and the 12th region of the southern jet might be caused by the overlap of both jets.
The spectral index is steadily steepening through the jet with the head being flat ($\alpha=-0.56$) and the most distant jet {region} showing very steep values up to -3.0. In the northern jet, it is steepening slowly until {the region marked by `Pn1'}, where we identify a plateau, here, the spectral index is found to be steady, and steepens rapidly thereafter.
The same {behaviour} can be see{n} in the southern jet {(see `Ps1')}. {The spectral index flattening at the end of the jet is consistent with the idea that source `2' is a background compact radio source with flat spectrum} (the contaminated values are highlighted with a grey shaded area in Fig. \ref{fig:trackHTI}).

\subsubsection{Linear polarized radio emission and fraction}

Figure \ref{fig:PolarizedIntensityHT1} shows the polarization, fraction, RM, and ordered magnetic field lines images of HTI at the same resolution as in total intensity from {left to right and top to bottom}, respectively. With the field-of-view in polarized intensity being reduced (due to the channel-wise imaging in which we use the smallest common field of view), the galaxy jets are located very close to the primary beam edges (left side in Fig.~\ref{fig:PolarizedIntensityHT1}, top left) such that the faintest, most distant part becomes more affected by the increasing noise at the primary beam edges.
Again, because we cannot distinguish the two different jets, we used the same circular apertures corresponding to the northern and southern jet as for the total intensity. Three significant differences in morphology can be 
{observed with respect to} 
the total intensity images (Fig.~\ref{fig:TotalIntensityHT1}): 1) the northern jet is lacking some emission (probably due to the noise-cut) {within `Pn1'}, 2) the jet becomes more clumpy with distance resulting in two substructures that cannot be identified with optical counterparts (one at {the blue square} {`Ps1'} and the other corresponds to `3', see Fig.~\ref{fig:PolarizedIntensityHT1}), and 3) no significant jet emission can be identified for the southern jet{ after `Ps1'}, while the northern jet is affected by the previously mentioned substructure. The total length of the jets is only 15\,kpc shorter than in total intensity. {The difference in length} can either be caused by the increased noise level at the primary beam edge, which makes the detection of diffuse faint emission more challenging or more probably by the non-detection of the background source (see `1' blue square in Fig.~\ref{fig:PolarizedIntensityHT1}, top left).

We find the polarized intensity of HTI to reach $4.98\pm0.25$\,mJy from which $0.52\pm0.03$\,mJy are generated by the head of the galaxy (see also Table~\ref{tab:2} for the individual head and jet properties). The galaxy shows a fractional polarization ($I/PI$ at 2.7\,GHz) of $12.4\pm4.8$\,\% and $14.0\pm5.7$\,\% at the head and jet, respectively. We find the RM to be mostly positive with an r.m.s of $\sim 23$\,rad~m$^{-2}$. The ordered magnetic field component is shown in Fig.~\ref{fig:PolarizedIntensityHT1} (bottom right) with the magnetic field lines shown in green, their length {being} proportional to the fractional polarization. We find the magnetic field lines to align with the jet direction showing only small deviations at a distance of 60\,kpc ({within `Pn1'}) from the head and clearly aligns again with the jet thereafter.

In the middle and bottom panels of Fig.~\ref{fig:trackHTI} the tracks along the polarization observable are shown (corresponding to the white circular regions superimposed in Fig.~\ref{fig:PolarizedIntensityHT1}) with the left panels corresponding to the northern and the right ones to the southern jet. The total intensity track at 2.7\,GHz is shown for comparison. Generally, the total and polarized intensity decrease while the fractional polarization increases with distance to the galaxy head visible in both jets. In the southern jet the fractional polarization increases and decreases between the first 7 regions by about 7\,\%. We identify a fractional polarization plateau at 12\,\% {within region `Pn1'}. Thereafter, the fractional polarization increases very fast reaching non-physically high values of 90\,\% at the 12th region. Here, we already identified a substructure in the jet that cannot be observed in total intensity. Generally, such values can be explained by the fact that the polarized emission is recovered on smaller scales and therefore track more diffuse emission than detected in total intensity. This phenomenon is based on the ``rotation'' of the polarised emission that we can recover using RM synthesis \citep[see Sect.~7.2 in][]{debruyn}. 
The RM is found to be purely positive (bottom left panel of Fig.~\ref{fig:trackHTI}) indicating a magnetic field pointing towards the observer. It increases within the first 5 regions, it drops quickly and is almost steady between the 7th and 10th region. A steep increase can be observed thereafter, again corresponding to the clumpy substructure seen in polarized intensity (see blue square marked with `3' in Fig.~\ref{fig:PolarizedIntensityHT1}). However, while the RM maps are corrected for the Galactic foreground (see methods in \cite{Mueller}) we did not separate the signal of the ICM from the intrinsic contribution by the galaxy and therefore only report on their combined characteristics.

In the southern jet, the fractional polarization reaches its highest value of 70\,\% at region 16. The final steep decrease is likely due to the already identified possible background source (`1' blue square in Figs.~\ref{fig:TotalIntensityHT1} and \ref{fig:PolarizedIntensityHT1}). The RM (Fig.~\ref{fig:trackHTI}, bottom right) is found to be almost steady within the first 6 regions and decreases to the 11th region reaching about $-18$\,rad~m$^{-2}$ corresponding to a magnetic field that is pointing in the opposite direction, away from the observer. It increases quickly and drops to the 16th region where the fractional polarization is found to reach its maximum. As mentioned before, the southern jet is sparsely detected between region{s} 11 and 15 and {is,} therefore{,} noise dominated. Substructures within these regions are therefore neglected in the discussion.

\subsection{HTII} \label{sec:HTII}
\begin{figure*}
    \begin{minipage}[t]{.49\textwidth}
        \centering
        \includegraphics[width=\textwidth]{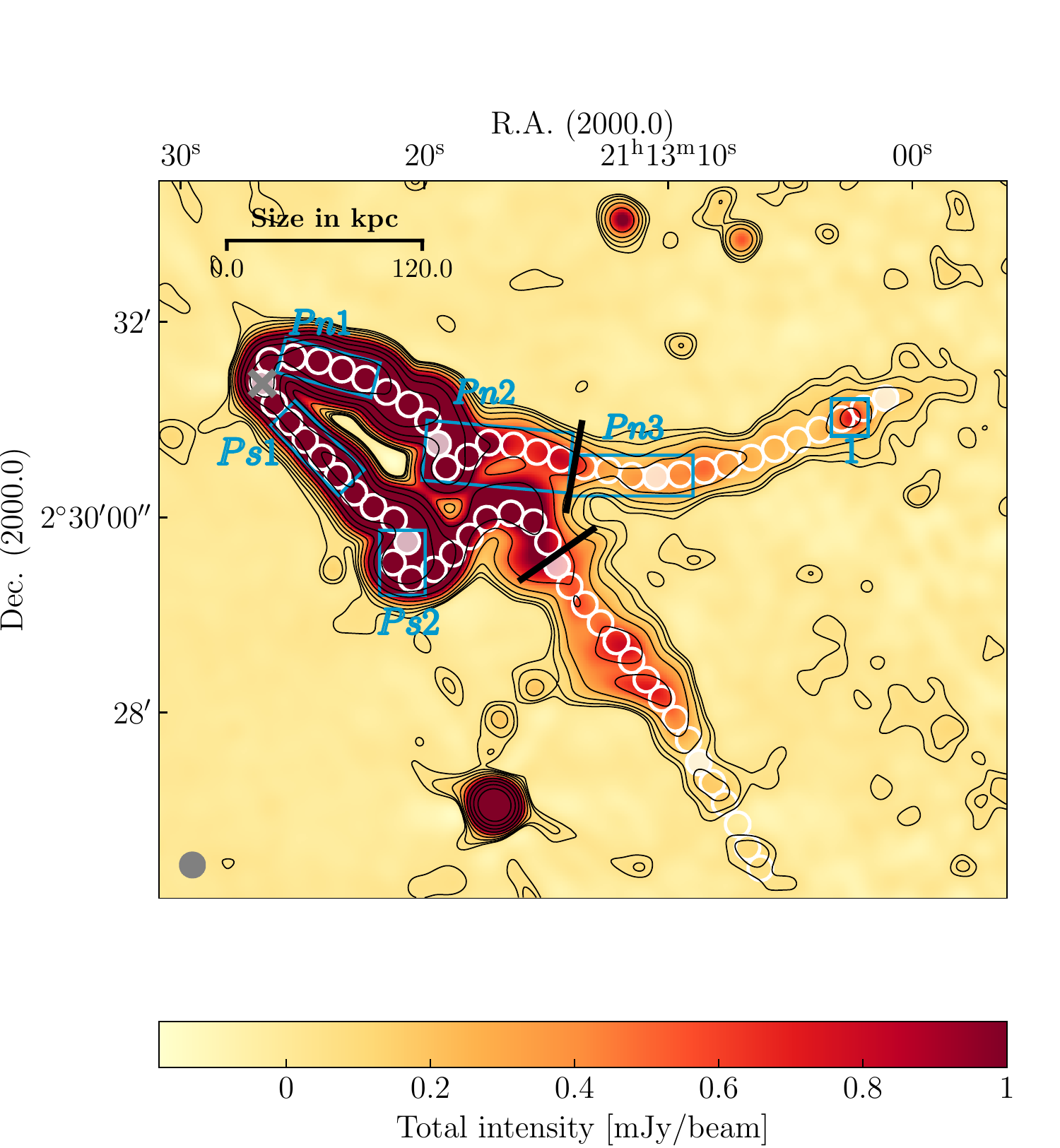}
    \end{minipage}  
    \hfill
    \begin{minipage}[t]{.49\textwidth}
        \centering
        \includegraphics[width=\textwidth]{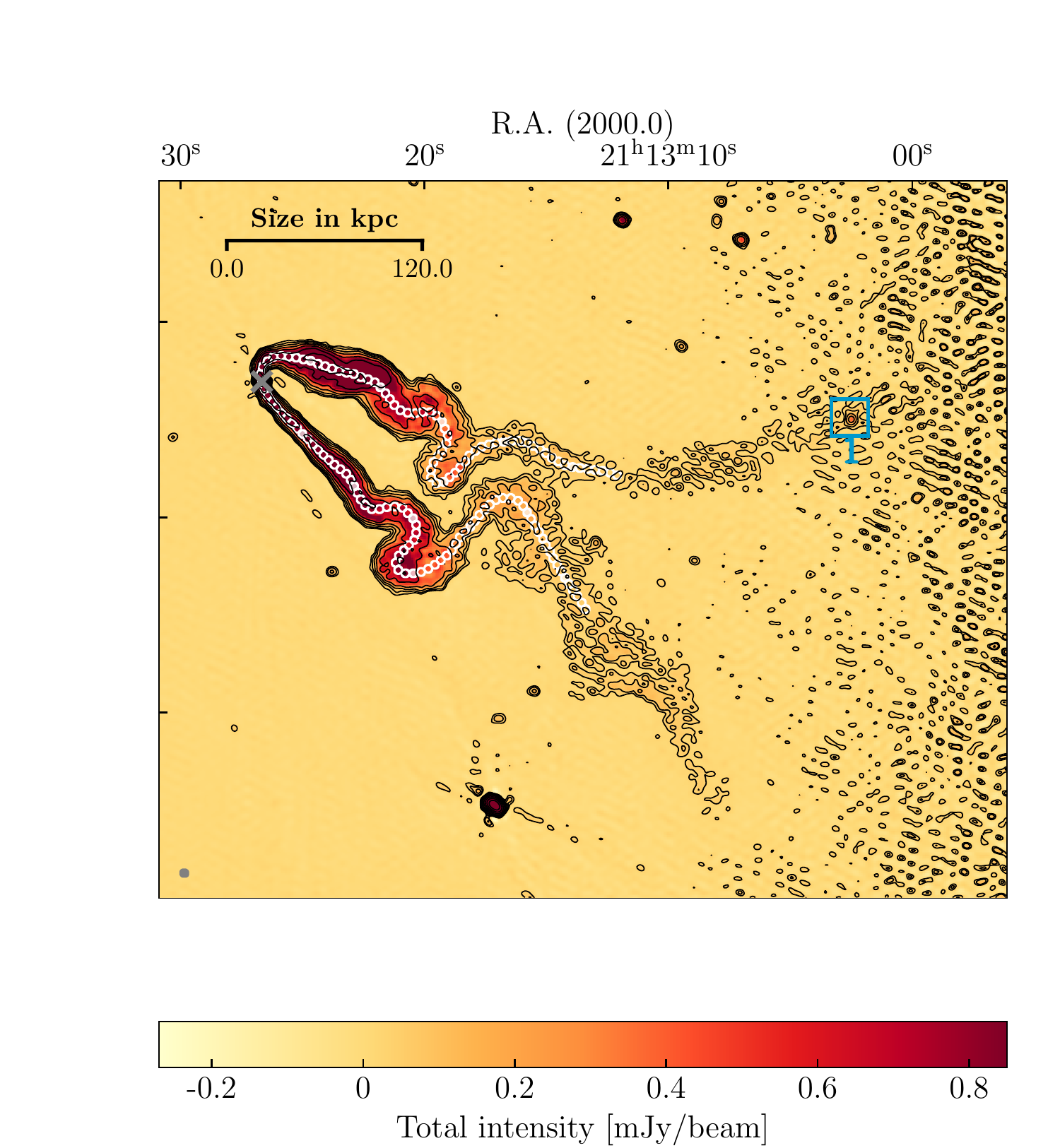}
    \end{minipage} 
     \begin{minipage}[t]{.49\textwidth}
        \centering
        \includegraphics[width=\textwidth]{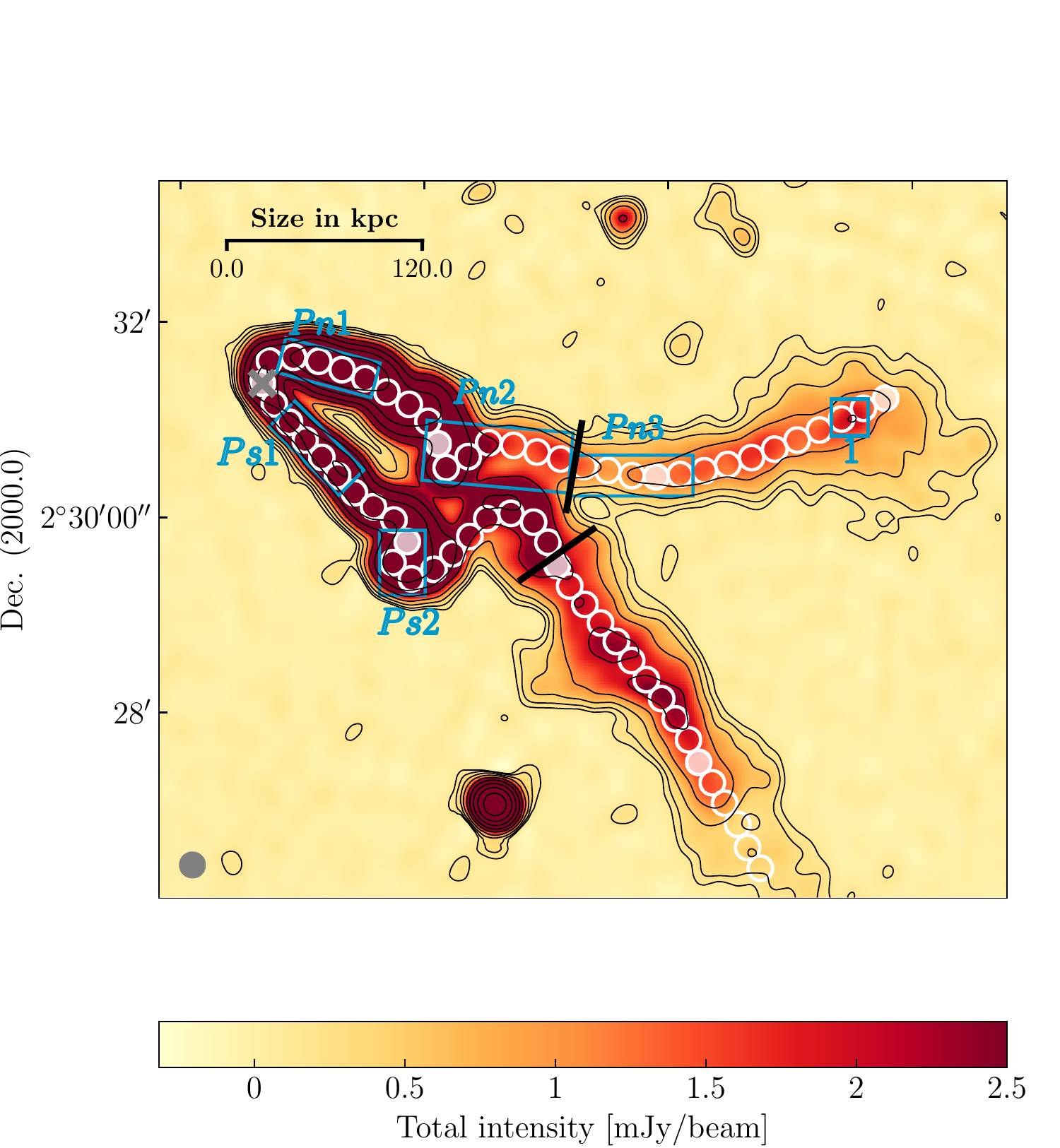}
    \end{minipage}  
    \hfill
    \begin{minipage}[t]{.49\textwidth}
        \centering
        \includegraphics[width=\textwidth]{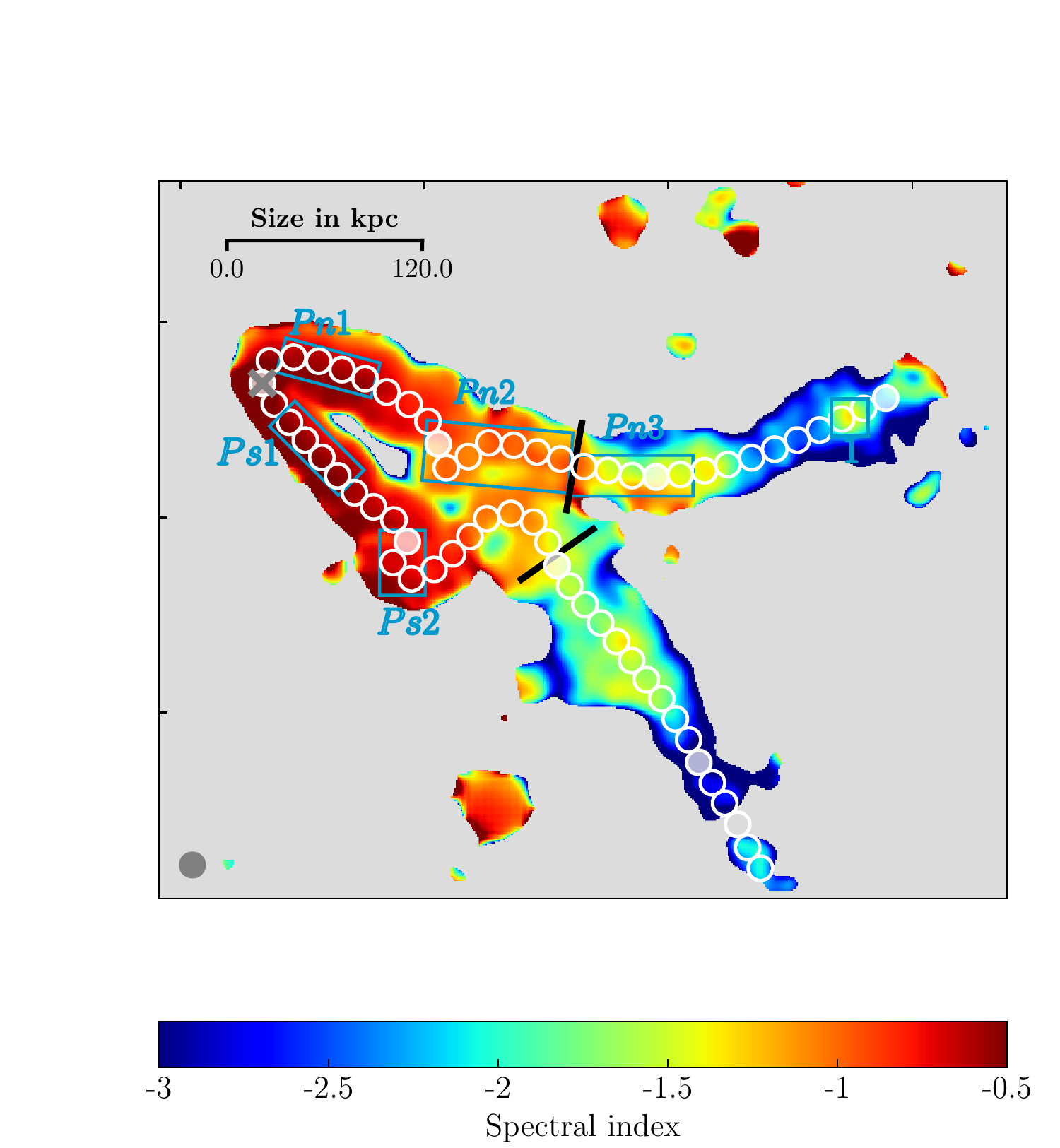}
    \end{minipage}
    \caption{HTII. Continuum results of the 2.7\,GHz (top) and 1.4\,GHz (bottom) data at a resolution of $15\arcsec\times15\arcsec$ if not stated differently. White circular beam-sized regions are superimposed along the jet to track different properties centred on the central AGN marked by a grey cross. Each 10th region is filled in white to ease the comparison with Fig.~\ref{fig:trackHTII}. Regions discussed as possible optical sources producing line-of-sight emission {(`1')} and {plateaus (`Pn1', `Ps1', `Pn2', `Ps2', `Pn3')} are marked with blue {squares and rectangles, respectively}. 
    The beam is shown in grey in the bottom left corner and {lightgrey} area contains no information. \textbf{Top left}: 2.7\,GHz total intensity map with superimposed contour levels of $\epsilon \times (1,\;2,\;4,\;8,\;16,\;32,\;64,\;128,\;256$) with $\epsilon = 35\,\upmu$Jy/beam. \textbf{Top right}: 2.7\,GHz total intensity map with superimposed contour levels of $\epsilon \times (1,\;2,\;4,\;8,\;16,\;32,\;64,\;128,\;256$) with $\epsilon = 33\,\upmu$Jy/beam at a resolution of $4.1\arcsec\times4.6\arcsec$. \textbf{Bottom left}: 1.4\,GHz total intensity map with superimposed contour levels of $\epsilon \times (1,\;2,\;4,\;8,\;16,\;32,\;64,\;128,\;256$) with $\epsilon = 130\,\upmu$Jy/beam. \textbf{Bottom right}: Spectral index map derived from the 1.4\,GHz and 2.7\,GHz data.} \label{fig:TotalIntensityHT2}
\end{figure*}

\begin{figure*}
    \begin{minipage}[t]{.49\textwidth}
        \centering
        \includegraphics[width=\textwidth]{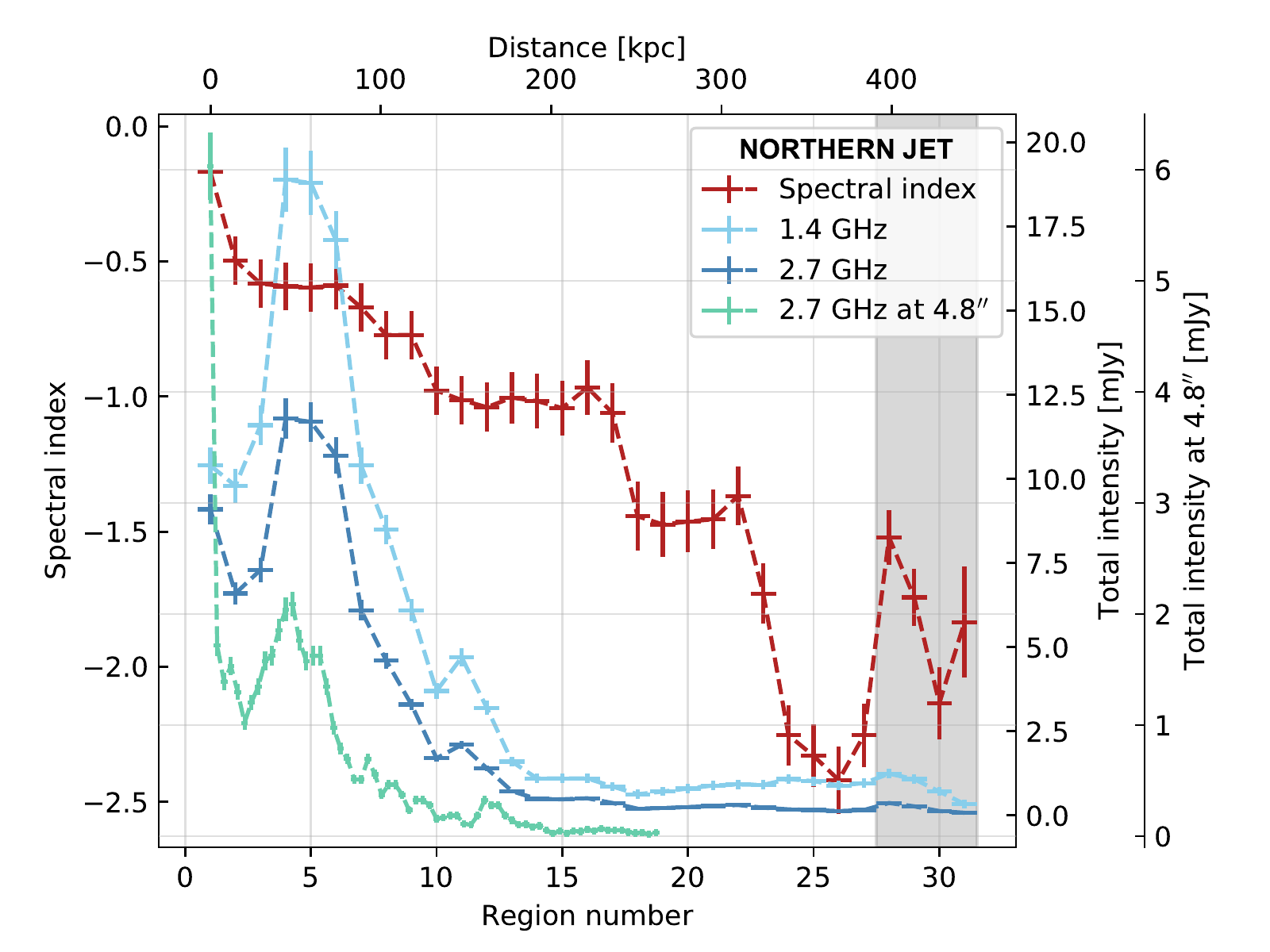}
    \end{minipage}  
    \hfill
    \begin{minipage}[t]{.49\textwidth}
        \centering
        \includegraphics[width=\textwidth]{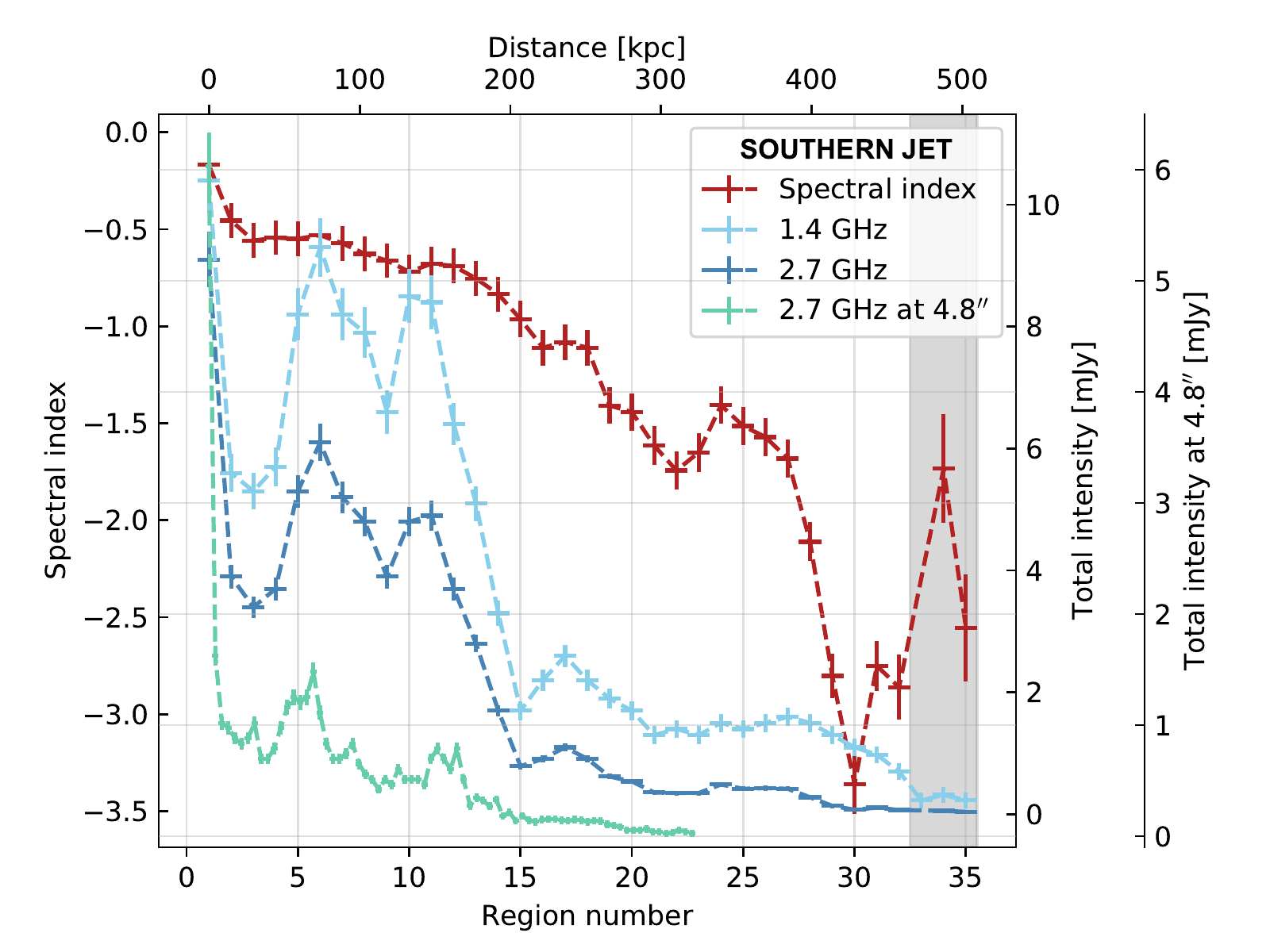}
    \end{minipage} 
     \begin{minipage}[t]{.49\textwidth}
        \centering
        \includegraphics[width=\textwidth]{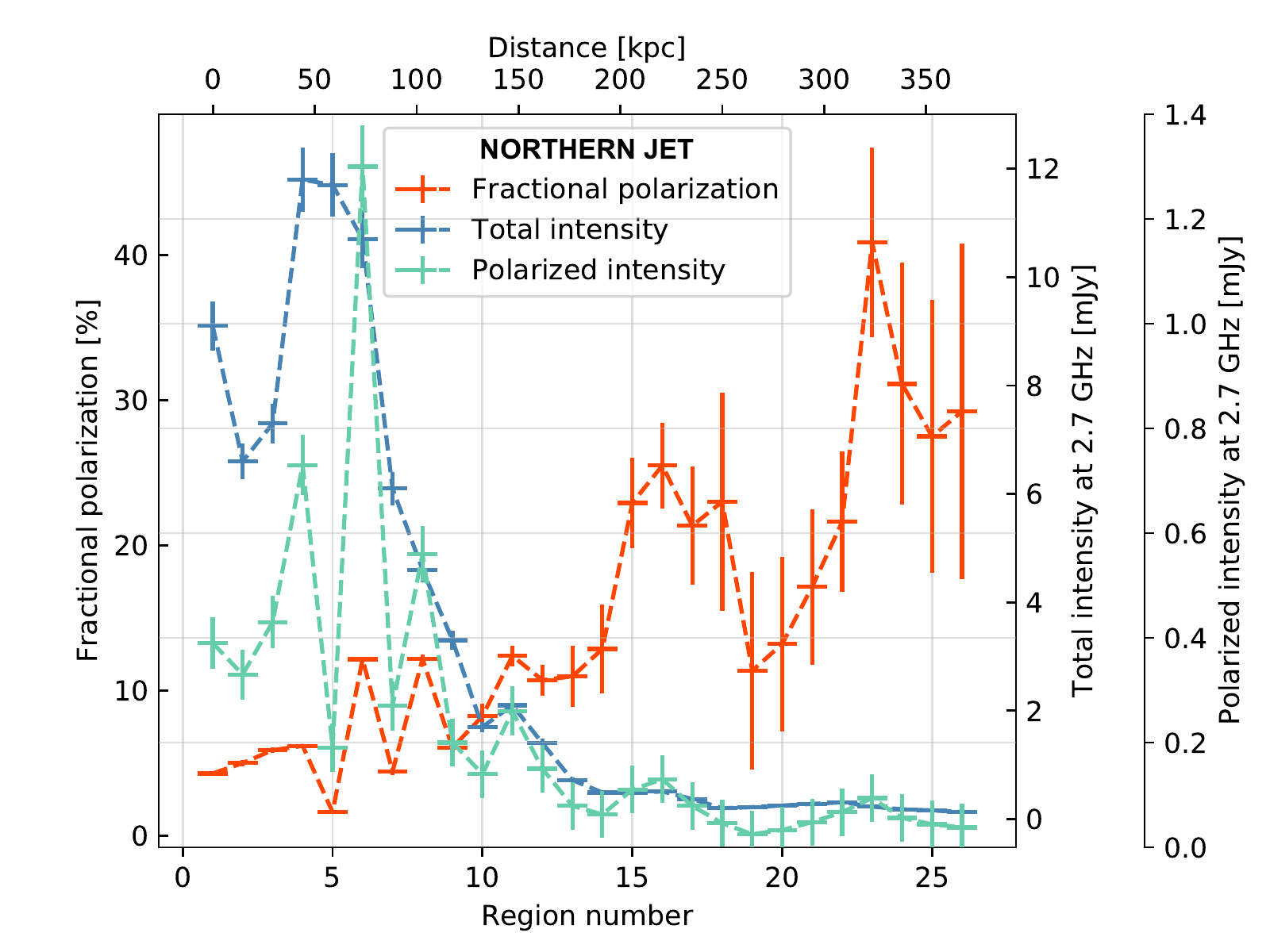}
    \end{minipage}  
    \hfill
    \begin{minipage}[t]{.49\textwidth}
        \centering
        \includegraphics[width=\textwidth]{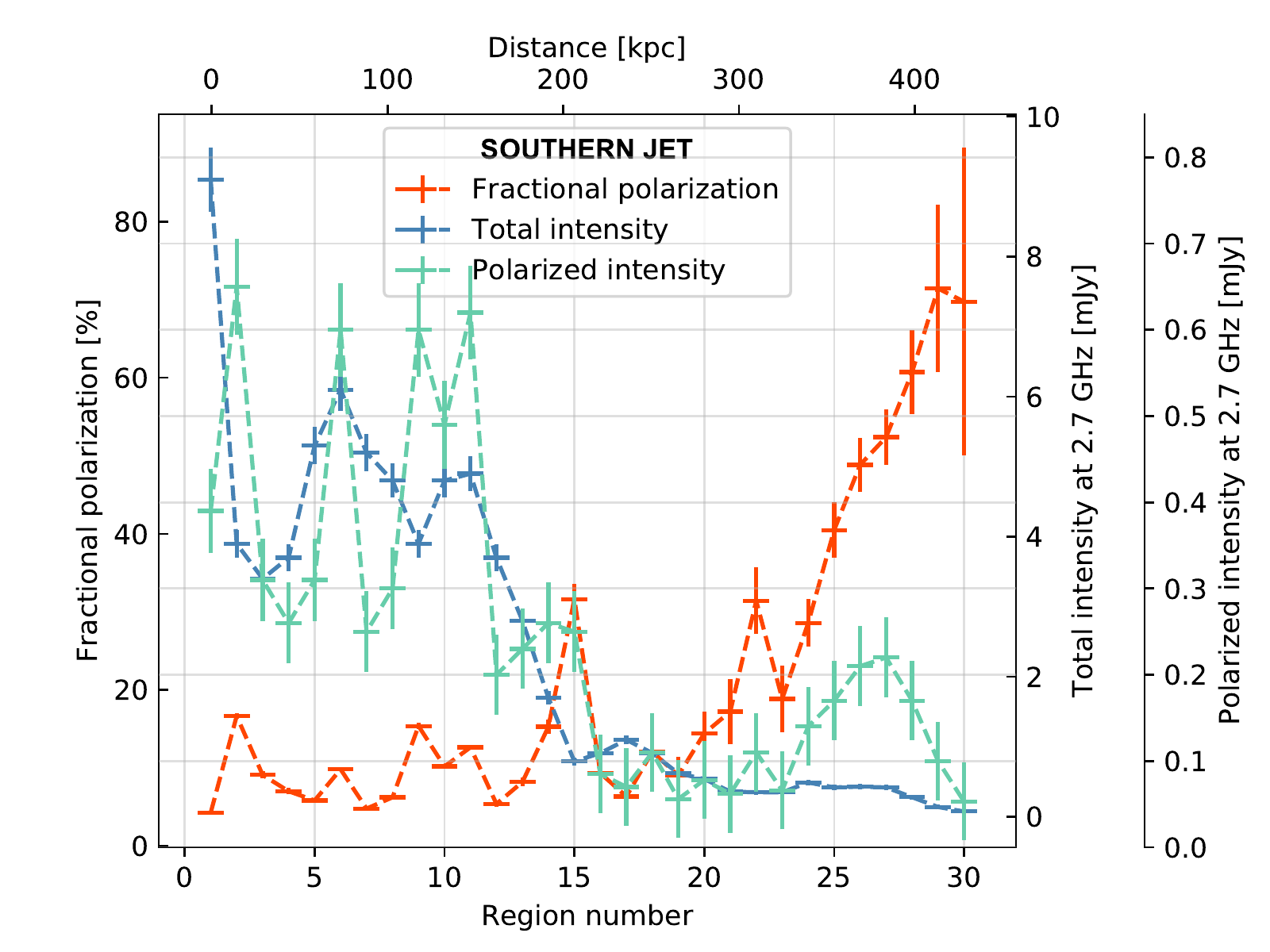}
    \end{minipage}
        \begin{minipage}[t]{.49\textwidth}
        \centering
        \includegraphics[width=\textwidth]{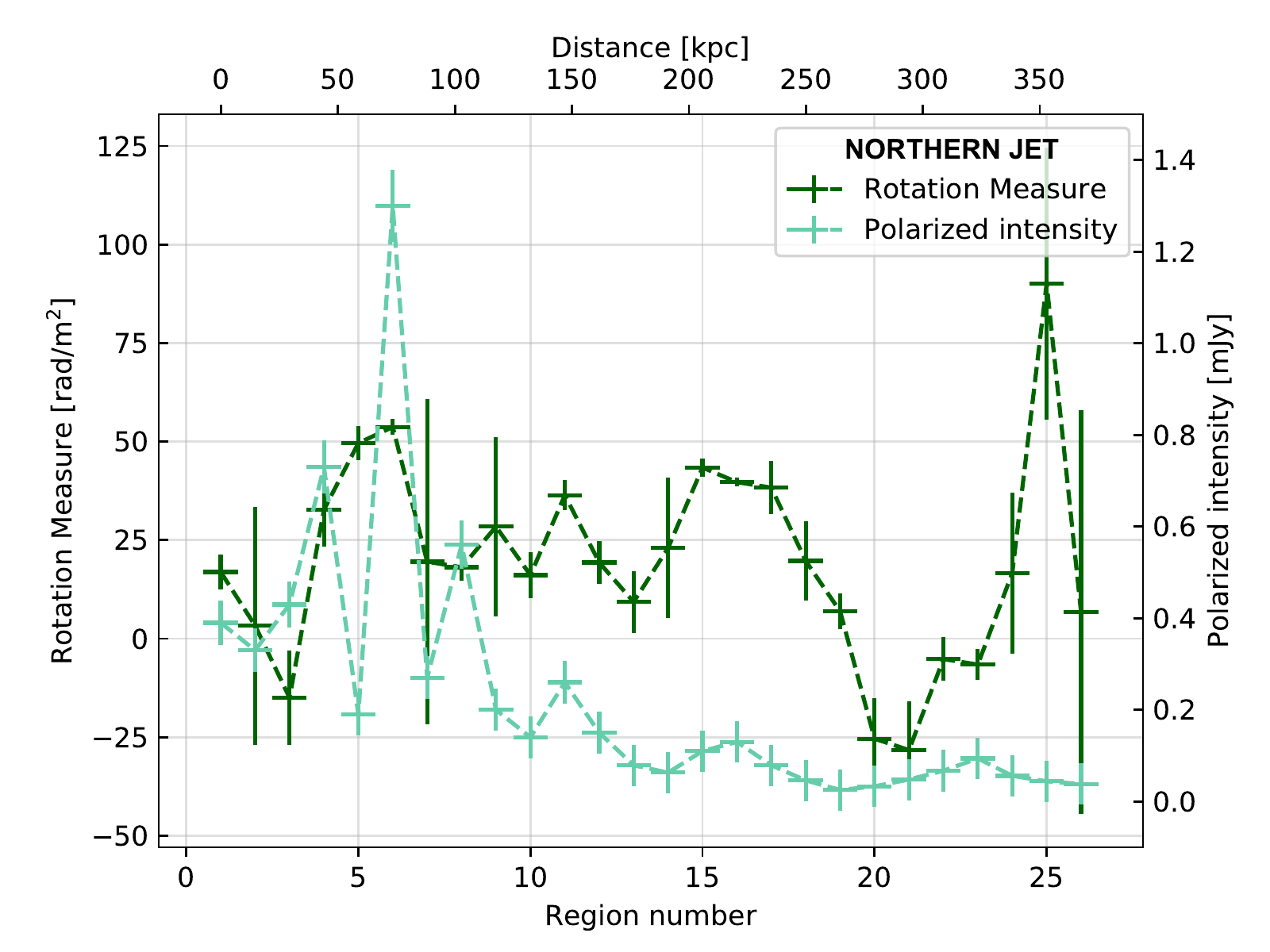}
    \end{minipage}  
    \hfill
    \begin{minipage}[t]{.49\textwidth}
        \centering
        \includegraphics[width=\textwidth]{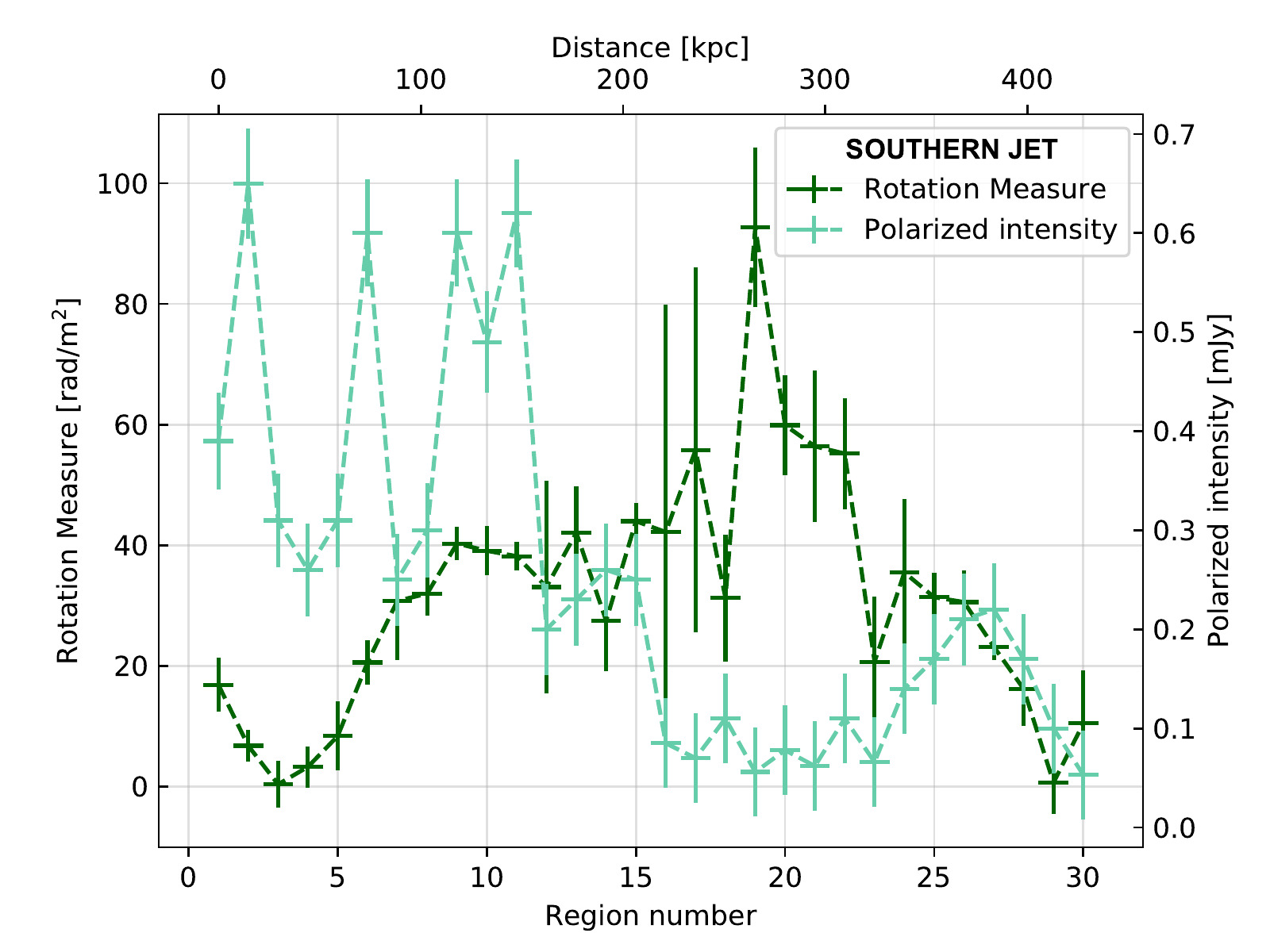}
    \end{minipage}
    \caption{HTII. Tracks of different observables along the galaxy jets with higher region numbers corresponding to larger distances to the head. The region number refer to the maps with a resolution of $15\arcsec\times15\arcsec$, the higher resolution regions are converted to that scale by taking the physical distances into account. For comparison, the head property is shown in each graph (1st region). \textbf{Left panels:} Northern tail. The values within the grey area are expected to be contaminated by a background source. \textbf{Right panels:} Southern tail. Values within the grey area might not belong to the galaxy jet or can be noise dominated. \label{fig:trackHTII}}
\end{figure*}

\begin{figure*}
    \begin{minipage}[t]{.49\textwidth}
        \centering
        \includegraphics[width=\textwidth]{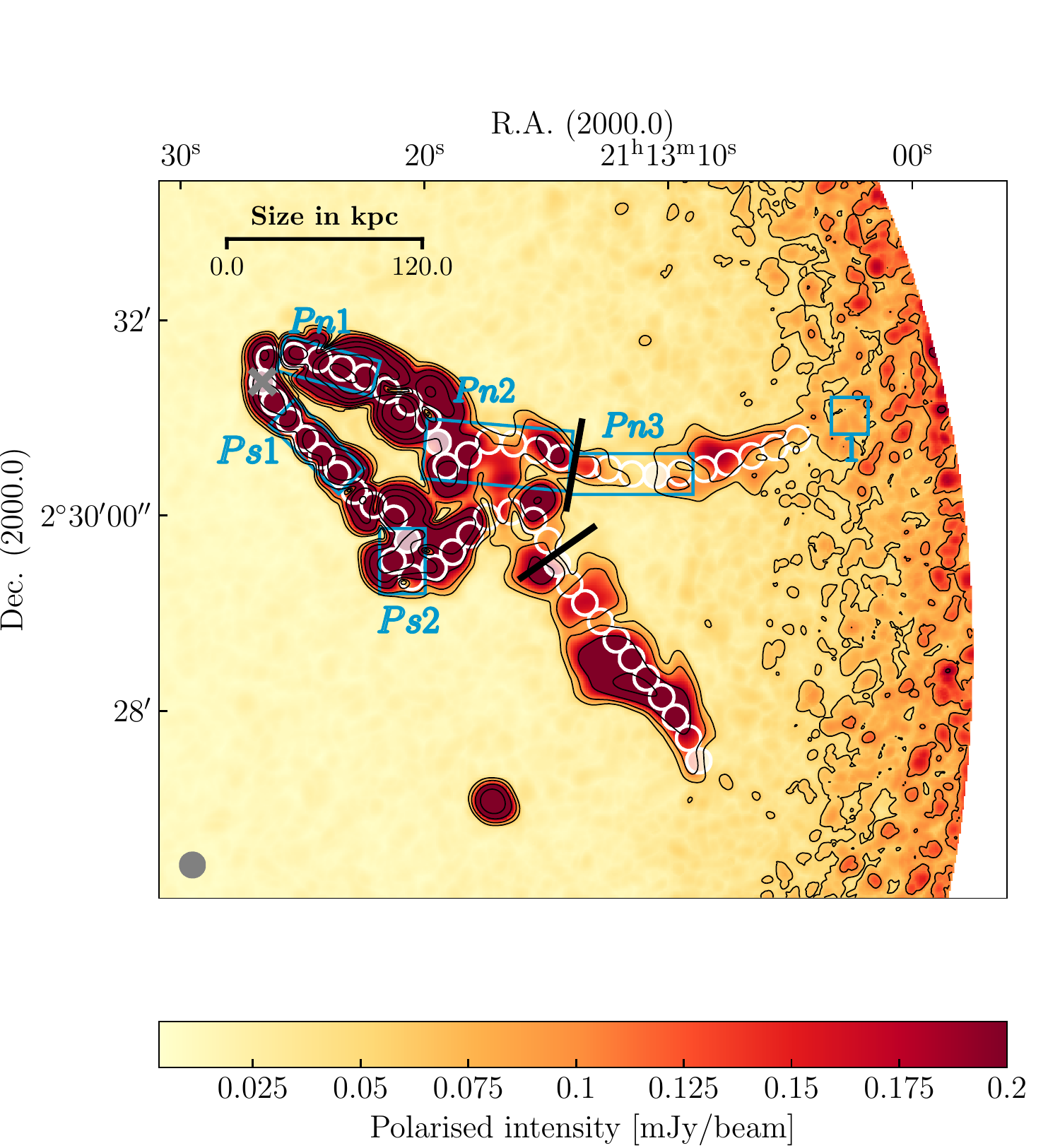}
    \end{minipage}  
    \hfill
    \begin{minipage}[t]{.49\textwidth}
        \centering
        \includegraphics[width=\textwidth]{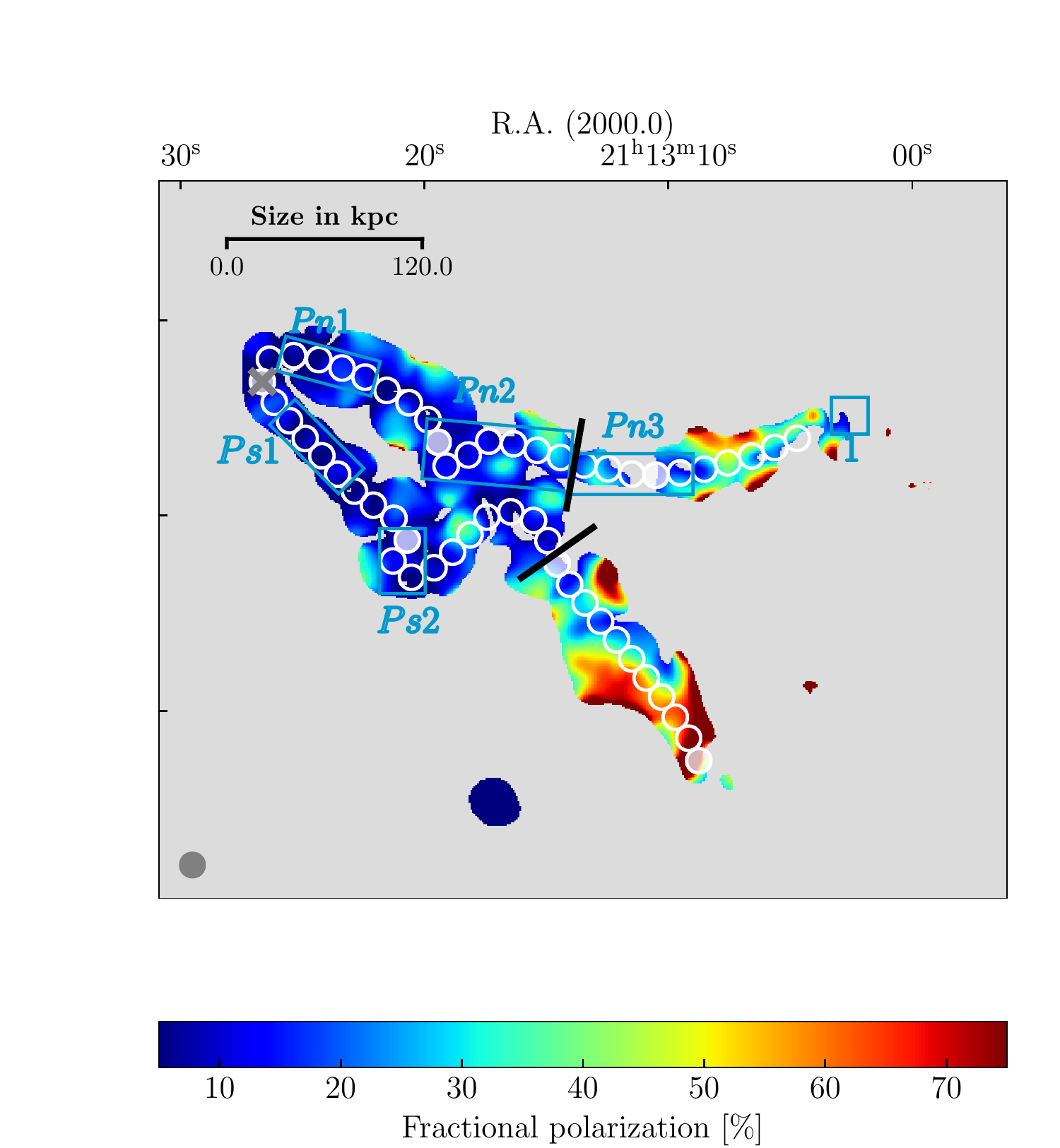}
    \end{minipage} 
     \begin{minipage}[t]{.49\textwidth}
        \centering
        \includegraphics[width=\textwidth]{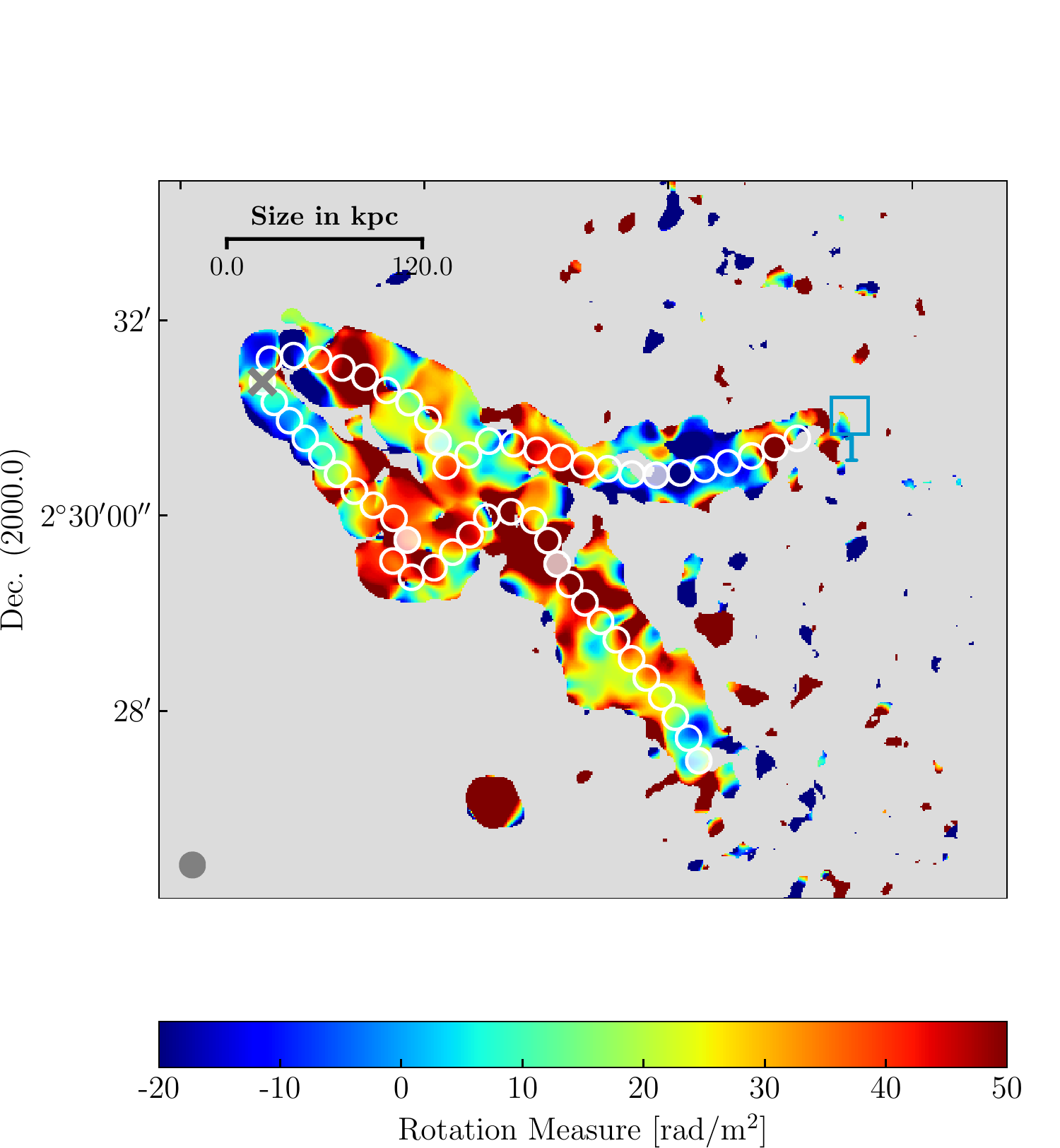}
    \end{minipage}  
    \hfill
    \begin{minipage}[t]{.49\textwidth}
        \centering
        \includegraphics[width=\textwidth]{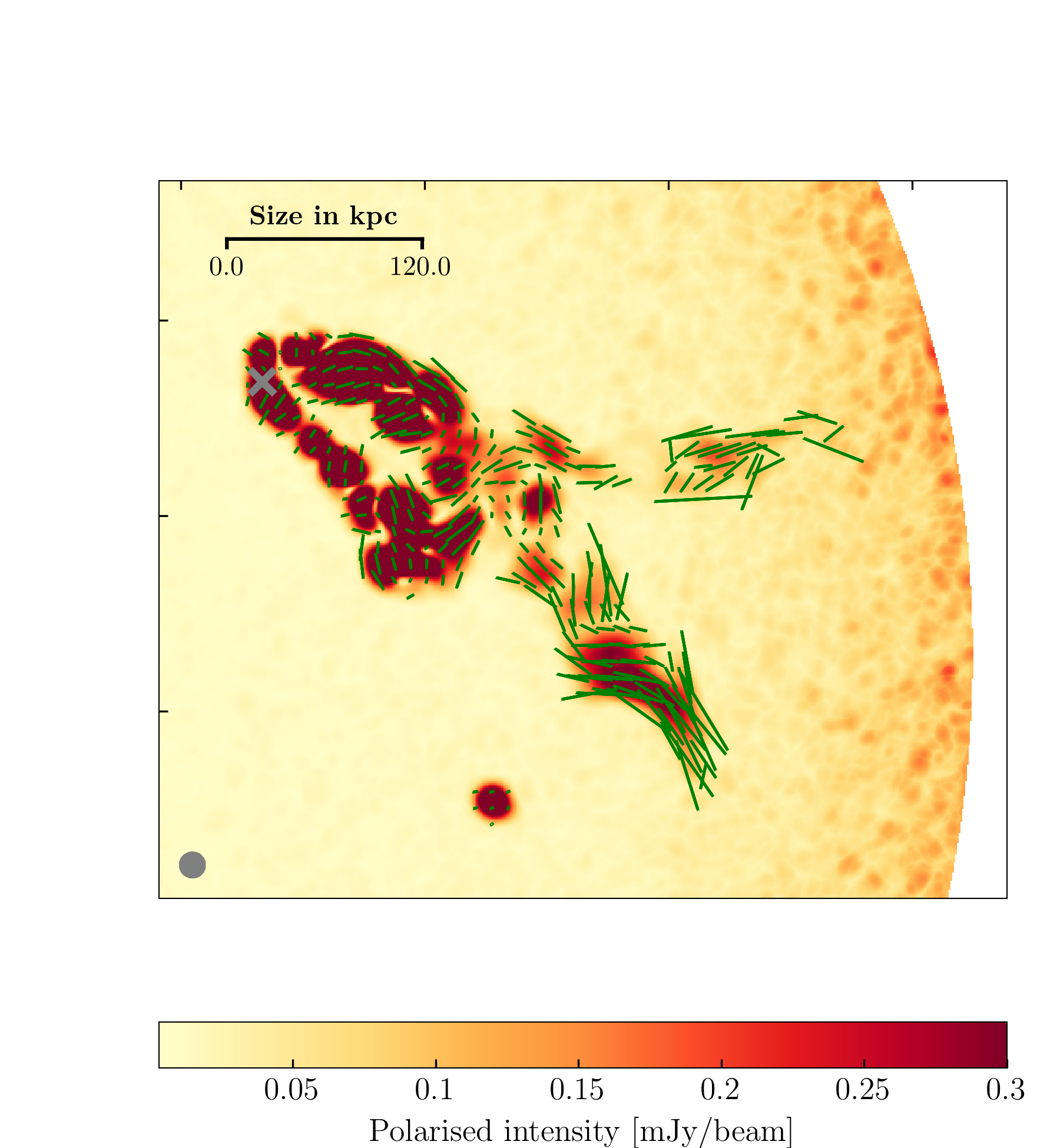}
    \end{minipage}
    \caption{HTII. Polarization results of the 2.7\,GHz data at a resolution of $15\arcsec\times15\arcsec$. White circular beam-sized regions are superimposed along the jet to track different properties centred on the central AGN marked by a grey cross. Each 10th region is filled in white to ease the comparison with Fig.~\ref{fig:trackHTI}. Regions discussed as possible optical sources producing line-of-sight emission {(`1')} and {plateaus (`Pn1', `Ps1', `Pn2', `Ps2', `Pn3')} are marked with blue {squares and rectangles, respectively}.
    The beam is shown in grey in the top right corner and {lightgrey} area contains no information. \textbf{Top left}: Polarized intensity map with superimposed contour levels of bias + $\epsilon \times (1,\;2,\;4,\;8,\;16,\;32,\;64,\;128,\;256$) with $\textrm{bias}=12\,\upmu$Jy/beam and $\epsilon = 45\,\upmu$Jy/beam. \textbf{Top right}: Fractional polarization map. \textbf{Bottom left}: RM map. \textbf{Bottom right}: Magnetic fields vectors weighted by $0.5\times$ (fractional polarization) on top of polarized intensity.} \label{fig:PolarizedIntensityHT2}
\end{figure*}

\subsubsection{Radio continuum and spectral index}

We show the radio continuum images at 2.7\,GHz in two different resolution{s}, at 1.4\,GHz, and the spectral index map derived between both frequencies in Fig.~\ref{fig:TotalIntensityHT2}.
HTII is showing two bent jets pointing away from the cluster cent{re}, in the direction of the main cluster axis, as indicated by the elongation of the BCG, the cluster galaxy distribution, and the X-ray surface brightness. 
The jets can be easily distinguished (white circular regions) at both resolutions (Fig.~\ref{fig:TotalIntensityHT2}, top left and right).
They are comparable in size and shape and their structure seems to become more complex at a distance of $\sim 120\,$kpc. In the top right panel of Fig.~\ref{fig:TotalIntensityHT2}, the jets seem to {broaden when they change direction}
(starting from the 20th white circular region superimposed on the emission) and weaken thereafter but at the 50th circular{,} region both jets broaden again. One can assume that the jets continue at a different location in the plane of the sky, where we see the weakened tail to be broadened again.
Several substructures (clumps) can be seen in the higher resolution jet that can partly also be identified in the lower resolution maps (Fig.~\ref{fig:TotalIntensityHT2}, top and bottom left), which is an indication for a more complex scenario in comparison to HTI.

At 1.4\,GHz and 15$\arcsec$ resolution the southern jet shows a projected extent of $\sim 560\,$kpc (unfolding the change of direction, see Fig.~\ref{fig:TotalIntensityHT2}, bottom left). At 1.4\,GHz the second part of the jet (15th circular region in the northern and 13th region in the southern jet, compare Fig.~\ref{fig:TotalIntensityHT2} bottom with top left) is found to be significantly broader than at 2.7\,GHz.
They show comparable substructures along the jets in both frequencies. These structures cannot be associated with detected optical counterparts (Fig.~\ref{fig:optical}, {top right}). However, the brightest point-like emission in the northern jet (marked with the `1' blue square in Fig.~\ref{fig:TotalIntensityHT2}) is probably caused by a background source (see Appendix \ref{App:optical} right panel of Fig.~\ref{fig:app_optical}) increasing the jet emission due to the line-of-sight integration.

HTII is reaching a total intensity (head + jets, for definition see Sect.~\ref{sec:data}) of $303.3\pm15.2$\,mJy {at 2.7\,GHz} and $609.0\pm30.5$\,mJy {at 1.4\,GHz} with the head contributing by $(9.1\pm0.5)$\,mJy and ($10.4\pm0.5)$\,mJy, respectively. The 2.7\,GHz flux is corrected for the contribution of two point sources, one east to the 10th and the other east to the 25th white circular region (Fig.~\ref{fig:TotalIntensityHT2}, top left). The spectrum at the head is found to be rather flat (spectral index of $-0.17\pm0.09$) and the spectral index is found to be decreasing through the jets with a per-circle mean slope of $-0.073\pm 0.006$, neglecting the two most distant regions in the southern tail.

Figure~\ref{fig:trackHTII} (top panels with the left one corresponding to the northern jet and the right one to the southern one) presents the total intensity at different frequenc{ies} and the spectral index along the jet{s} corresponding to the white circular regions in Fig.~\ref{fig:TotalIntensityHT2}. While the head is found to be the brightest region in the higher resolution data, we find a substructure next to the head that is brightest in both frequencies at lower resolution. Such an increase in total intensity is probably caused by a recollimation shock \citep{Perucho}, {which follows after} the jet pressure becomes lower than the pressure of the ambient medium.
Thereafter, the total intensity is generally decreasing except for the final part of the jets. While the increase in the northern tail can be associated with a possible background source detected in both frequencies, the southern tail can be affected by the low signal to noise ratio at 2.7\,GHz.
The spectral index is generally decreasing through the jets with the head being quite flat {while} the northern jet shows a steepening to -2.5 and the southern {one} to -3.5. The increase of the spectral index in the most distant jet regions{,} as well as the first plateau{s (regions `Pn1', `Ps1`)}{,} can be traced back to the before mentioned reasons. However, in the northern tail ({region `Pn2'}) and also partly in the southern tail ({region `Ps2'}) another plateau can be identified that can possibly be traced back to a change in motion (see Sect. \ref{sec:turbulence}). Moreover, a third plateau can be identified in the northern tail ({region `Pn3'}) or {a} kind of disruption in the decreasing spectral index in the southern tail ({circular regions after the black line in Fig. \ref{fig:TotalIntensityHT2}}). Possible scenarios will be discussed in Section
\ref{sec:AGNcycle}.

\subsubsection{Linear polarized radio emission and fraction} 

In Fig.~\ref{fig:PolarizedIntensityHT2} we show the images of the polarization observables of HTII at 2.7\,GHz.
Both, the head and jets of HTII{ (top left)}, show a comparable morphology in polarized and total intensity. It is noticeable that, even though the noise cut is chosen with care (4.5$\sigma$ of a {source-free} region close to the most extended jets, see Table~\ref{tab:3}) with respect to the most distant jet emission, the last dip of the northern jet is highly affected by the increased noise at the image edges (see Fig.~\ref{fig:PolarizedIntensityHT2}, top left). Here, we again see important differences in morphology: 1) Within the first 120\,kpc of the jet several clumps are detected that cannot be associated with the substructures in total intensity, 2) even though the clumps thereafter can be seen also in total intensity here they show a broader extent, and 3) the northern jet is divided into two ({region `Pn3', possibly due to the noise cut}).
The polarised emission of the southern jet of HTII is 75\,kpc less extended than its counterpart in total intensity. Here, again this might either be caused by the increased noise level at the edges or the substructures seen in the most distant southern jet part (31st to 35th circular region in Fig.~\ref{fig:TotalIntensityHT2}, top left) in total intensity {might not be a part of the galaxy jet emission}.

HTII has a polarized intensity of $0.39\pm0.02$\,mJy in the head and $34.0\pm1.7$\,mJy {in the} jet resulting in a fractional polarization of $4.3\pm5.5$~\% and $10.6\pm5.2$~\%, respectively. We find the RM to change its direction towards and away from the observer more than once but being mostly positive (i.e., the large-scale magnetic field points towards the observer) resulting in an r.m.s of $\sim 28$~rad~m$^{-2}$. The ordered magnetic field (presented by the green lines in the bottom right panel of Fig.~\ref{fig:PolarizedIntensityHT2} weighted by the fractional polarization) is found to be complex within the first 120\,kpc. It partly aligns with the jet direction but otherwise shows orthogonal direction. Some directional changes correspond to regions where the jet changes its direction in the plane of the sky. The more distant jet shows a clear alignment with the jet direction.

The quantitative analysis of the polarization properties along the northern and southern jet can be found in Fig.~\ref{fig:trackHTII} (middle and bottom left- and right-hand panels, respectively corresponding to the white circular regions shown in Fig.~\ref{fig:PolarizedIntensityHT2}). Overall, the polarized intensity decreases with distance from the {head} and the fractional polarization is increasing, but a lot of substructures are found through the tracks. The radio continuum jet appears to be smooth while the jet in polarized intensity shows a clumpy structure that could be caused by depolarization effects. Only slight changes in structure at 8.7$\arcsec$ (see Appendix \ref{App:high} Fig.~\ref{fig:app_high}) can be detected, however, higher resolution data {are} needed to test whether the clumpy structure is purely physical or not since depolarization effects can occur on much smaller scales.

In the northern jet, we find {three} peaks arising within the first 9 regions corresponding to substructures resolved in the higher resolution total intensity track. These peaks show a comparable width reaching a brightness much higher than the galaxy head. These structures are identified as rapid increases in fractional polarization. We find the fractional polarization to be almost steady within region `Pn1' and `Pn2' while it is generally steepening in between, increasing up to 40\,\% at region 23. Then, we find the fractional polarization to decrease again being consistent with the substructure that we already identified in the total intensity map. The RM is found to vary around 25\,rad~m$^{-2}$ within the first 15 regions showing a broader substructure at a distance where we find the first two polarization peaks. It decreases thereafter showing a significant magnetic field component pointing away from the observer ($-25$~rad~m$^{-2}$). We find the RM increasing at a location where the fractional polarization is decreasing.

Within the first 12 regions{,} the polarized intensity in the southern jet is peaking {four} times showing again a very bright emission with respect to the head, which again results in increased fractional polarization. {This} peaky substructure continues along the whole jet but with less intensity such that the fractional polarization {generally} begin to increase from about 10\,\% in region 16 to 75\,\% in region 30, {which also corresponds} to a very steep spectral index. The RM is found to be almost positive, increasing between the 3rd and 19th region up to 90\,rad~m$^{-2}$ and decreasing thereafter to 0\,rad~m$^{-2}$.

\subsection{X-ray properties}\label{sec:xray}

\begin{table*}
\def\arraystretch{1.2}
    \centering
\begin{tabular}[t]{r|r|r|r|r|r|r|}
name & $n_\mathrm{e}$ [cm$^{-3}$] & $k_\rmn{B}T$ [keV] & $P$ [erg\,cm$^{-3}$] & $\rho$ [g\,cm$^{-3}$] & $c_s$ [km/s] & $\mathcal{M}_\textrm{s}$ \\
  \hline
 HTI & $4.6\times10^{-4}$ & 3.2 & $4.3\times10^{-12}$ & $8.9\times10^{-28}$ & 898 & 1.1\\
 HTII & $3.9\times10^{-4}$ & 2.0 & $2.3\times10^{-12}$ & $7.5\times10^{-28}$ & 710 & 1.4\\
\end{tabular}
    \caption{Thermodynamical properties of the ICM estimated by extrapolating the de-projected spectral analysis presented in \citet{Mueller} at the galaxy locations: name used through the text, electron density, temperature, pressure, density, speed of sound, and sonic Mach number.}
    \label{tab:4}
\end{table*}
\begin{figure*}
    \centering
    \includegraphics[width=1.0\textwidth]{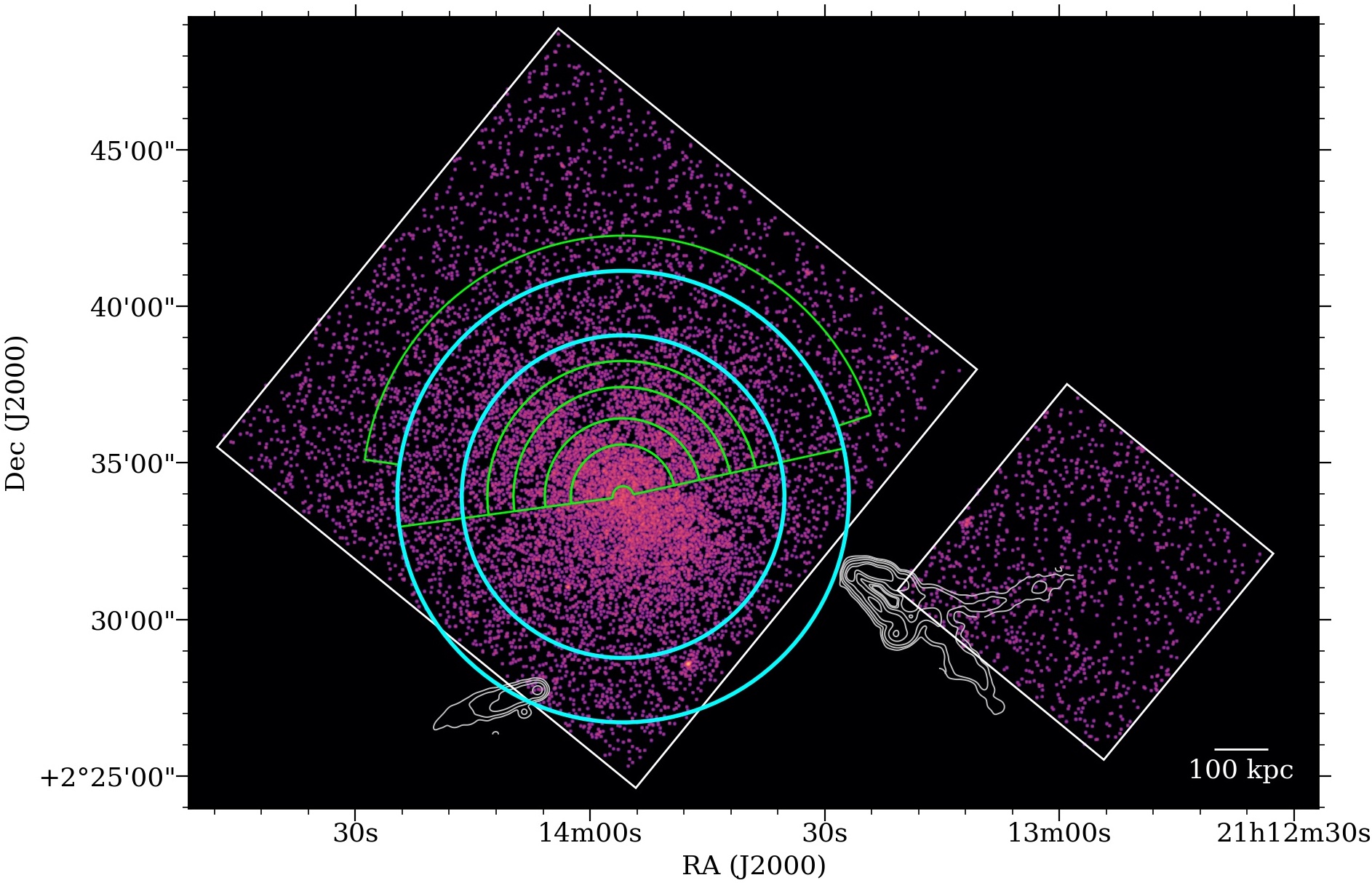}
    \caption{ {\it Chandra} image of  the  cluster in the  (0.5–2.0)\,keV band smoothed with a 2.5'' Gaussian. We show the contours of the radio emission of the head-tail galaxies (white) and the annuli used for the projected (cyan) and de-projected (green) spectral analysis presented in \citet[][Supplementary information]{Mueller}.}
    \label{X-ray}
\end{figure*}
We aim to use the Chandra observation to study the HT environment by analysing the {intracluster} medium at the location of the two galaxies individually.
In Fig.~\ref{X-ray} we show the annuli that were used to investigate the X-ray properties of the cluster at the position of the jellyfish galaxy JO206 \citep{Mueller} as well as the total power contours of the {HT} galaxies with respect to the Chandra mosaic. Despite the low photon statistics of the observation, we were able to successfully fit the de-projected electron density with a $\beta$-model profile \citep[][]{Cavaliere_1976}:
\begin{eqnarray}
    n_\textrm{e}(r)=n_0\left[1+\left(\frac{r}{r_\textrm{c}} \right)^2 \right]^{-3\beta/2},
\end{eqnarray}
where $n_0=2.61\times 10^{-3}$ cm$^{-3}$ is the central density, $r_\textrm{c}=112.14\arcsec$ is the core radius{,} and $\beta=0.44$ describes the ratio between thermal and gravitational energy of the plasma.

Here, we extend these results to extrapolate the thermal properties of the ICM surrounding the HT galaxies. Because the galaxies are actually outside the CCD, we can only infer the ICM properties from the de-projected spectral analysis, which was carried out under the assumption of spherical symmetry of the northern part of the cluster, and from the best-fitting $\beta$-model profile. However, we note that the cluster appears to be slightly asymmetric, with a putative over-density towards {southwest} in the direction of HTII (optical{ly} identified filament, Sect. \ref{sec:optical}), thus the assumption of spherical symmetry could not hold for the southern part of the cluster at the galaxies' cluster-centric distances. Unfortunately, the current data are too shallow to carry out a more reliable analysis, hence we stress that the results presented here should be considered approximate for the following estimates.

Considering the projected distance of the {HT} galaxies to the cluster {centre} (Table~\ref{tab:1}), the $\beta$-model yields electron densities, $n_e$, of $4.6\times10^{-4}$\,cm$^{-3}$ and $3.9\times10^{-4}$\,cm$^{-3}$ at the head of HTI and HTII, respectively. In Table \ref{tab:4} further cluster ICM properties with respect to the galaxies head can be found. The de-projected temperatures are measured within the green annuli (Fig.~\ref{X-ray}) with the second {outermost} corresponding to HTI and the outermost to HTII, the gas pressure via $P=1.83\, n_\textrm{e}kT$, the gas density $\rho=1.9\, \mu n_\textrm{e} m_\textrm{H}$ with the mean molecular weight $\mu\simeq0.61$ and the proton rest mass $m_\textrm{H}$, and the speed of sound $c_\textrm{s}=\sqrt{\gamma P/\rho}$, where $\gamma=5/3$ is the adiabatic index for a mono-atomic, thermal gas. Both HT galaxies have small line-of-sight velocities with respect to the cluster. However, we can see a strong in-plane component so that we assume the galaxy velocity to be equal to the {three-dimensional} cluster velocity dispersion (Table~\ref{tab:1}). This assumption is used to calculate the sonic Mach numbers $\mathcal{M}_\textrm{s}$ presented in Table~\ref{tab:4}. Both galaxies are found to move supersonically through the ICM ($\mathcal{M}_\textrm{s}>1$).

\section{Discussion}
\label{sec:discussion}

The observational findings suggest the following picture. HTI and HTII can be classified as standard NATs, experiencing jet bending by ram pressure as a result of the fast motion of the radio galaxies in the ICM potential (Sect.~\ref{sec:HT}). The plateau regions in {the} spectral index and slower decrease of total and polarized intensity suggest a transition of the gas flow from laminar to turbulent motions, which is accompanied by a re-energizing process of the electrons along the jet in HTI and HTII (Sect.~\ref{sec:turbulence}). While this explains the entire appearance of the jets of HTI the morphology of HTII calls for an additional explanation: we propose two epochs of AGN jet outbursts where the recent AGN activity connects to the earlier outburst via the spectral index plateaus seen in both tails and steeply declines in {the} spectral index along the jets towards larger distances (Sect.~\ref{sec:AGNcycle}).

\subsection{Bent head-tail morphology}\label{sec:HT}

Both galaxies show the brightest emission at their heads, as expected for FRI galaxies. They most probably fall into {the} cluster for the first time as indicated by the cluster filaments. Assuming that they move with velocities of order 1000\,km~s$^{-1}$ with respect to the cluster {centre} (corresponding to the cluster three-dimensional velocity dispersion) and {face} ambient ICM densities of about $4\times 10^{-4}$\,cm$^{-3}$ {we find} both {galaxies to be} consistent with the findings for typical NATs \citep{Venkatesan}. At the galaxy jets, we find a dip in total intensity (fast decrease and increase) within the first 6 analysis regions. Such a {behaviour} is also found in other HT galaxies \citep[e.g.,][]{Srivastava} and FRI sources \citep[e.g.,][]{Hada} and {follows as a direct consequence of} a recollimation shock \citep{Perucho}, which occurs {in response to the expansion} of the jets as they encounter a larger ambient ISM pressure.

We generally find the spectral index to steepen and the fractional polarization to increase with larger distance to the galaxy heads, which resembles other HT galaxies \citep[e.g.,][]{Srivastava, Gasperin}. Their ordered magnetic field component is aligned with the jet (and follows the bent jet morphology). Several substructures can be identified as plateaus for HTI and two for HTII in the spectral index tracks along the jets that will be addressed below. The increase in the spectral index map for HTI and the resulting decrease in fractional polarization is the result of the line-of-sight integration of the galaxy jet emission and a point source located at the very end of the jet. In case of HTII, another background source causes a comparable effect in the northern jet, while the fragmentation of the southern tail at 2.7\,GHz is causing the same {behaviour} possibly due to a bad signal-to-noise ratio.

The galaxies experience a ram pressure of $8.9\times10^{-12}$\,g\,cm$^{-1}$s$^{-2}$ (HTI) and $7.5\times10^{-12}$\,g\,cm$^{-1}$s$^{-2}$ (HTII), causing the bending of the galaxy jets in the opposite direction of the cluster {centre}. Using the geometry of the head-tail radio galaxies, we can estimate the jet quantities and denote the mass density, velocity, and radius of the jet by $\rho_\rmn{jet}$, $\varv_\rmn{jet}$, and $r_\rmn{jet}$, respectively. Initially, the two jets emerging from the supermassive black hole are assumed to be a cylinder of length $l_\rmn{jet}$. The ram pressure wind of mass density $\rho_\rmn{ICM}$ and velocity $\varv$ causes each jet to bent over a bending radius $r_\rmn{b}$. Equating the jet momentum $\rho_\rmn{jet} \varv_\rmn{jet} \uppi r_\rmn{jet}^2 l_\rmn{jet}$ with the transverse force due to the ram pressure wind that operates on a jet propagation timescale (along the bended path in steady state), $\rho_\rmn{ICM}\varv^2 2 r_\rmn{jet} l_\rmn{jet} \uppi r_\rmn{b} /(2\varv_\rmn{jet})$, we obtain \citep{PfrommerII}
\begin{equation}
  \label{eq:bend}
  \rho_\rmn{ICM} \varv^2\,\frac{\uppi r_\rmn{b}}{2 \varv_\rmn{jet}} = 
  \rho_\rmn{jet} \varv_\rmn{jet} r_\rmn{jet}\,\frac{\uppi}{2}.
\end{equation}
Hence, the background-to-jet density ratio can be estimated as 
\begin{eqnarray}
x_\rmn{jet}^{-1}\equiv\frac{\rho_\textrm{ICM}}{\rho_\textrm{jet}} = \frac{r_\textrm{jet}}{r_\textrm{b}} \left(\frac{\varv_\textrm{jet}}{\varv}\right)^2 \sim 0.067\times \left(\frac{0.7 c}{10^3\,\textrm{km}~\textrm{s}^{-1}}\right)^2 \sim 3000,
\end{eqnarray}
where $r_\textrm{jet}\approx1$\,kpc is inferred from VLBI observations of nearby HTs and corresponds to the width of the tail at the bending radius $r_\textrm{b}$ for a resolution of $1\arcsec$ (at a distance of 200\,Mpc), $r_\textrm{b}\approx15\,$kpc is the estimated curvature radius of the northern tail of HTII {(estimated within the first 6 white circular regions in the high resolution image shown in the top right panel of Fig. \ref{fig:TotalIntensityHT2})}, and $\varv_\textrm{jet}\approx0.7c$ \citep{Laing}. 

Entrainment of ambient ICM into the jet via Kelvin-Helmholtz instabilities increases $\rho_\textrm{jet}$ with distance to the galaxy head and we adopt the following two cases: (i) $\rho_\textrm{jet} \sim 0.01\rho_\textrm{ICM}$ (entrainment was at work and reduced the jet density by a factor of 30, which we consider a realistic case) and (ii) $\rho_\textrm{jet} \sim 0.1 \rho_\textrm{ICM}$ (a very effective entrainment).

\subsection{Turbulence in AGN jets} \label{sec:turbulence}

As a next step, we explore whether a transition from laminar to turbulent flow can explain the fanning out of the high-resolution 2.7\,GHz total intensity maps in the top right-hand panels in Figs.~\ref{fig:TotalIntensityHT1} and \ref{fig:TotalIntensityHT2}.
In particular the winding morphology of both jets of HTII after the 20th high-resolution region (top right-hand panel in Fig.~\ref{fig:TotalIntensityHT2}) are reminiscent of a firehose instability which first grows linearly and develops into a non-linear turbulent state. The transition to turbulence in {a} pipe flow takes place at a critical Reynolds number around \citep{Eckhardt}
\begin{eqnarray}
\label{eq:Re}
  \textrm{Re} \sim \frac{L}{\lambda_\textrm{eff}}\frac{\varv}{\varv_\textrm{th}} \gtrsim \rmn{Re}_\rmn{crit}\approx2300,
\end{eqnarray}
where $L$ corresponds to the width of the jet, $\lambda_\textrm{eff}$ is the effective mean free path, $\varv$ is the velocity of the jet, and $\varv_\textrm{th}$ the thermal velocity of the particles in the jet. At the 20th region of the high-resolution 2.7\,GHz total intensity map (Fig.~\ref{fig:TotalIntensityHT2}{, top right}), we infer $L\approx24$\,kpc (30\,kpc) for the southern (northern) jet of HTII. While we aim to investigate a general picture, in the following we adopt numerical values specific to HTII and find them to be generally also valid for HTI. Adopting $\varv=0.7c$ and assuming that the jet is in pressure equilibrium with the ICM, we obtain
\begin{eqnarray}
\label{vth}
\varv_\textrm{th} = \sqrt{2 k_\textrm{B}T / m_\textrm{p}} = \sqrt{2 P_\textrm{ICM} / \rho_\textrm{jet}}
\end{eqnarray}
so that we can estimate the effective mean free path
\begin{eqnarray}
\label{eq:Re2}
  \lambda_\textrm{eff} \lesssim \frac{L}{\textrm{Re}_\rmn{crit}}\frac{\varv}{\varv_\textrm{th}} \sim 0.3~\rmn{kpc},
\end{eqnarray}
for $x_\rmn{jet}\equiv\rho_\rmn{jet}/\rho_\rmn{ICM} = 0.01$ and $\lambda_\textrm{eff} \lesssim1~\rmn{kpc}$ for $x_\rmn{jet}=0.1$. By contrast, the standard Spitzer mean free path is given by 
\begin{eqnarray}
\label{Spitz}
\lambda_\textrm{mfp,Spitzer} \sim 400 \left(\frac{n_\textrm{jet}}{7.5\times10^{-6}\,\textrm{cm}^{-3}}\right)^{-1} \left(\frac{k_\textrm{B}T_\textrm{jet}}{170\,\textrm{keV}}\right)^2\,\textrm{Mpc}, 
\end{eqnarray}
where $n_\textrm{jet}$ and $T_\textrm{jet}$ are the density and temperature for our jets with $x_\rmn{jet}=0.01$, more than $10^6$ times larger than $\lambda_\textrm{eff}$. Even if we adopted $x_\rmn{jet}=0.1$, we find $\lambda_\textrm{mfp,Spitzer} \sim450$\,kpc, which is also significantly larger than the corresponding effective mean free path. 

Thus, we find in both cases that turbulent magnetic fields or plasma instabilities are essential to reduce the Spitzer mean free path by three to six orders of magnitude so that the flow can transition to turbulence over the travel length of the jet. We emphasize that such compressible turbulence is required in order to re-energi{z}e electrons and hence to (partially) balance the strong radiative cooling losses along the jet and to explain the (i) spectral index plateaus and (ii) larger jet lengths in comparison to the combined (synchrotron and inverse Compton) cooling length (see Sect.~\ref{sec:L_cool}). Turbulent re-acceleration via wave-particle energy exchange by means of transit time damping \citep{Brunetti2007,Brunetti2011,Miniati2015,Pinzke2017} is preferred over diffusive shock acceleration at medium to strong internal jet shocks, which would instantaneously flatten the radio spectral index to -0.5 while cooling would cause the spectrum to steepen afterwards, which is not observed. While those works apply the process of turbulent re-acceleration to radio halos, we adopt the same concept for the extended radio emission along the faint radio tails, similarly to the finding by \citet{Gasperin}. Indeed, we observe a gentle re-acceleration process that is able to temporarily balance cooling, a natural characteristic of stochastic second-order Fermi (Fermi-II) acceleration. 

We note that this strong case for magnetic or weakly collisional plasma effects to {extend} the turbulent cascade to much smaller scales in comparison to the Spitzer value is complementary to the inference of a smaller mean free path in the bulk of the ICM by analysing {surface density fluctuations in deep \textit{Chandra} observation of the Coma cluster \citep{Zhuravleva_2019} and} velocity structure functions of H$\alpha$ filaments in three nearby galaxy clusters \citep{Li2020,Wang2021}.

\subsection{Does the complex HTII morphology reveal two AGN outbursts or a previous shock passage?} \label{sec:AGNcycle}

In HTII, we observe a continuous decrease of the total intensity starting from the head in the top panels of Fig.~\ref{fig:trackHTII} until region 15, after which the radio emission is maintained at a rather constant level. Associated with this finding are spectral index plateaus in the further evolution of the jets (Fig.~\ref{fig:TotalIntensityHT2}, region {`Pn3'} for the northern tail and {subsequent regions after black line} for the southern tail) and an increasing fractional polarization with the magnetic field vectors primarily aligned with the bent jets. Taken together, this suggests a physical mechanism, such as Fermi-II acceleration that balances cooling and is able to re-accelerate electrons, which effectively offsets/slows down electron {ageing} and maintains the electron spectral index in the radio-emitting energy band with Lorentz factor $10^3$ to $10^4$ \citep[e.g., Figure~7 of][]{Winner2019}.

In particular, we observe a minimum of the total radio intensity around high-resolution region 50 (top right-hand panel of Fig.~\ref{fig:TotalIntensityHT2}) after which the emission is getting stronger and shows a more turbulent morphology in line with being gently re-energi{z}ed by interacting with compressible magnetic modes. Thereafter, the electrons quickly age ({lose} their energy), which could be interpreted as an earlier AGN outburst of those {pairs} of AGN jets. The change in morphology observable in both jets can be the result of two AGN outbursts, with the in-active jets are dislocated (behind the active jets) due to the ram-pressure the jets are facing. There is a weak bridge emission connecting these inactive lobes with the active jets. This suggests that the AGN jet is never completely off for a long time but rather varies in jet luminosity as it passes through a period of low luminosity between the two strong AGN jet outbursts. This would be consistent with the high AGN jet duty cycles found in a complete sample of radio-mode feedback in cool core clusters \citep{Birzan2012}.

There is an alternative interpretation for our findings: the passage of an ICM shock across the NAT could give rise to shock compression of cosmic ray electrons and the jet magnetic field. In particular, shock-induced vertical motions in the tails can drive coherent turbulent dynamo processes that amplify the magnetic fields significantly after initial shock compression \citep{O'Neill2019shocks}, which could possibly explain the turbulent morphology seen towards the late-time jet tail structures. In fact, synthetic radio observations of three-dimensional MHD jet simulations reveal an extended time period after the formation of the NAT in which it displays a nearly steady-state morphology and has approximately constant integrated fluxes and a self-similar, curved integrated spectrum \citep{O'Neill2019}, not unlike what we observe in our active jet region. After {the} passage of an ICM shock across the NAT, the shocked radio {emission} is a function of the Mach number, which implies strong variations in intensity, spectral, and polarization properties throughout the shocked tail. In particular, the radio emission can be rejuvenated due to shock-compression and magnetic field amplification as a result of a small-scale magnetic dynamo \citep{O'Neill2019shocks,Nolting2019b}.

{However, if the shock were still interacting with the tails, then those regions overrun by the shock should show clear morphological differences in total intensity and as well as in the spectral index \citep[see Figure~7 of][]{O'Neill2019shocks} for which we have no evidence in Fig.~\ref{fig:TotalIntensityHT2}. If the shock had already passed the tails the result depends on the geometry: (i) if the shock propagated perpendicular to the tails, then it would have first re-accelerated one tail and then the other so that the first tail would have had time to cool and therefore should show a steeper spectral index in comparison to the second tail \citep[see Figures~8, 11, and 14 of][]{O'Neill2019shocks}, which is inconsistent with our spectral index findings in Fig.~\ref{fig:TotalIntensityHT2}; and (ii) if the shock propagated along the tail, then it would have produced a torus \citep{Jones2017,Nolting2019a}, which is also inconsistent with our observed morphology and thus renders these shock interaction scenarios implausible.} Moreover, the current {\it Chandra} observation does not allow to properly investigate the presence of shocks in the ICM in correspondence of the radio tails. Deeper X-ray observations are needed to further test this scenario.

Thus, this comparison points to a scenario with a varying jet luminosity and two strong outbursts. In particular, the interaction of the NAT with a turbulent ICM \citep{Heinz2006,Ehlert2021} can explain the complex morphology of HTII, possibly also aided by magnetic draping of ICM magnetic fields across these jet lobes \citep{Lyutikov2006,2007Ruszkowski,Dursi,Ehlert2018}, which could explain part of the strongly coherent magnetic field structures seen in radio polarisation \citep{Pfrommer2010}.

\subsection{Electron cooling length}
\label{sec:L_cool}

The estimate of the electron cooling length along the AGN jets is complicated by the uncertainty of the transport velocity of the radio-emitting relativistic electrons. Initially, the jet transports the electrons close to relativistic speeds. Entrainment processes as a result of Kelvin-Helmholtz instabilities at the jet boundaries and magnetic re-connection of jet and ICM magnetic field slow the jet down and may cause a larger hadronic composition, which is characteristic of FRI jets \citep{Croston2018}. Once the jet fades out, it will eventually asymptotically approach the ICM velocity (in addition to buoyancy effects that can be neglected for HT galaxies). Hence, we concentrate on the HTII system because it shows two possible jet outbursts (see Sect.~\ref{sec:AGNcycle}) so that the old radio lobe has approximately a relative velocity $\varv$ of the ram pressure wind in the rest frame of the galaxy. 

First, we investigate the magnetic field strength in the galaxy jets that are generated by the central AGN and possibly amplified in the jet as an order of estimate to restrict the tail length neglecting reacceleration processes of the electrons. 
The magnetic field strength in the jets can be approximated by 
\begin{eqnarray}
  B=\sqrt{8\uppi P_{B}},
\end{eqnarray}
where $P_\textrm{B}$ is the magnetic pressure. Assuming that the surrounding ICM and the jets are in pressure equilibrium and equipartition of the cosmic-ray and magnetic pressure ($P_\textrm{ICM}=P_\textrm{jet}=P_\textrm{B}+P_\textrm{CR}\simeq2P_\textrm{B}$) we find the magnetic field strength in the jets to be $\sim 5\,\upmu$G for HTII. This simplistic assumption provides us with a good estimate {of} the magnetic field strength of the galactic jets. While they are generally found to be highly dynamic, this estimate is nevertheless not too far from results obtained with full three-dimensional MHD simulations \citep{Ehlert2021}.

The cooling time of electrons that emit into a radio synchrotron frequency $\nu_\mathrm{syn}$ is defined by \citep[e.g.,][]{PfrommerII}
\begin{eqnarray}
  t_\mathrm{cool} =
  \frac{\sqrt{54\pi m_\mathrm{e} c\, e B \nu_\mathrm{syn}^{-1}}}
  {\sigma_\mathrm{T}\,(B_\mathrm{cmb}^2+B^2)},
\end{eqnarray}
where $m_\mathrm{e}$ and e is the mass and charge of the electrons, c the speed of light, $\sigma_\rmn{T}$ is the Thompson cross section, $B_\textrm{cmb}=3.2 (1+z)^2~\mu$G is the equivalent magnetic field of the cosmic microwave background (at redshift $z$), and $B$ the magnetic field strength of the jet. Generally, the cooling time reaches its maximum at $B=B_\mathrm{cmb}/\sqrt{3} \simeq 2.06\,\upmu\mathrm{G}$ corresponding to a cooling length of $\sim200$\,kpc at 1.4\,GHz while assuming a HT velocity of $\varv\sim1000\,\rmn{km~s}^{-1}$ as explained above.
To achieve a more realistic value for the cooling length we adopt the equipartition values estimated above.
We find, at an emission frequency of 1.4\,GHz, a cooling time of $t_\textrm{cool}=9.5\times 10^7$\,yr ($B=5\,\umu$G for HTII) which corresponds to cooling length of $\sim 100$\,kpc. Most importantly, the cooling length of the electrons is found to be shorter than the maximum jet length of 560\,kpc found at 1.4\,GHz (Table~\ref{tab:1}). This makes a strong case for re-acceleration processes and is in line with our evidence for a turbulent flow in the relic jet fluid as explained in Sect.~\ref{sec:turbulence}.

\section{Conclusion}
\label{sec:conclusion}

In this work{,} we have {studied two HT galaxies members of the same cluster IIZW108 presenting observations} in radio continuum and polarization, optical photometry and spectroscopy, and X-ray. We presented general properties of the head and tail in total and polarized intensity, their spectral index, fractional polarization, RM, and ordered magnetic field component as well as the evolution of these properties by building track{s} through the galaxy jets. A summary of the values for the head and jets can be found in Table~\ref{tab:2}. We estimated important properties of the cluster ICM at the location of the HT galaxies, which are found to be in agreement with other HT galaxies from the literature (see Table~\ref{tab:4}). 

We found very extended jets pointing away from the cluster {centre} {bent} by the ram-pressure the galaxy is facing {while} falling into the cluster. We were therefore able to classify both galaxies as NATs. Both galaxies are found to have bright heads and diffuse jets and show an increasing flux at the most distant regions, which coincides with optical background sources and is therefore expected to be caused by line-of-sight projection effects such that the morphology is found to be in agreement with a typical FRI source, expected for NATs. We found the spectral index to 
{steepen along} the jets reaching values of -2.5 for HTI and -3.5 for HTII, while the fractional polarization was found to increase along the jets up to 70\,\%. The RM is found to be quite low in both galaxies with an r.m.s. of $\sim 25$\,rad\,m$^{-2}$ and the ordered magnetic field component is generally found to be aligned with the jet directions. Nevertheless, we were able to identify exceptions, that we explained as follows:

(i) The galaxy jets experience a transition from laminar to turbulent flow, which causes gentle re-acceleration of the electrons that were firstly accelerated by the central AGN. We suggest turbulent re-energization as a possible process to explain the spectral index plateaus and the extended jets in both HT galaxies.  

(ii) The complex morphology of HTII can be explained by two AGN outbursts (duty cycle) and a varying jet luminosity. The jets can be divided into two jets that are found to be dislocated due to the ram-pressure the galaxy is facing (the inactive jet behind the active jet), each corresponding to one AGN {outburst}. However, we find a weak bridge connecting the {two-component} jets suggesting that the AGN is never turned off completely. The optical spectra complement this picture where we found that both galaxies show no current AGN activity. Will another AGN outburst occur or might the AGN feeding material be shattered or stripped from the galaxy {centre}? Such questions need to be studied in further higher resolution data for HT galaxies and also in theoretical simulations.

\section*{Acknowledgements}

We are grateful to the anonymous referee for her/his
constructive comments. We acknowledge Paolo Serra, Mpati Ramatsoku, and Jacqueline van Gorkom for their contribution to the L-band (1.4\,GHz) data products used for this research.
This research has made use of JVLA (project 17A-293, PI Poggianti and project 18B-018, PI Poggianti) and Chandra observation (obsID 10747, PI Murray). 
CP acknowledges support by the European Research Council under ERC-CoG grant CRAGSMAN-646955.
AI, RP, MG, and BV acknowledge the Italian PRIN-Miur 2017 (PI A. Cimatti).
AlMo and BV acknowledge the financial contribution from the agreement ASI-INAF n.2017-14-H.0 (PI Moretti). 
AlMo, MG, BP, BV, and AB acknowledge funding from the INAF main-stream funding programme (PI B. Vulcani).
AL acknowledges the financial support of the National Agency for Research and Development (ANID) / Scholarship Program / DOCTORADO BECA NACIONAL/2019-21190049.
YJ acknowledges financial support from CONICYT PAI (Concurso Nacional de Inserci\'on en la Academia 2017) No. 79170132 and FONDECYT Iniciaci\'on 2018 No. 11180558.
Based on observations collected at the European Organization for Astronomical Research in the Southern Hemisphere under ESO programme 196.B-0578. This project has received funding from the European Research Council (ERC) under the European Union's Horizon 2020 research and innovation programme (grant agreement No. 833824).

\section*{Data Availability}
The data underlying this article will be shared at reasonable request to the corresponding author. The original data sets for the continuum analysis (JVLA) and X-ray studies ({\it Chandra}) can be downloaded from the corresponding archives. 




\bibliographystyle{mnras}
\bibliography{literature} 




\appendix

\section{Optical counterparts}\label{App:optical}
\begin{figure*}
    \centering
    \includegraphics[width=0.49\textwidth]{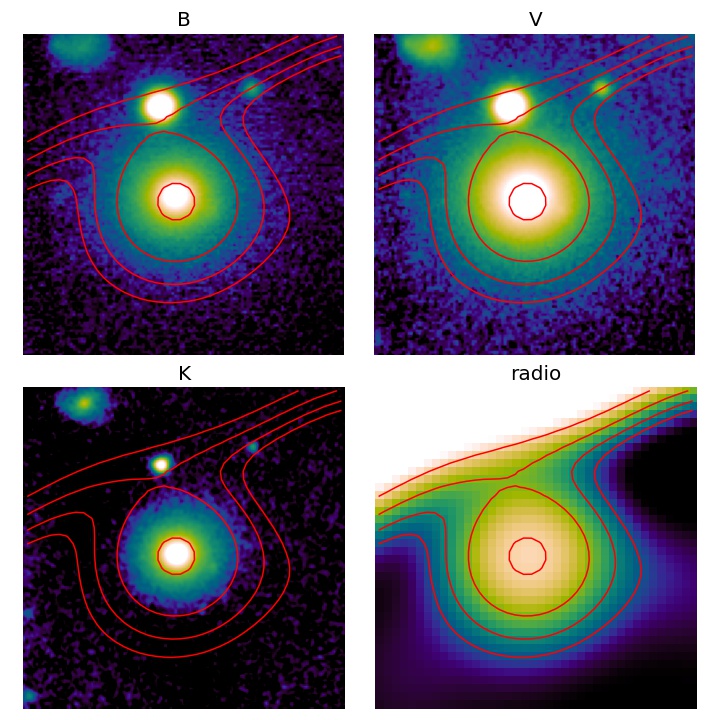}
    \includegraphics[width=0.49\textwidth]{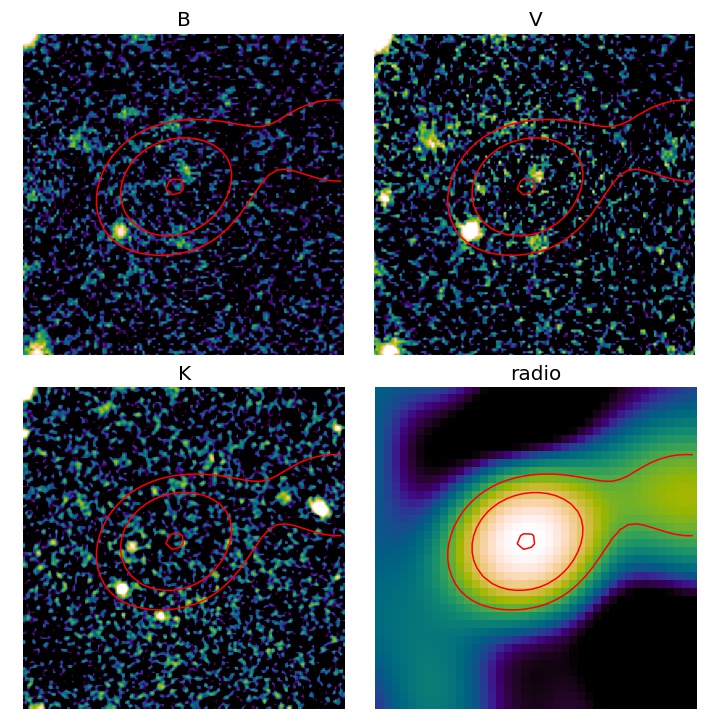}
    \includegraphics[width=0.49\textwidth]{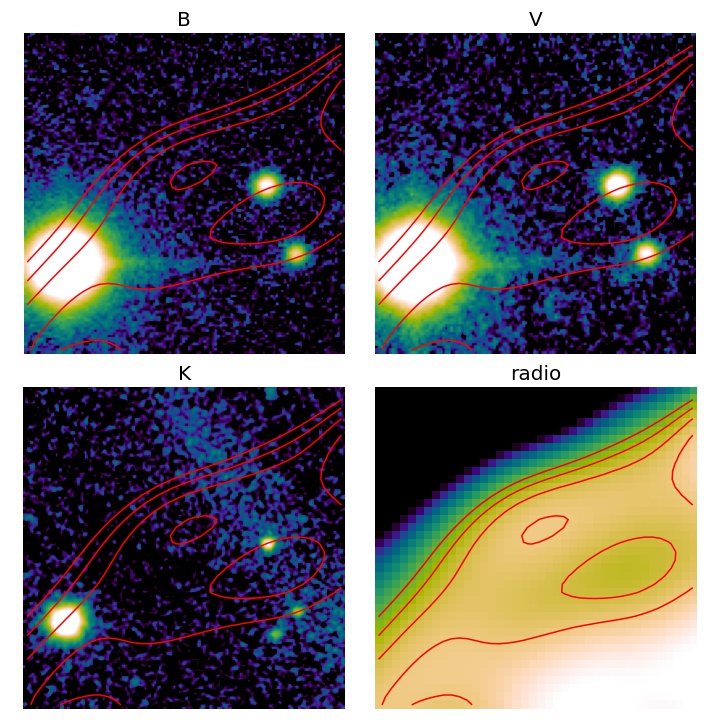}
    \includegraphics[width=0.49\textwidth]{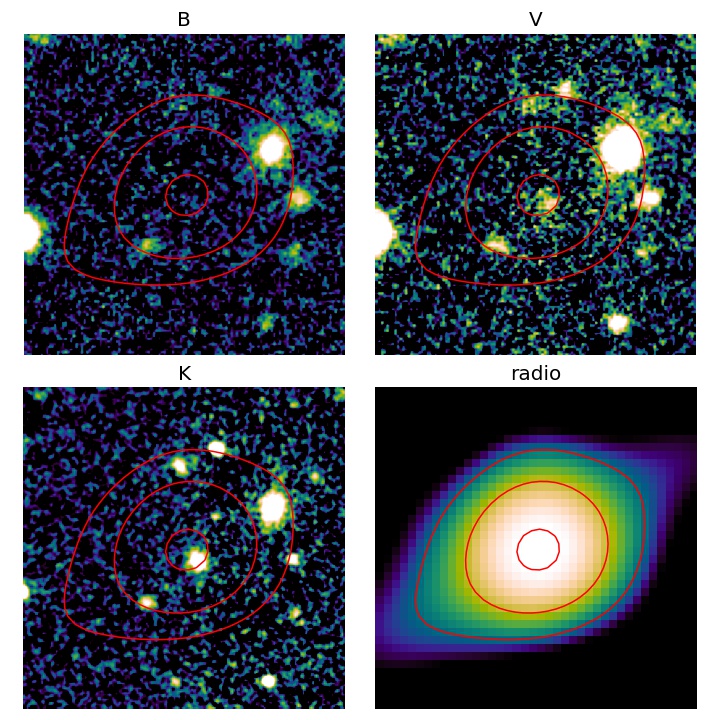}
    \caption{Optical images (B, V, and K) at the location of four blue squares shown in the main text Figures (`1' top left, `2' top right, and `3' bottom left of HTI and `1' bottom right of HTII,) found to be in superposition with the diffuse emission of the galaxy jets. The radio contours are shown in red and the last panel of each source shows the continuum emission for comparison.}
    \label{fig:app_optical}
\end{figure*}

Figure~\ref{fig:app_optical} shows three different optical filter observation{s} of a region where we detect bright radio sources at 2.7\,GHz corrupting the total intensity, spectral index, and/or fractional polarization findings of the diffuse jet emission. The optical source in the top left panels (corresponds to the `1' square in all figures of HTI, Figs.~\ref{fig:optical}, \ref{fig:TotalIntensityHT1}, and \ref{fig:PolarizedIntensityHT1}) coincides with the peak emission in radio continuum and is therefore expected to be its optical counterpart. Overall properties are corrected for this source by modelling its contribution by a circular mask, the tracks through the jets are not affected by this source. The top and {bottom-right} panels (corresponding to `2' of HTI, Figs.~\ref{fig:optical}, \ref{fig:TotalIntensityHT1}, and \ref{fig:PolarizedIntensityHT1} and `1' of HTII, Figs.~\ref{fig:optical}, \ref{fig:TotalIntensityHT2}, and \ref{fig:PolarizedIntensityHT2}, respectively) show two sources that are found to be in superposition with the jet emission. Both can be identified with an optical source, the first being quite faint, also in {the} V band, where it is found to be brightest while the second stated is found to be strong in V and K band. Their peak in continuum emission does not coincide with the probable optical counterpart, however, taking the low resolution in radio continuum as well as the superposition with the diffuse jet emission into account the small coordinate offset is negligible. We identify these two sources as {the} origin for the bright continuum emission and discussed our findings respectively. The bottom left panel of Fig.~\ref{fig:app_optical} coincides with the `3' square in Figs.~\ref{fig:optical}, \ref{fig:TotalIntensityHT1}, and \ref{fig:PolarizedIntensityHT1}, where we find very bright polarized emission but no associated bright continuum emission. Its origin cannot be traced back to one specific optical background source and remain unclear.

\section{Depolarization?} \label{App:high}

\begin{figure*}
    \centering
    \includegraphics[width=0.49\textwidth]{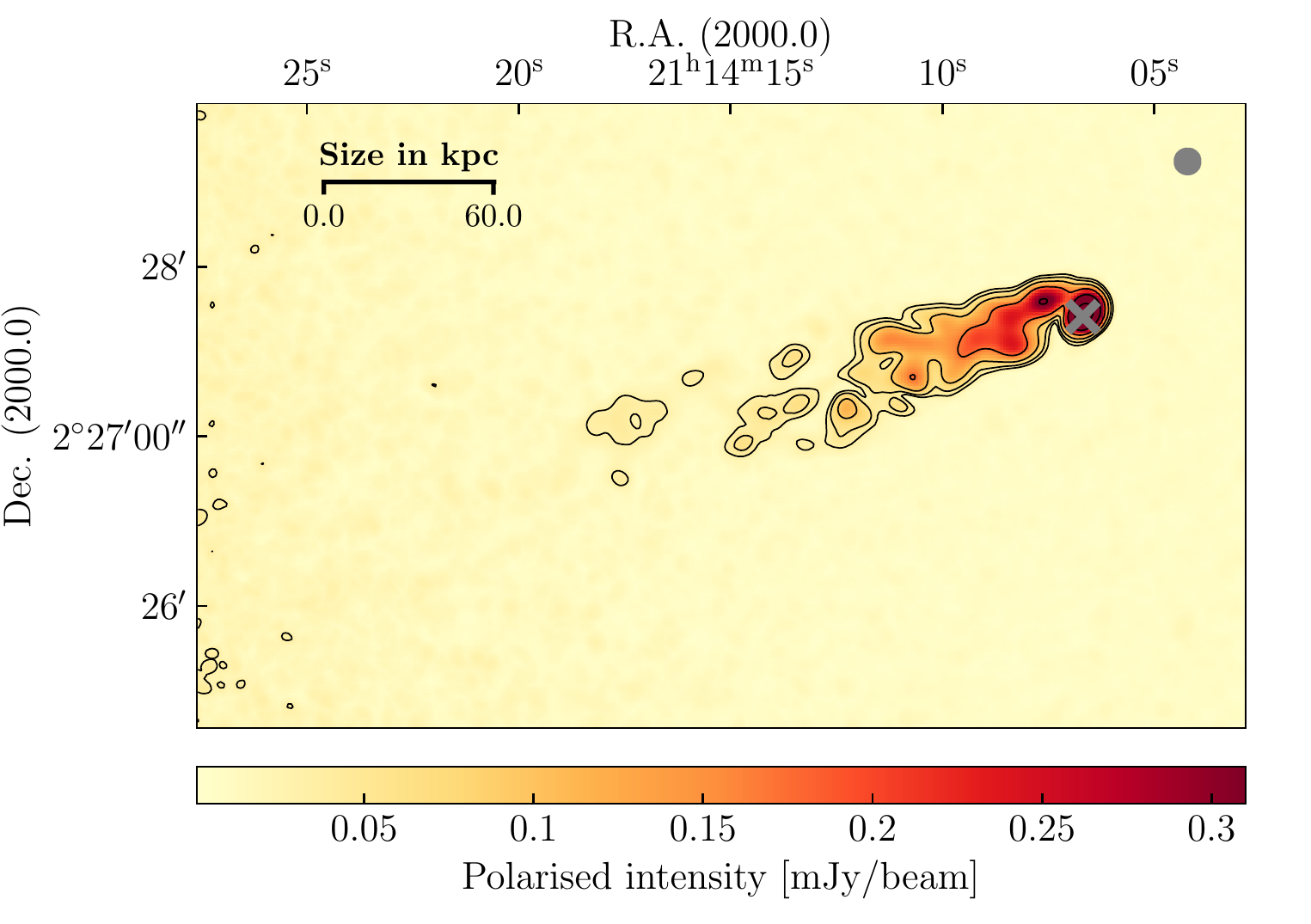}
    \includegraphics[width=0.49\textwidth]{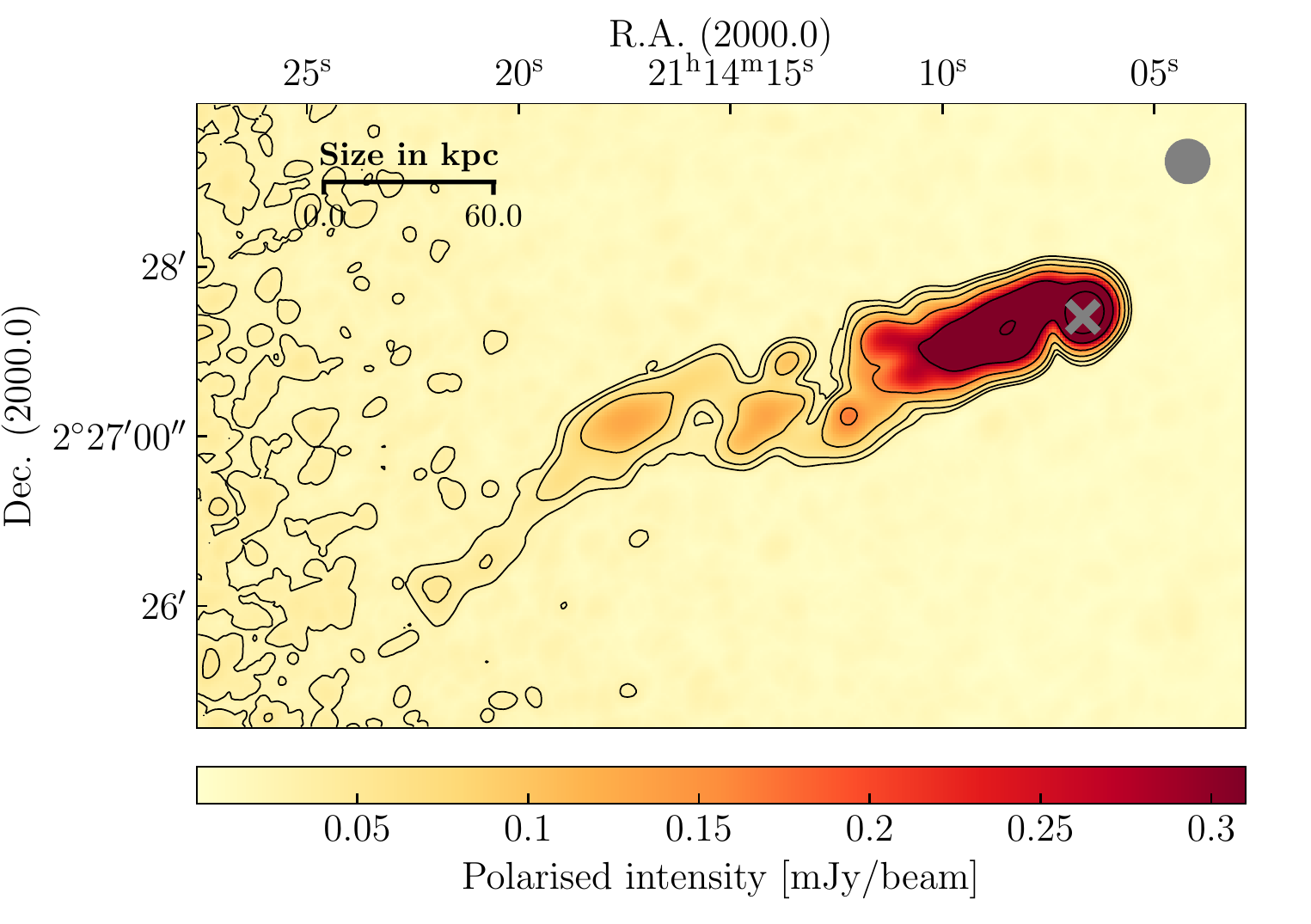}
        \includegraphics[width=0.49\textwidth]{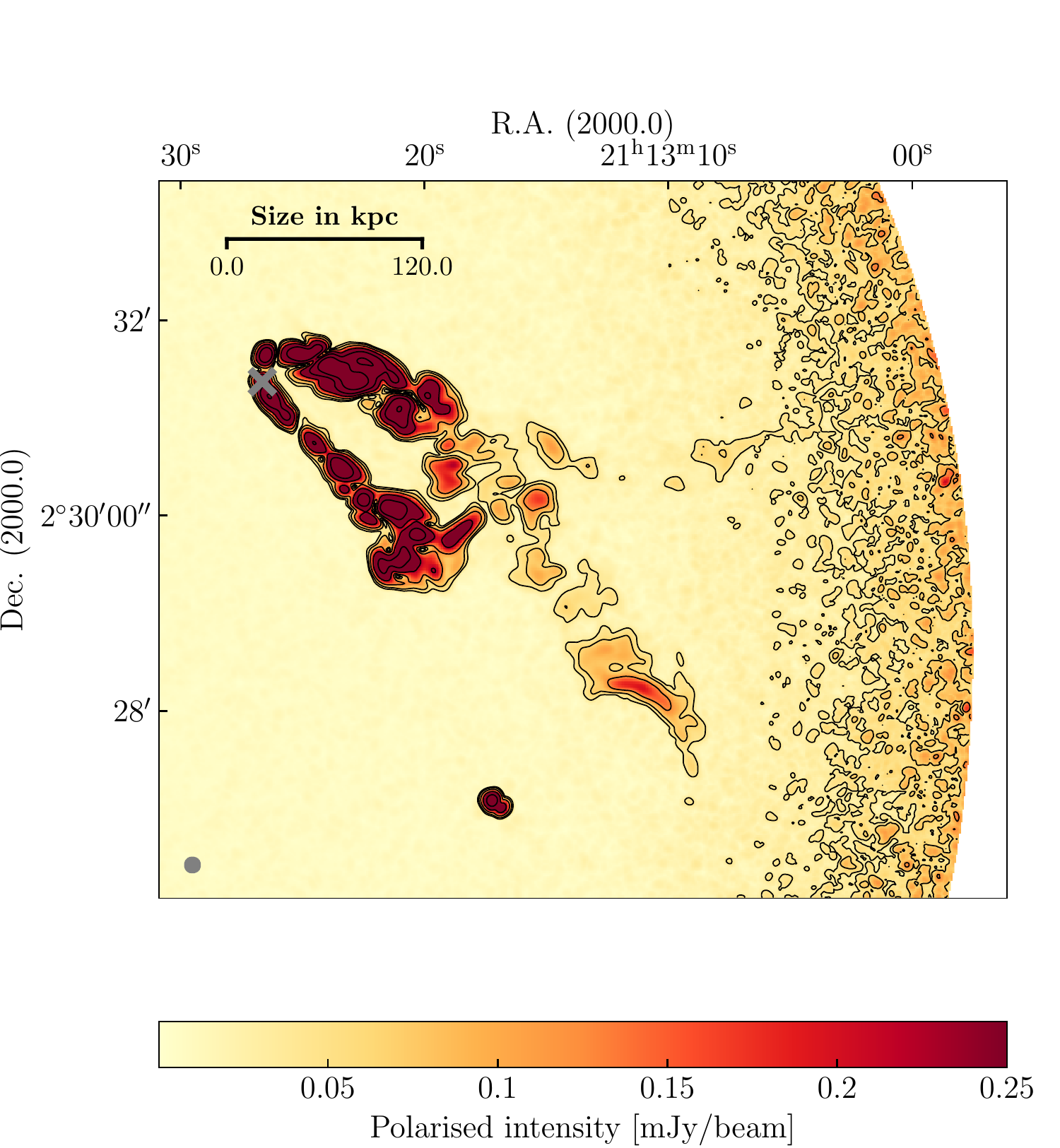}
    \includegraphics[width=0.49\textwidth]{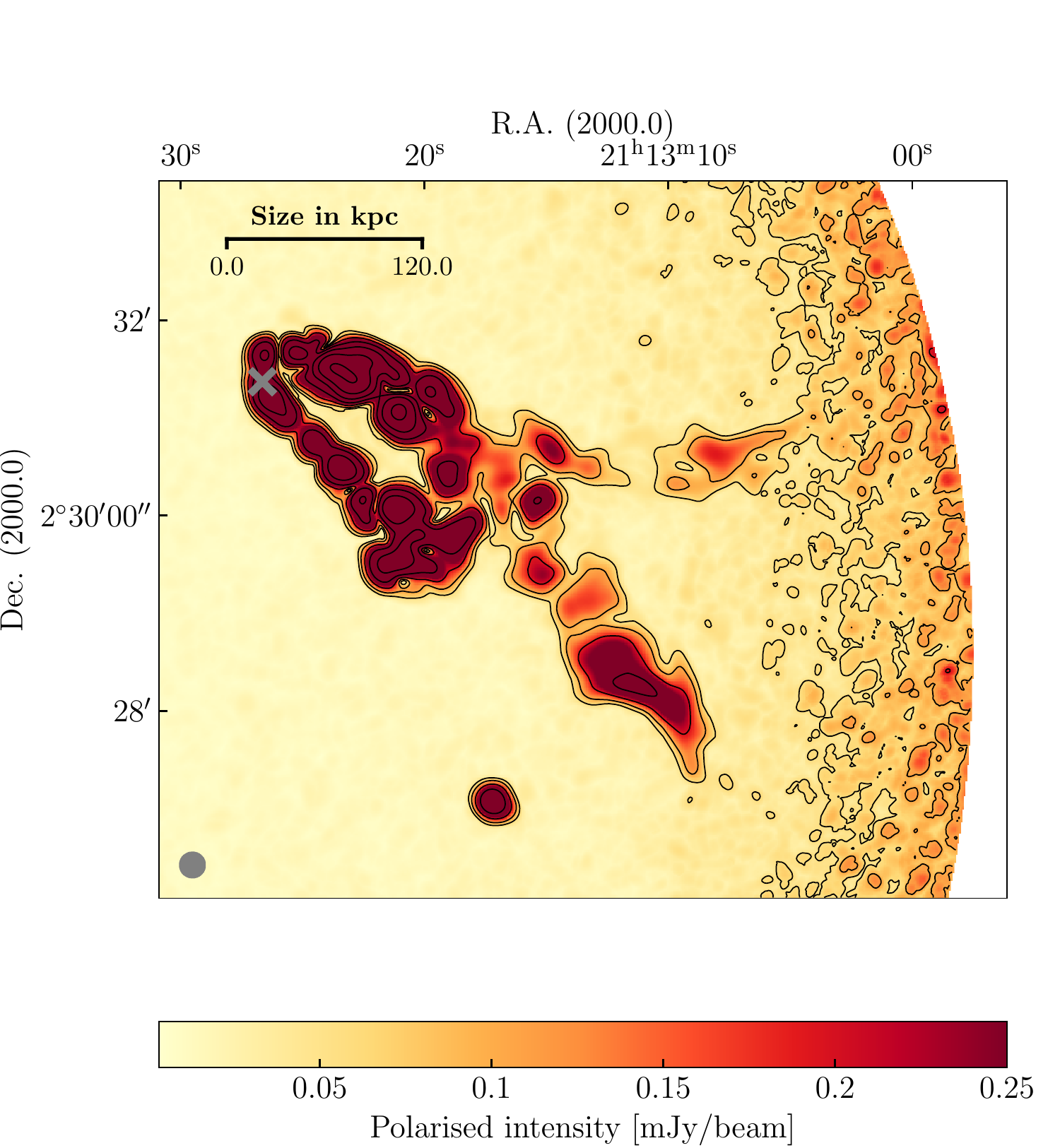}
    \caption{Higher resolution polarized intensity maps of HTI (top left) and HTII (bottom left) at $8.7\arcsec$ circular beam (grey circle) and lower resolution maps on the right for comparison. The position of the head is marked with a grey cross and the polarization contours of the higher and lower resolution map are shown in black.}
    \label{fig:app_high}
\end{figure*}

The polarized emission at $15\arcsec\times15\arcsec$ resolution (Figs.~\ref{fig:PolarizedIntensityHT1} and \ref{fig:PolarizedIntensityHT2}) is found to show several substructures that could not be identified in total intensity (Figs.~\ref{fig:TotalIntensityHT1} and \ref{fig:TotalIntensityHT2}). To test whether this is a{n} intrinsic characteristic of the galaxy jets we produced the highest possible resolution with a{n} $8.7\arcsec$ circular beam. The polarized emission results are shown in Fig.~\ref{fig:app_high} and the contours of the lower resolution images are shown in green for comparison. By increasing the resolution by a factor of two we {lose} a lot of sensitivity in the galaxy jets but nevertheless{,} the bright, clumpy features identified in the lower resolution data are also detected at higher resolution. Only marginal changes can be identified, however, the depolarization effects possibly occur at even smaller scales and higher resolution deep observations are needed to exclude the depolarization as reasoning.



\bsp	
\label{lastpage}
\end{document}